\documentclass[journal]{IEEEtran}

\usepackage{cite}
\usepackage{amsmath,amssymb,amsfonts}
\usepackage{bm}
\usepackage{graphicx}
\usepackage{textcomp}
\usepackage{booktabs}
\usepackage{multirow}
\usepackage{makecell}
\usepackage{array}
\usepackage[caption=false,font=footnotesize]{subfig}
\usepackage{url}
\usepackage{cleveref}
\usepackage{placeins}
\usepackage{xcolor}

\graphicspath{{Figs/}}

\crefname{figure}{Fig.}{Figs.}
\Crefname{figure}{Fig.}{Figs.}
\crefname{table}{Table}{Tables}
\Crefname{table}{Table}{Tables}
\crefname{section}{Section}{Sections}
\Crefname{section}{Section}{Sections}
\crefname{equation}{equation}{equations}
\Crefname{equation}{Equation}{Equations}

\newcommand{\vect}[1]{\mathbf{#1}}
\newcommand{\mat}[1]{\mathbf{#1}}
\newcommand{\norm}[1]{\left\lVert#1\right\rVert}
\renewcommand{\j}{\mathrm{j}}
\newcommand{\rx}{\text{Rx}}
\newcommand{\tx}{\text{Tx}}
\newcommand{\RIS}{\text{RIS}}
\newcommand{\ele}{\text{ele}}
\newcommand{\tot}{\text{tot}}
\newcommand{\gen}{\text{gen}}
\newcommand{\cali}{\text{cal}}
\newcommand{\tgt}{\text{tgt}}
\newcommand{\sca}{\text{sca}}
\newcommand{\quant}{\mathrm{Q}}
\newcommand{\FW}{\text{ref}}
\newcommand{\BG}{\text{bg}}
\newcommand{\by}{\text{bypass}}
\newcommand{\pathset}{\mathcal{P}}
\newcommand{\pathsetB}{\mathcal{Q}}
\newcommand{\pathsetRIS}{\mathbb{P}_{\RIS}}
\newcommand{\pathsetBY}{\mathbb{P}_{\by}}
\newcommand{\varMethod}{\mathcal{M}}
\newcommand{\robs}{\vect{r}}

\begin{document}

\title{Full-Wave-Calibrated Element-Wise RIS Modeling With Cross-Aperture Coefficient Transfer for Multipath Channel Prediction}

\author{Yuxuan Ding,~\IEEEmembership{Member,~IEEE}, and Minseok Kim,~\IEEEmembership{Senior Member,~IEEE}
\thanks{This work was supported by the Ministry of Internal Affairs and Communications (MIC)/FORWARD under Grant JPMI240410003, Japan. (\textit{Corresponding author: Minseok Kim.})}
\thanks{Yuxuan Ding and Minseok Kim are with the Institute of Science and Technology, Niigata University, Niigata 950-2181, Japan (e-mail: yuxuan.ding.25@niigata-u.ac.jp; mskim@eng.niigata-u.ac.jp).}%
}

\maketitle

\begin{abstract}
Practical reconfigurable intelligent surfaces (RISs) can exhibit deterministic parasitic scattering that is not captured by idealized element-wise models. As a result, such models may overestimate the gain of the intended RIS-assisted path and bias multipath prediction. This paper develops a full-wave-calibrated element-wise model using three Bragg-order basis functions to represent the intended and dominant parasitic scattering components. Environmental multipath is incorporated by unfolding ray-tracing (RT)-identified RIS--Rx reflection sequences with image theory into path-dependent image points, allowing direct and reflected RIS-assisted paths to be evaluated by the same per-element kernel and coherently combined with Tx--Rx bypass paths. 
To reduce the full-wave calibration burden for large RISs under a prescribed focusing configuration, only the three Bragg-order coefficients are transferred from a 25$\times$25 calibration aperture, while the target-aperture basis functions and geometry are recomputed. 
At 154~GHz, transferred coefficients keep the intended-order errors within 0.8~dB for the $50\times50$ and $100\times100$ RISs, while reducing the one-time calibration from 7.56 to 1.08~h relative to direct $100\times100$ calibration. For multipath validation with the $50\times50$ RIS, the calibrated model with the same transferred coefficients keeps the nominal-Rx gain error within 0.88~dB across four PEC-reflector configurations, against 1.75--5.51~dB for the general model, and reduces it from 6.49 to 0.23~dB in the scaled indoor environment.
\end{abstract}


%===========================================================================================%
%                                        New Section                                        %
%===========================================================================================%

\section{Introduction}
\label{sec:introduction}

\IEEEPARstart{R}{econfigurable} intelligent surfaces (RISs) provide engineered electromagnetic interfaces for coverage enhancement, blockage mitigation, and site-specific radio-environment design~\cite{basar2019wireless,direnzo2020smart,wu2020smart}. RISs are particularly attractive at millimeter-wave and sub-terahertz frequencies, where blockage, wavefront curvature, and small path-length variations become increasingly important while electrically large apertures can be realized within a compact physical area~\cite{kim2026thz}. Predictive deployment analysis therefore requires propagation models that preserve the relevant electromagnetic scattering behavior while remaining computationally tractable over many receiver (Rx) locations and environmental configurations~\cite{wang2026terahertz}.

A coherent element-wise formulation provides a natural basis for RIS-assisted channel modeling~\cite{ozdogan2020intelligent,tang2021wireless}. By evaluating the transmitter (Tx)-to-element and element-to-Rx distances separately for every RIS element, it accounts for path-length variation and wavefront curvature across the finite aperture without a far-field RIS approximation; the same formulation can therefore be evaluated in both radiative near- and far-field regions. In its conventional form, referred to here as the \emph{general model}, the RIS response is prescribed by idealized local reflection coefficients according to the designed quantized phase profile. This formulation captures the intended coherent response but does not, in general, reproduce the complete scattering behavior of a practical RIS.

Practical RIS implementations can exhibit deterministic scattering components beyond the intended anomalous-reflection beam. In particular, the recurrent electromagnetic structure of a configured RIS can support coherent phase-matched diffraction orders, while embedded-element interactions, mixed-state coupling, structural currents, and finite-aperture effects modify their complex excitation~\cite{degli-esposti2022reradiation,li2024allangle,merluzzi2024anomalous,cardoso2025indoor}. In the authors' preliminary study~\cite{dingimproved}, electromagnetically distinct RIS realizations exhibited a qualitatively similar three-component structure comprising the intended anomalous-reflection beam, a residual specular component, and a dominant higher-order Bragg-type component. In multipath environments, the residual components can illuminate surrounding objects outside the intended beam and form additional RIS-assisted paths toward the Rx~\cite{mao2026indoor}. Neglecting them can therefore bias both the intended-path gain and the coherent multipath channel.

Recent studies have addressed complementary aspects of electromagnetic fidelity and propagation efficiency in RIS channel modeling. Wijekoon \emph{et al.}~\cite{wijekoon2026physicallyconsistent} formulate a non-local RIS using multiport $S$-parameters to account for mutual coupling, multiple reflections, radiation characteristics, and tunable states, although residual scattering is not retained in the subsequent system-level analysis. At the propagation level, Pyhtil\"a \emph{et al.}~\cite{pyhtilae2023ray} integrate an idealized phased-array RIS model into a GPU-accelerated deterministic ray-tracing (RT) framework for urban coverage prediction. The RIS response is prescribed by element phase shifts with inter-element coupling neglected, and the reported simulations employ a far-field RIS representation. Hao \emph{et al.}~\cite{hao2025analysis} incorporate realistic anomalous reflectors into RT using imported far-field scattering patterns and validate the resulting channel predictions experimentally. While this approach preserves the characterized angular response, it requires pattern data over the frequencies and observation angles of interest and does not directly represent radiative near-field interactions.
Vitucci \emph{et al.}~\cite{vitucci2024efficient} develop a macroscopic ray-based RIS model supporting multiple reradiation modes in the radiative near and far fields. The RIS is represented as a slowly modulated homogenized surface rather than by its individual physical unit cells, and the reradiated field is represented by reflected and diffracted rays for propagation analysis.
This provides an efficient near-field-capable representation for RT, but implementation-dependent parasitic scattering is not predicted directly from the actual discrete element-state arrangement.

A complementary need therefore remains for a physically grounded element-wise model that can incorporate implementation-dependent scattering while retaining the actual RIS state distribution and element-specific propagation.
The calibrated RIS response should also be reusable across different RIS--Rx environments without environment-specific recalibration. Moreover, when full-wave data are used for calibration, direct full-wave characterization of an electrically large RIS can become computationally prohibitive. These considerations motivate the framework developed in this paper.

\begin{figure*}[t]
\centering
\includegraphics[width=16.8cm]{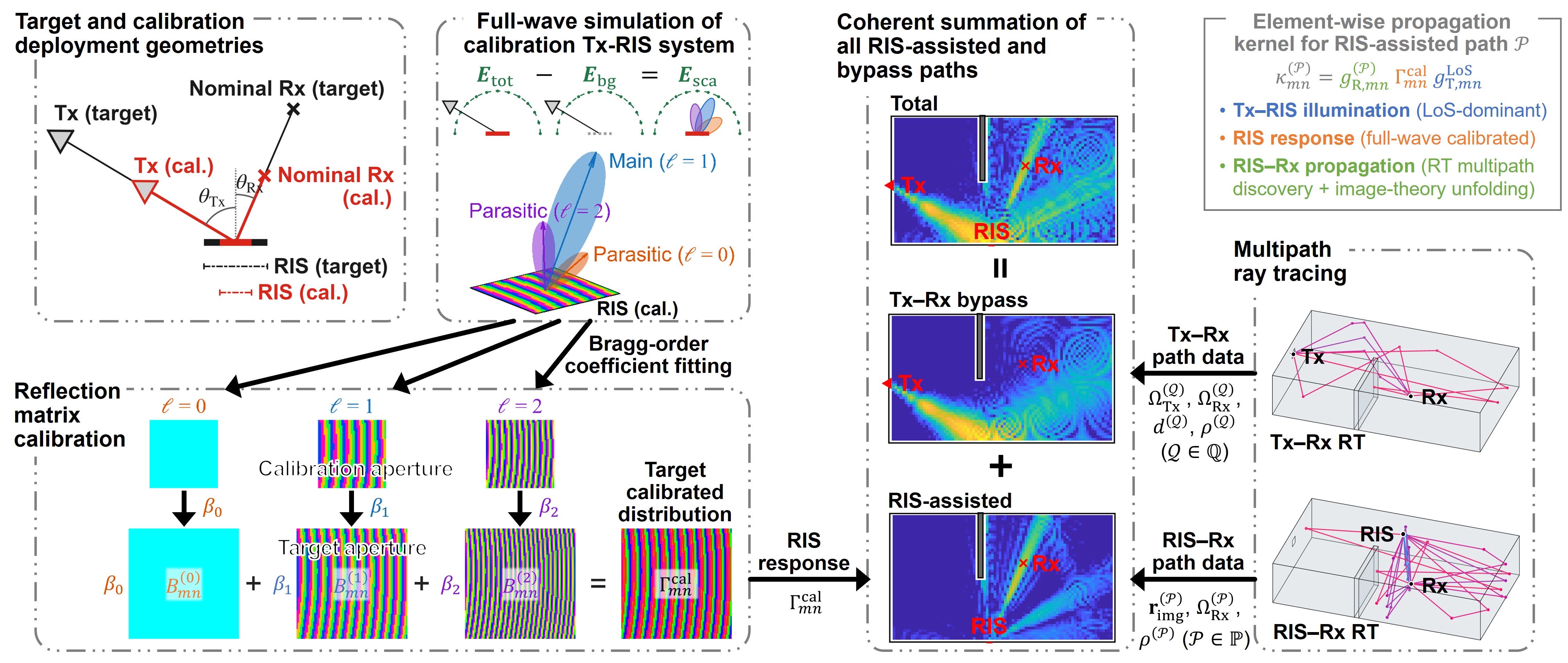}
\caption{Workflow of the proposed model. A reduced calibration geometry is constructed from the target deployment by preserving the Tx/Rx directions and distance-to-aperture ratios. A compact full-wave simulation of the calibration aperture yields the Bragg-order coefficients $\beta_0$, $\beta_1$, and $\beta_2$, which are used to construct the calibrated reflection distribution $\Gamma_{mn}^{\cali}$ for the target aperture. Ray tracing (RT) identifies the valid RIS--Rx and Tx--Rx reflection sequences, denoted by $\pathset$ and $\pathsetB$, respectively. Each RIS--Rx sequence is unfolded by image theory to obtain the corresponding image point $\vect r_{\mathrm{img}}^{(\pathset)}$ for element-wise RIS-assisted channel evaluation. The resulting RIS-assisted contributions are then coherently combined with the Tx--Rx bypass contributions.}
% Suggested labels inside workflow.jpg:
%   kappa_{mn}^{(P)} = g_{T,mn}^{LoS} Gamma_{mn}^{cal} g_{R,mn}^{(P)}
%   Tx--RIS illumination (LoS-dominant)
%   RIS response (full-wave calibrated)
%   RIS--Rx propagation (RT multipath discovery + image-theory unfolding)
%   Unfolded RIS--Rx path data: r_img^{(P)}, rho^{(P)}
%   Tx--Rx bypass path data: d^{(Q)}, Omega_Tx^{(Q)}, Omega_Rx^{(Q)}, rho^{(Q)}
\label{fig:workflow}
\end{figure*}

This paper considers a prescribed RIS state matrix under fixed, highly directional Tx--RIS illumination, where the line-of-sight (LoS) component is dominant. 
For full-wave characterization, the configured RIS is implemented in HFSS as an equivalent fixed-state anomalous reflector with the same unit-cell geometry and phase-state matrix. 
Isolated-RIS simulations are used to quantify the accuracy of the calibrated scattering representation and to assess cross-aperture transfer. Perfect-electric-conductor (PEC)-reflector configurations and an approximately $1/16$-scale reinforced-concrete (RCC) room are then used to validate the multipath formulation without environment-specific recalibration. \Cref{fig:workflow} summarizes the overall procedure.

The main contributions beyond the authors' preliminary study~\cite{dingimproved} are summarized as follows.
\begin{itemize}
    \item \textit{Full-wave-calibrated element-wise RIS model:}
    A three-Bragg-order representation is introduced to correct the dominant deterministic scattering components missed by the ideal phase-only general model. The complex order coefficients are extracted from isolated-RIS full-wave fields, while the RIS phase distribution, finite-aperture geometry, and Tx-to-element and element-to-observation propagation terms remain explicit in the element-wise channel model.
    \item \textit{Image-theory integration of environmental multipath:}
    Reflection sequences obtained from deterministic RT are mapped by image-theory unfolding~\cite{esposti2021ray,yang20266g,ding2026twoway} to path-dependent image points. This allows direct and reflected RIS--Rx paths to be evaluated by the same element-wise formulation, with per-element propagation distances retained, and enables coherent combination with Tx--Rx bypass contributions.
    \item \textit{Cross-aperture coefficient transfer for scalable calibration:}
	The three complex Bragg-order coefficients are separated from the aperture-dependent basis functions and propagation geometry. A reduced calibration geometry preserves the RIS-center-to-Tx and RIS-center-to-nominal-Rx directions and distance-to-aperture ratios of the target deployment. Only the coefficient vector is transferred to the larger target RIS, while the target-aperture phase profile, basis functions, and element-wise geometry are regenerated. This reduces the full-wave calibration burden for electrically large RISs while retaining substantial accuracy improvements over the general model.
\end{itemize}

The remainder of this paper is organized as follows. \Cref{sec:model} establishes the element-wise RIS formulation and the Bragg-order calibrated response. \Cref{sec:calibration} presents coefficient extraction and then examines cross-aperture transfer. \Cref{sec:rt} develops the image-theory coupling between RT-discovered reflection paths and the element-wise RIS model. \Cref{sec:validation} validates the resulting multipath channel-gain prediction for a $50\times50$ RIS with PEC reflectors and in the scaled RCC room, and compares the computational cost with the corresponding HFSS references. \Cref{sec:conclusion} concludes the paper.

%%%%%%%%%%%%%%%%%%%%%%%%%%%%%%%%%%%%%%%%%%%%%%%%%%%%%%%%%%%%%%%%%%%%%%%
%%%%%%%%%%%%%%%%%%%%%%%%%%%%%%%%%%%%%%%%%%%%%%%%%%%%%%%%%%%%%%%%%%%%%%%

%===========================================================================================%
%                                        New Section                                        %
%===========================================================================================%
\section{Element-Wise RIS Model and Bragg-Order Representation}
\label{sec:model}

\subsection{General Element-Wise Formulation}
\label{sec:general_model}

Consider a planar RIS on the $z=0$ plane with $M\times N$ elements and spacings $d_x$ and $d_y$ along the $x$- and $y$-directions, respectively. The position of element $(m,n)$ is $\vect r_{mn}=[x_{mn},y_{mn},0]^{\mathsf T}$. The physical Tx position and the nominal Rx position used to synthesize the RIS phase profile are denoted by $\vect r_{\tx}$ and $\vect r_{\rx}$, respectively. An arbitrary physical observation point is denoted by $\robs$. Throughout this work, the Tx--RIS link is assumed LoS-dominant, so the physical Tx position $\vect r_{\tx}$ is used directly in all Tx-side element-wise propagation terms~\cite{ding2026twoway}.

The continuous phase required to focus the RIS-reflected field at the nominal Rx $\vect r_{\rx}$ under illumination from $\vect r_{\tx}$ is
\begin{equation}
\Phi_{mn}
=
k\left(
\norm{\vect r_{\tx}-\vect r_{mn}}
+
\norm{\vect r_{\rx}-\vect r_{mn}}
\right),
\label{eq:continuous_phase}
\end{equation}
where $k=2\pi/\lambda$ and $\lambda$ is the free-space wavelength. For $N_{\mathrm{bit}}$-bit phase quantization,
\begin{equation}
\Phi_{mn}^{\quant}
=
\frac{2\pi}{2^{N_{\mathrm{bit}}}}
\left[
\operatorname{round}\!\left(
\frac{2^{N_{\mathrm{bit}}}}{2\pi}\Phi_{mn}
\right)
\bmod 2^{N_{\mathrm{bit}}}
\right],
\label{eq:quantized_phase}
\end{equation}
and the ideal local reflection coefficient is
\begin{equation}
\Gamma_{mn}^{\gen}
=
\Gamma_0\exp\!\left(\j\Phi_{mn}^{\quant}\right),
\label{eq:general_gamma}
\end{equation}
where $\Gamma_0$ is a uniform reflection magnitude.

For an arbitrary physical observation point $\robs$, the Tx--RIS--Rx channel coefficient associated with an aperture distribution $\Gamma_{mn}$, in the absence of additional environmental reflections, is
\begin{equation}
\begin{split}
h(\robs;\Gamma_{mn})
={}&
C\sum_{m,n}
\frac{A_{mn}(\robs)}
{\norm{\vect r_{\tx}-\vect r_{mn}}\norm{\robs-\vect r_{mn}}}
\,\Gamma_{mn}\\
&\times
\exp\!\left[
-\j k
\left(
\norm{\vect r_{\tx}-\vect r_{mn}}
+
\norm{\robs-\vect r_{mn}}
\right)
\right],
\end{split}
\label{eq:element_kernel}
\end{equation}
where
\begin{equation}
A_{mn}(\robs)
=
\sqrt{
F_{mn}^{\tx}(\vect r_{\tx})
F_{mn}^{\ele}(\vect r_{\tx})
F_{mn}^{\ele}(\robs)
F_{mn}^{\rx}(\robs)
}
\label{eq:pattern_factor}
\end{equation}
and
\begin{equation}
C=
\sqrt{
G_{\tx}G_{\rx}G_{\ele}d_xd_y\lambda^2/(4\pi)^3
}.
\label{eq:normalization}
\end{equation}
Here, $G_{\tx}$, $G_{\rx}$, and $G_{\ele}$ are reference power gains in linear scale, and $F_{mn}^{\tx}$, $F_{mn}^{\rx}$, and $F_{mn}^{\ele}$ are the corresponding normalized power-pattern factors of the Tx, Rx, and RIS element. Because the Tx-to-element and element-to-observation distances are evaluated separately for every element, \eqref{eq:element_kernel} accounts for distance variation and wavefront curvature across the finite aperture without invoking a far-field RIS approximation. Environmental reflection and visibility factors are introduced in \cref{sec:rt}. The conventional general model follows by setting $\Gamma_{mn}=\Gamma_{mn}^{\gen}$.

\subsection{Practical RIS Realization}
\label{sec:ris_implementation}

An RIS changes its scattering response by reconfiguring the discrete states of its elements. 
Once a particular RIS state matrix is fixed, its electromagnetic response at the operating frequency can be characterized using a static surface implementing the same unit-cell states. Accordingly, the configured RIS is realized in HFSS as a fixed-state anomalous reflector with the same unit-cell geometry and phase-state matrix. This static realization is used only for full-wave characterization; throughout the subsequent modeling and channel analysis, the surface is referred to as the RIS.

The practical RIS realization uses the grounded metal-patch unit cell reported in~\cite{dingimproved}. Each element consists of a square metallic patch on a 0.14-mm-thick polyethylene terephthalate (PET) substrate with relative permittivity $\varepsilon_{\mathrm r}=3.28$ and loss tangent $\tan\delta=0.00136$, backed by a metallic ground plane. The element period is $\lambda/4$. At 154~GHz, four physical states implement the 2-bit phase library used by all apertures; their patch dimensions and nominal reflection phases are listed in \cref{tab:parameters}.

The ideal 2-bit phase states in \eqref{eq:quantized_phase} are $\{0^\circ,90^\circ,180^\circ,270^\circ\}$, whereas periodic-boundary-condition simulations of the implemented unit cell yield $\{39^\circ,124^\circ,214^\circ,299^\circ\}$. After removing the common offset of $34^\circ$, the residual state-dependent deviations are only $\{+5^\circ,0^\circ,0^\circ,-5^\circ\}$. Such small deviations cannot account for the pronounced discrete parasitic lobes observed in the full-wave results, so the calibration introduced below addresses the additional coherent Bragg-order scattering of the practical RIS rather than the imperfection of the individual states.

The RIS comprises a periodic element lattice carrying a spatially structured quantized phase distribution, which enables its full-wave response to couple energy into discrete phase-matched Bragg diffraction orders. Embedded-element interactions, mixed-state coupling, structural currents, local illumination, and edge-associated effects affect the complex excitation of these orders~\cite{wijekoon2026physicallyconsistent}. The graphene realization reported in~\cite{dingimproved} is not used in the environmental simulations; it provides independent prior evidence that deterministic Bragg-type parasitic scattering is not unique to the metal-patch implementation.

\begin{figure}[t]
\centering
\subfloat[]{\includegraphics[width=3.8cm]{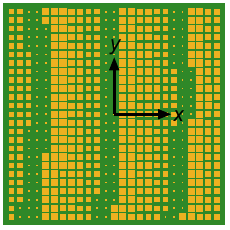}}
\hfil
\subfloat[]{\includegraphics[width=3.8cm]{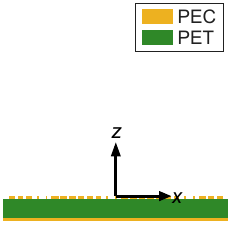}}
\caption{Structure of the 2-bit-quantized variable-size metal-patch $25\times25$ RIS: (a) front view and (b) side view.}
\label{fig:RISstructure}
\end{figure}

%Three square RIS apertures, $25\times25$, $50\times50$, and $100\times100$, are characterized. They share the same unit-cell implementation, carrier frequency, polarization, and Tx/Rx angular geometry. \Cref{fig:RISstructure} shows the $25\times25$ realization. The Tx--RIS and RIS--nominal-Rx distances are scaled in proportion to the aperture size, as listed in \cref{tab:coefficients}. This scaling preserves the angular geometry and, for an ideal coherently reflecting RIS, keeps the nominal-Rx path gain approximately invariant, because the increase in aperture area is compensated by the increased propagation distances. For each aperture, the isolated Tx--RIS subsystem is simulated in HFSS using the finite-element boundary-integral (FE-BI) method.
Three square RIS apertures, $25\times25$, $50\times50$, and $100\times100$, are characterized. They share the same unit-cell implementation, carrier frequency, polarization, and Tx/Rx angular geometry. \Cref{fig:RISstructure} shows the $25\times25$ realization. The three geometries are constructed according to the reduced-aperture calibration rule introduced in \cref{sec:transfer_condition}.

%================================================================================%
\begin{table}[t]
\caption{Common parameters for isolated-RIS characterization and channel calculations.}
\label{tab:parameters}
\centering
\small
\begin{tabular}{@{}p{0.413\columnwidth}p{0.532\columnwidth}@{}}
\toprule
Parameter & Value \\
\midrule
Carrier frequency & 154~GHz ($\lambda=1.947$~mm) \\
Phase quantization & $N_{\mathrm{bit}}=2$ \\
Element type & Variable-size metal patch on PET substrate~\cite{dingimproved} \\
Reflection states & $1\angle39^\circ,1\angle124^\circ,1\angle214^\circ,1\angle299^\circ$~\cite{dingimproved} \\
Patch side lengths & $0.45$, $0.38$, $0.33$, $0.02$~mm~\cite{dingimproved} \\
Element spacing & $d_x=d_y=\lambda/4$ \\
RIS element gain in model & $G_{\ele}=0$~dBi \\
RIS aperture in multipath validation & $50\times50$ \\
Tx antenna & $y$-polarized pyramidal horn antenna, $G_{\tx} = 26$~dBi\\
Rx antenna & $y$-polarized omnidirectional antenna, $G_{\rx} = 0$~dBi \\
Tx power & $P_{\mathrm t}=1$~W \\
RIS-to-Tx direction & $(\theta,\phi)=(59.25^\circ,175.39^\circ)$ \\
RIS-to-Tx distance & $45.23$~cm for $50\times50$ RIS \\
RIS-to-nominal-Rx direction & $(\theta,\phi)=(23.41^\circ,-8.13^\circ)$ \\
RIS-to-nominal-Rx distance & $33.37$~cm for $50\times50$ RIS \\
\bottomrule
\end{tabular}
\end{table}
%================================================================================%

%%%%%%%%%%%%%%%%%%%%%%%%%%%%%%%%%%%%%%%%%%%%%%%%%%%%%%%%%%%%%%%%%%%%%%%
%%%%%%%%%%%%%%%%%%%%%%%%%%%%%%%%%%%%%%%%%%%%%%%%%%%%%%%%%%%%%%%%%%%%%%%

\subsection{Bragg-Order Basis and Calibrated Reflection Distribution}
\label{sec:bragg_basis}

\begin{figure*}[t]
\centering
\newcommand{\transferfigheight}{4.0cm}
\newcommand{\transferfigcola}{0.26\textwidth}
\newcommand{\transferfigcolb}{\transferfigcola}
\newcommand{\transferfigcolc}{0.34\textwidth}
\subfloat[$25\times25$: full-wave reference.]
{\makebox[\transferfigcola]{\includegraphics[height=\transferfigheight]{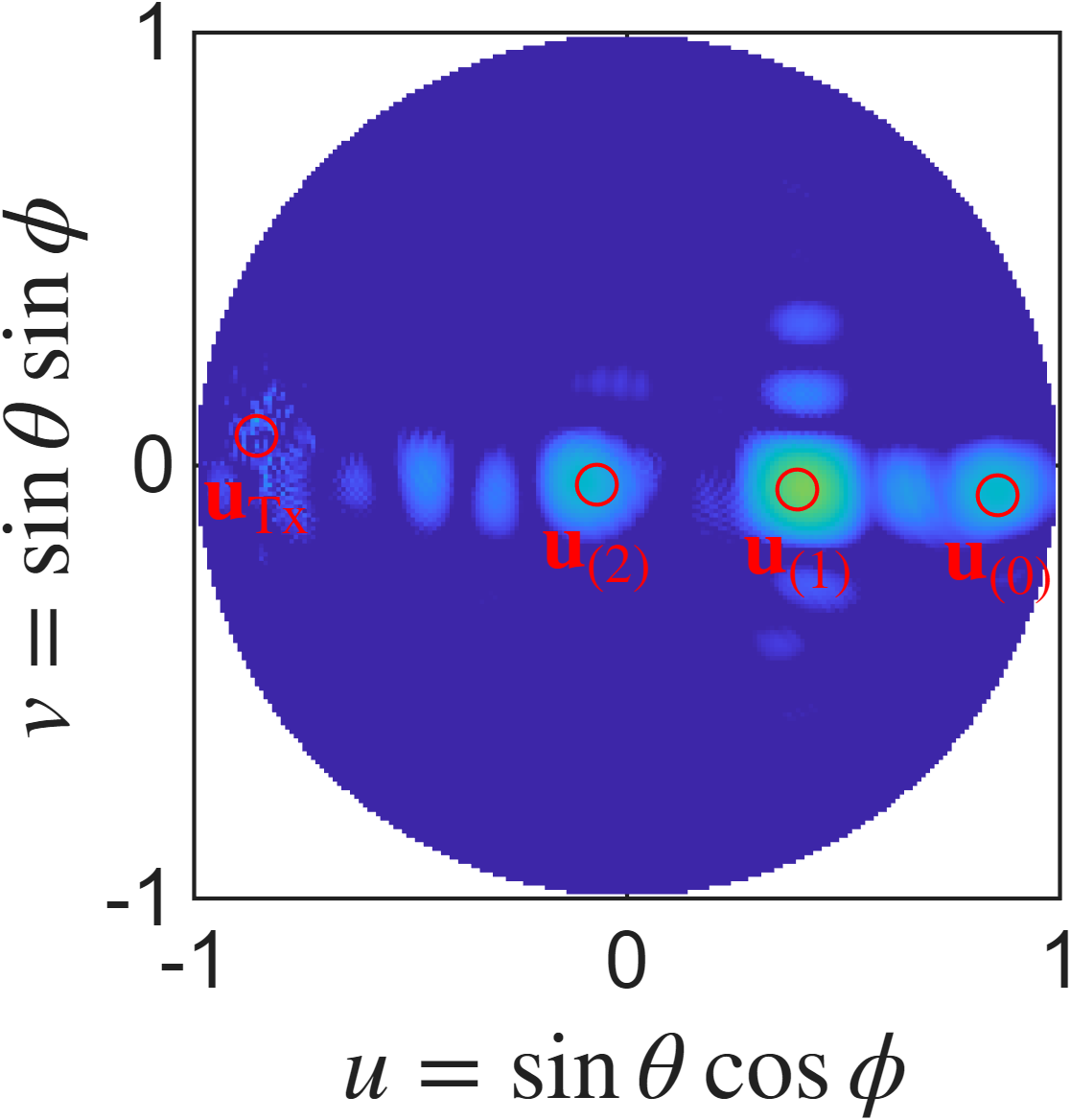}}}
\hfil
\subfloat[$25\times25$: general model.]
{\makebox[\transferfigcolb]{\includegraphics[height=\transferfigheight]{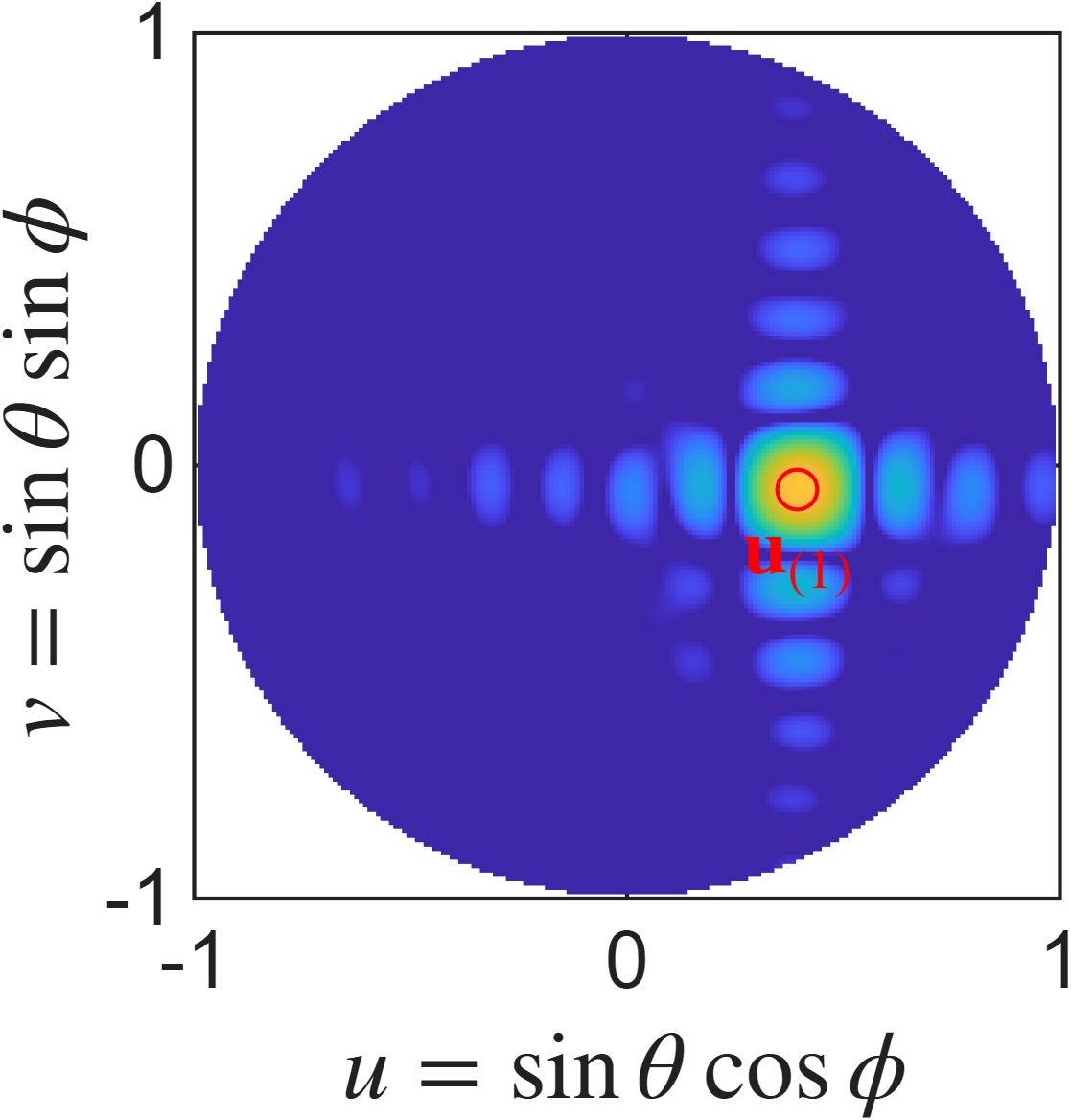}}}
\hfil
\subfloat[$25\times25$: calibrated ($25\times25$ coefficients).]
{\makebox[\transferfigcolc]{\includegraphics[height=\transferfigheight]{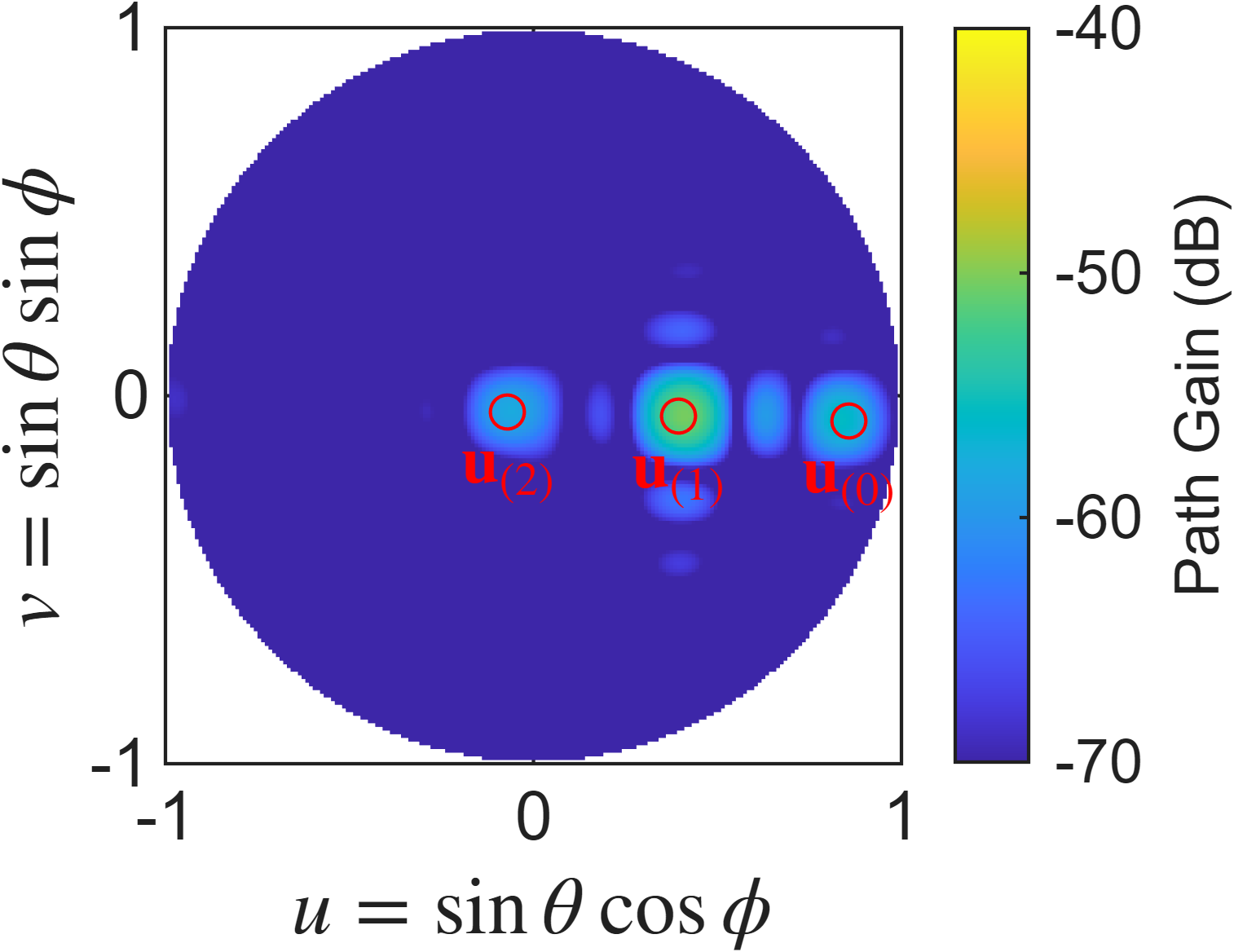}}}
\\
\subfloat[$50\times50$: full-wave reference.]
{\makebox[\transferfigcola]{\includegraphics[height=\transferfigheight]{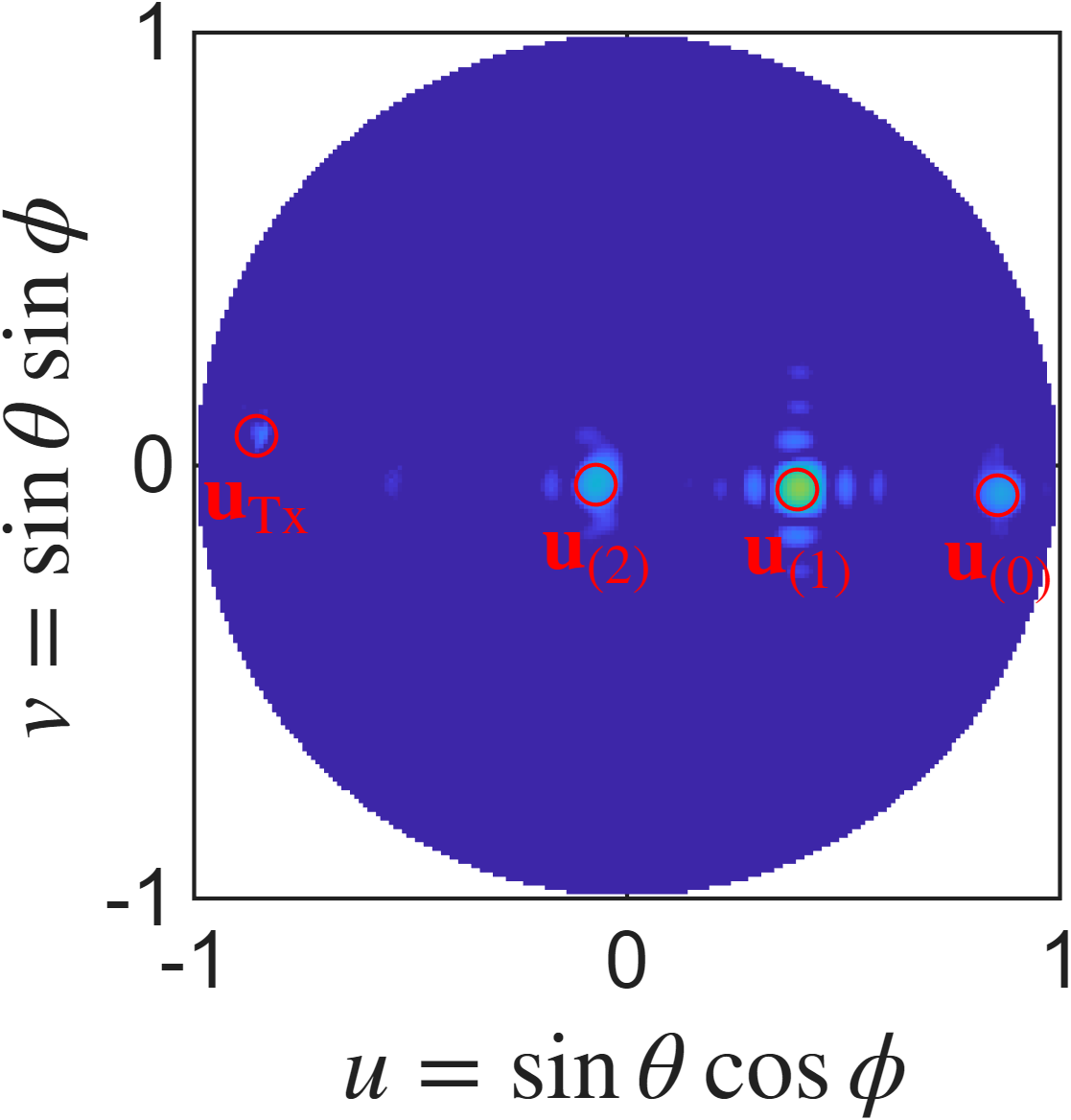}}}
\hfil
\subfloat[$50\times50$: general model.]
{\makebox[\transferfigcolb]{\includegraphics[height=\transferfigheight]{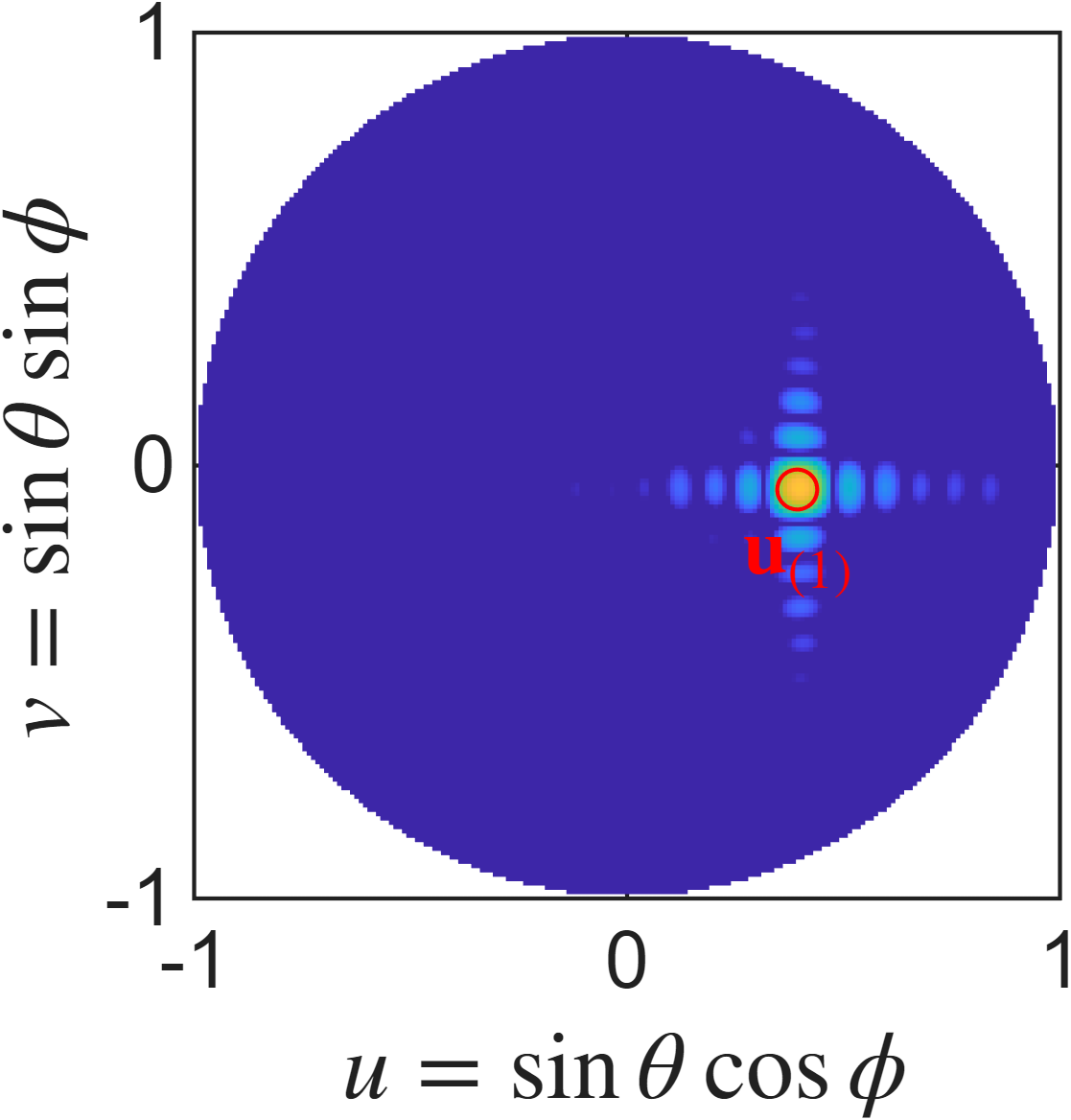}}}
\hfil
\subfloat[$50\times50$: transferred calibration.]
{\makebox[\transferfigcolc]{\includegraphics[height=\transferfigheight]{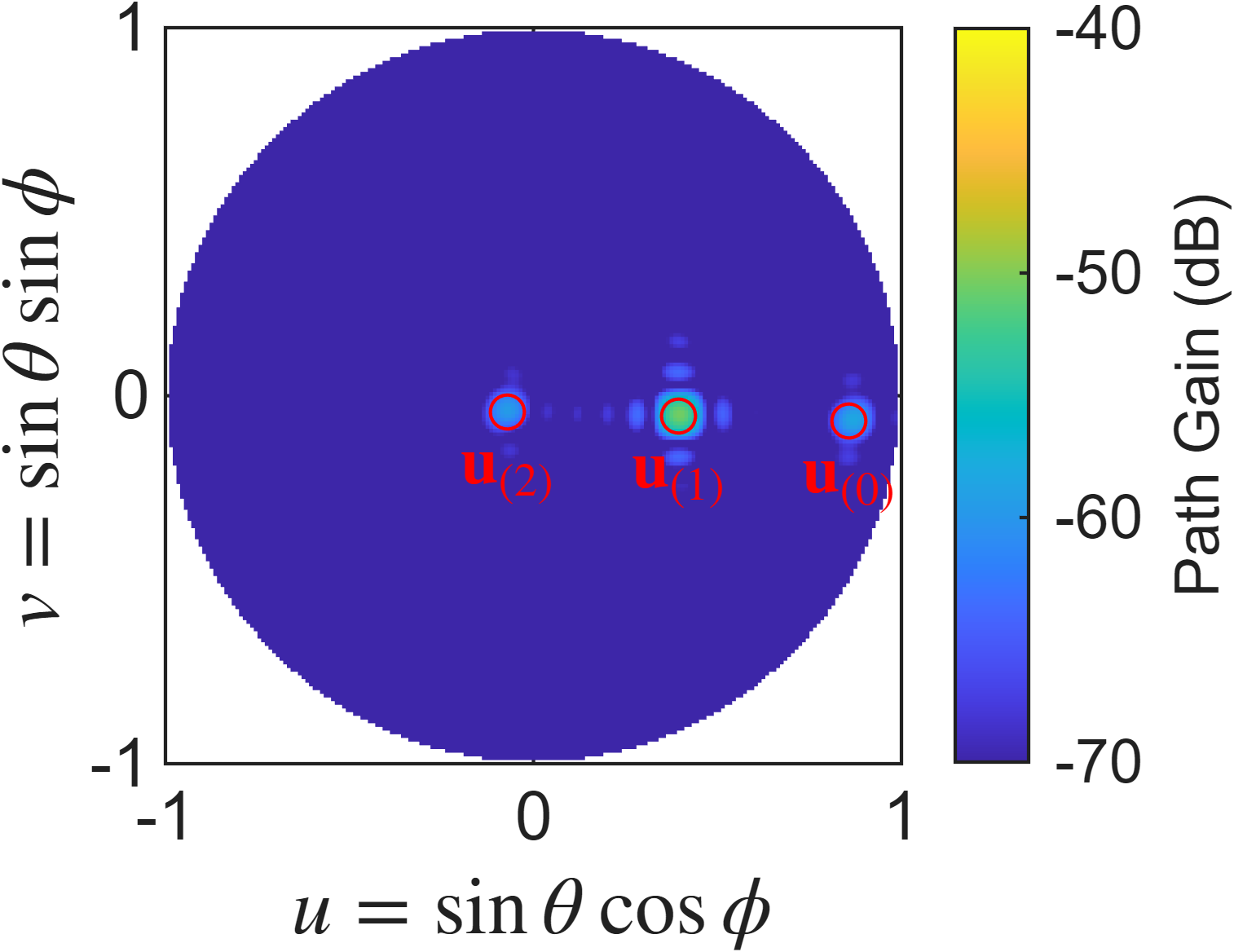}}}
\\
\subfloat[$100\times100$: full-wave reference.]
{\makebox[\transferfigcola]{\includegraphics[height=\transferfigheight]{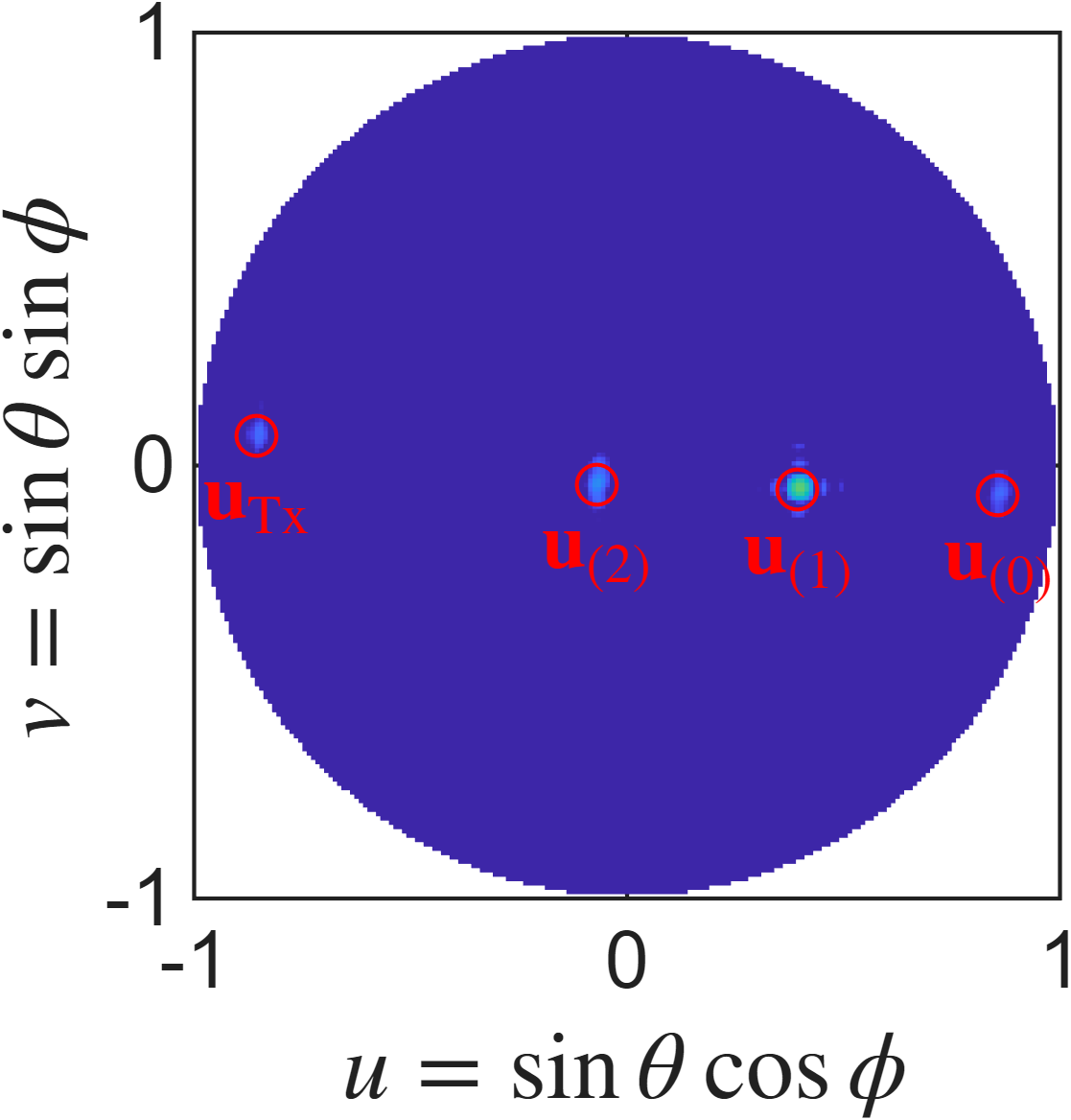}}}
\hfil
\subfloat[$100\times100$: general model.]
{\makebox[\transferfigcolb]{\includegraphics[height=\transferfigheight]{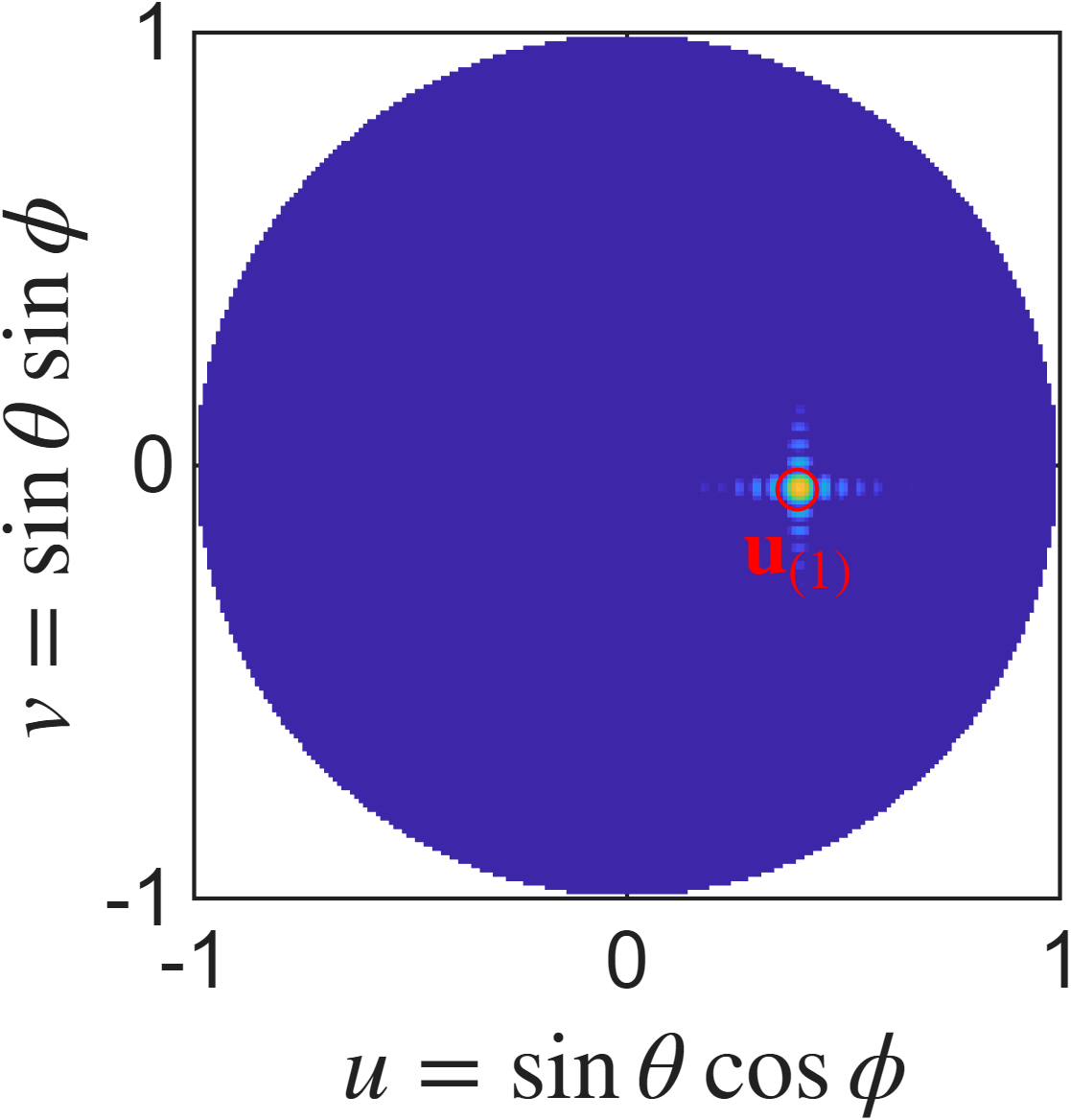}}}
\hfil
\subfloat[$100\times100$: transferred calibration.]
{\makebox[\transferfigcolc]{\includegraphics[height=\transferfigheight]{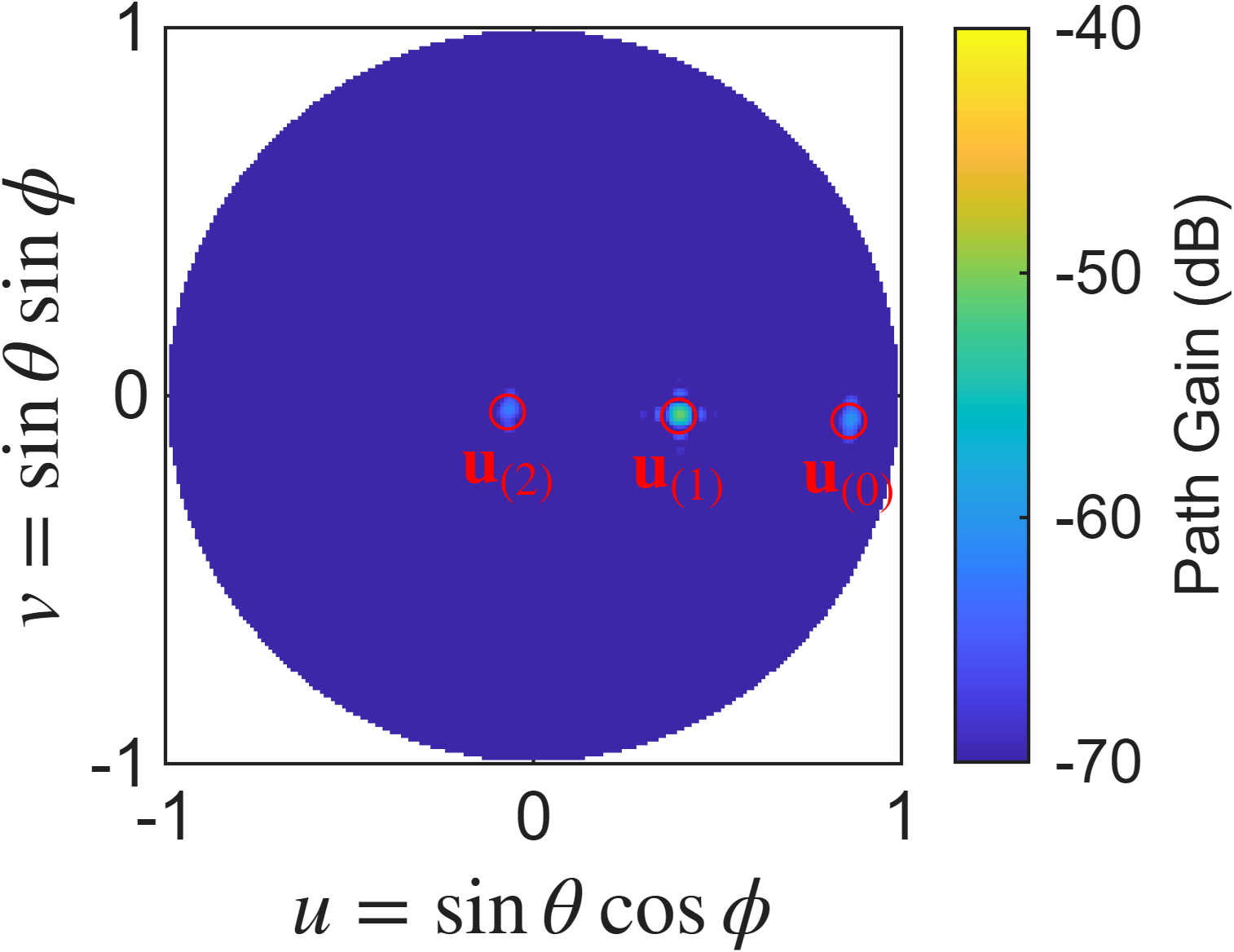}}}
\caption{Isolated-RIS scattered path gain for the three apertures. Markers denote the Bragg-order directions $\vect u_{(0)}$, $\vect u_{(1)}$, and $\vect u_{(2)}$ predicted by \eqref{eq:order_direction} and the Tx direction $\vect u_{\tx}$.}
\label{fig:aperture_maps}
\end{figure*}

The general model in \eqref{eq:general_gamma} already incorporates the quantized phase distribution used to synthesize the intended RIS response, so parasitic scattering caused solely by ideal phase quantization is inherently contained in the general-model pattern. The first two columns of \cref{fig:aperture_maps} nevertheless show that the general model reproduces the intended beam while missing two pronounced parasitic lobes present in the full-wave response, one near the geometrical-specular direction and one at a higher diffraction order.

Spatially irregular or weakly correlated amplitude and phase errors primarily produce diffuse scattering distributed over angle, whereas periodic or quasi-periodic aperture errors can add coherently at discrete phase-matched directions and produce high-level Bragg diffraction lobes~\cite{ruze1966antenna,holm1992analysis}. The pronounced parasitic lobes observed here and in~\cite{dingimproved,cardoso2025indoor} are therefore consistent with recurrent spatial deviations across the practical RIS aperture and motivate a Bragg-order representation. The embedded response of an element can depend on its neighboring state arrangement, mixed-state mutual coupling, structural currents, and local illumination~\cite{wijekoon2026physicallyconsistent}. Because similar local state configurations recur across the aperture, the resulting deviations can contain recurrent spatial components that add coherently at discrete spatial frequencies. A correction based only on independent, fixed offsets for the individual phase states would not, in general, represent this collective aperture behavior.

Accordingly, the calibration represents the dominant recurrent aperture error by a truncated Bragg-order basis. For the considered RIS, three components dominate the full-wave scattering~\cite{dingimproved}: the residual specular, intended anomalous-reflection, and dominant higher-order components. The calibrated aperture distribution is
\begin{equation}
\Gamma_{mn}^{\cali}=\sum_{\ell=0}^{2}\beta_\ell B_{mn}^{(\ell)},
\label{eq:calibrated_gamma}
\end{equation}
where $\beta_\ell$ is the complex coefficient of the $\ell$th Bragg-order component and $B_{mn}^{(\ell)}$ is the corresponding aperture basis function,
\begin{equation}
B_{mn}^{(\ell)}=\Gamma_{mn}^{\gen}\exp\!\left[\j(\ell-1)\Phi_{mn}\right].
\label{eq:bragg_basis}
\end{equation}
The continuous design phase $\Phi_{mn}$ generates the spatial phase progression of successive Bragg orders, while the discrete phase-state implementation remains contained in $\Gamma_{mn}^{\gen}$. Explicitly,
\begin{equation}
\begin{aligned}
B_{mn}^{(0)} &= \Gamma_{mn}^{\gen}\exp(-\j\Phi_{mn}), \\
B_{mn}^{(1)} &= \Gamma_{mn}^{\gen}, \\
B_{mn}^{(2)} &= \Gamma_{mn}^{\gen}\exp(\j\Phi_{mn}).
\end{aligned}
\label{eq:order_basis}
\end{equation}
The zeroth-order basis contains primarily the quantization residual $\Phi_{mn}^{\quant}-\Phi_{mn}$ and therefore has an approximately zero average transverse phase gradient, producing a residual component near the geometrical-specular direction. The first-order basis is the prescribed quantized focusing distribution, whereas the second-order basis contains approximately twice the focusing-phase progression and produces the dominant higher-order component.
%The three bases therefore differ only in phase, i.e., $|B_{mn}^{(\ell)}|=\Gamma_0$ for every $\ell$, $m$, and $n$, so the calibrated distribution obeys
%\begin{equation}
%\left|\Gamma_{mn}^{\cali}\right|\le\Gamma_0\sum_{\ell=0}^{2}\left|\beta_\ell\right|,
%\label{eq:passivity_bound}
%\end{equation}
%which is used later to verify that no passivity rescaling is required.
All three bases have the same magnitude, $|B_{mn}^{(\ell)}|=\Gamma_0$, and differ only in phase.

These components are analogous to the Floquet diffraction orders of an infinite periodic aperture and are referred to here as \emph{Bragg diffraction orders}~\cite{2005introduction,barnett2011new}. For the finite RIS considered here, their approximate directions follow from transverse phase matching. In RIS-centered spherical coordinates, with $\theta$ measured from the $+z$ axis and $\phi$ from $+x$ toward $+y$, the transverse direction-cosine vector is
\begin{equation}
\vect u=(u,v)=(\sin\theta\cos\phi,\sin\theta\sin\phi).
\label{eq:direction_cosine}
\end{equation}
Since the near-field focusing phase $\Phi_{mn}$ is generally nonlinear over the finite aperture, its spatial variation is characterized by the average transverse phase gradient
\begin{equation}
\bar{\vect K}=\frac{1}{MN}\sum_{m,n}\nabla_{xy}\Phi_{mn},
\label{eq:average_gradient}
\end{equation}
where $\nabla_{xy}$ denotes the transverse gradient on the RIS plane. The average transverse phase gradient of each aperture harmonic determines its approximate scattering direction
\begin{equation}
\vect u_{(\ell)}\simeq\vect u_{(0)}+\ell\,\frac{\bar{\vect K}}{k},
\label{eq:order_direction}
\end{equation}
where $\vect u_{(0)}$ is the geometrical-specular direction. Because the first-order basis is designed for the intended anomalous-reflection direction, $\bar{\vect K}\simeq k[\vect u_{(1)}-\vect u_{(0)}]$, so that $\ell=0$, 1, and 2 correspond to the geometrical-specular, intended, and dominant higher-order directions, with $\vect u_{(2)}\simeq2\vect u_{(1)}-\vect u_{(0)}$. For the geometry of \cref{tab:parameters}, \eqref{eq:order_direction} gives $\vect u_{(0)}=(0.86,-0.07)$, $\vect u_{(1)}=(0.39,-0.06)$, and $\vect u_{(2)}=(-0.07,-0.04)$, which are the marked positions in \cref{fig:aperture_maps} and coincide with the observed lobe centers. These are the nominal directions of the three basis responses. They are determined by the prescribed phase profile and do not depend on the fitted coefficients $\beta_\ell$; practical full-wave lobe centers may deviate slightly because of finite-aperture and implementation effects. \Cref{eq:order_direction} is used only to estimate the lobe centers.

A weaker lobe also appears near the Tx direction $\vect u_{\tx}=(-0.86,0.07)$ in the full-wave maps of \cref{fig:aperture_maps}. It is attributed to RIS backscattering toward the Tx and may additionally contain a numerical residual of the background subtraction, since the Tx radiation field is strong in that direction. This component is represented by neither the general nor the calibrated model.

%%%%%%%%%%%%%%%%%%%%%%%%%%%%%%%%%%%%%%%%%%%%%%%%%%%%%%%%%%%%%%%%%%%%%%%
%%%%%%%%%%%%%%%%%%%%%%%%%%%%%%%%%%%%%%%%%%%%%%%%%%%%%%%%%%%%%%%%%%%%%%%

%===========================================================================================%
%                                        New Section                                        %
%===========================================================================================%
\section{Full-Wave Calibration and Cross-Aperture Characterization}
\label{sec:calibration}

\subsection{Reduced-Aperture Calibration Condition}
\label{sec:transfer_condition}

Consider a target deployment characterized by the physical RIS side length $D_\tgt$, RIS-center-to-Tx distance $\norm{\vect r_{\tx,\tgt}}$, and RIS-center-to-nominal-Rx distance $\norm{\vect r_{\rx,\tgt}}$. To reduce the full-wave characterization cost, a smaller calibration aperture $D_\cali$ is introduced while retaining the same RIS-center-to-Tx and RIS-center-to-nominal-Rx directions. The calibration geometry is selected according to
\begin{equation}
\frac{D_\tgt}{D_\cali}
=
\frac{\norm{\vect r_{\tx,\tgt}}}
     {\norm{\vect r_{\tx,\cali}}}
=
\frac{\norm{\vect r_{\rx,\tgt}}}
     {\norm{\vect r_{\rx,\cali}}}.
\label{eq:reduced_calibration_rule}
\end{equation}
Under this scaling, an element at the same fractional position on the two apertures sees the Tx and the nominal Rx at the same angles. The calibration problem therefore reproduces the spread of local incidence and departure angles across the target aperture, which is the illumination-dependent condition governing how the practical embedded response deviates from the ideal element response. 
The coefficient transfer therefore preserves this normalized Tx/Rx geometry, while the target-aperture phase profile, basis functions, and element-specific propagation distances are recomputed for the actual target geometry. This scaling does not preserve the normalized near-field distance, which changes with aperture size; the resulting wavefront-curvature variation is retained explicitly through the recomputed element-wise propagation terms.
%As a consequence of this proportional scaling, the ideal coherently reflected gain at the nominal Rx also remains approximately invariant, since the increase in aperture area is compensated by the increased Tx--RIS and RIS--Rx distances.

In this work the $50\times50$ RIS is the target of the multipath validation, and \eqref{eq:reduced_calibration_rule} is applied with $D_\cali$ corresponding to the $25\times25$ aperture; the $100\times100$ RIS is included as a larger target to test the same rule. The resulting distances are listed in \cref{tab:coefficients}.

\subsection{Bragg-Order Coefficient Extraction}
\label{sec:coefficient_extraction}

\begin{figure}[t]
\centering
\includegraphics[width=0.9\linewidth]{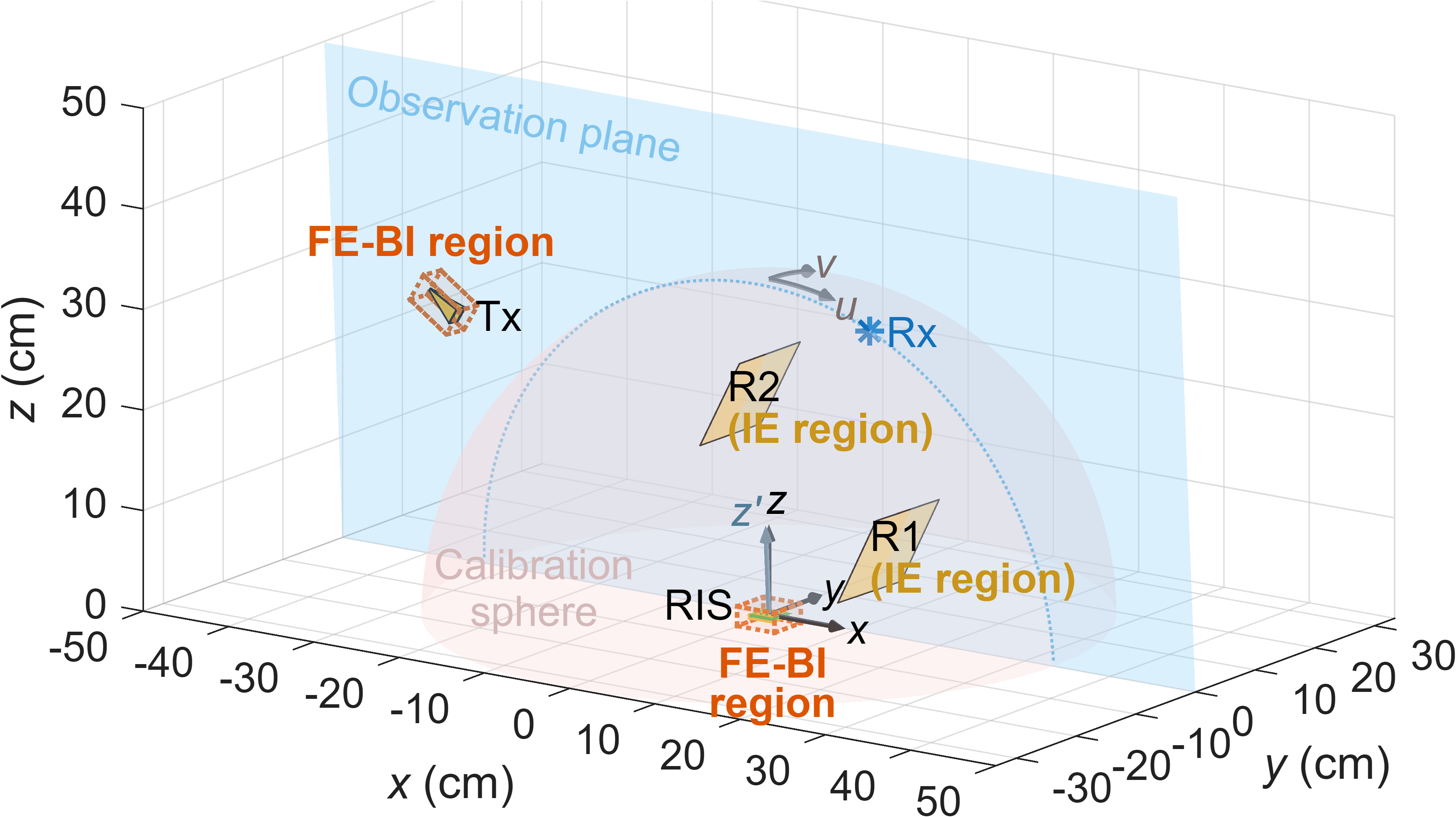}
\caption{HFSS setup shared by the isolated-RIS characterization and the PEC-reflector validation. The calibration surface is the upper hemisphere of the RIS-centered sphere with radius $R_{\cali}$. FE-BI regions enclose the Tx and the RIS; reflectors R1 and R2 are treated by the integral-equation (IE) domain and are removed during characterization.}
\label{fig:setup_PECreflectors}
\end{figure}

For the isolated-RIS characterization and PEC-reflector validation, the RIS-induced reference field is obtained by coherent background subtraction: a simulation containing the RIS is paired with a phase-consistent background simulation in which the RIS is removed while every other object and setting is retained, and the complex difference is taken before any magnitude conversion. For the isolated-RIS characterization, the background contains the Tx only. The corresponding HFSS setup is shown in \cref{fig:setup_PECreflectors}: separate finite-element boundary-integral (FE-BI) regions enclose the Tx and the RIS, and the scattered field is sampled on the upper hemisphere of the RIS-centered calibration sphere with radius $R_{\cali}=\norm{\vect r_{\rx}}$, so the surface passes through the nominal Rx position. The reflectors and the observation plane also visible in the figure belong to the PEC-reflector validation of \cref{sec:pec_results} and are absent from the isolated-RIS characterization.

Both solves export the upper-hemisphere field on a $1^\circ\times1^\circ$ grid in $(\theta,\phi)$. These physical samples remain on the calibration sphere but are represented for fitting by the direction cosines $\vect u=(u,v)$ of \eqref{eq:direction_cosine}. The difference field is interpolated onto a uniform $(u,v)$ grid over the unit disk $u^2+v^2\leq1$; the corresponding physical point is $\vect r(\vect u)=R_{\cali}\left[u,v,\sqrt{1-u^2-v^2}\right]^{\mathsf T}$ on the upper-hemispherical calibration surface. Using a uniform direction-cosine grid avoids the zenith clustering of an equiangular $(\theta,\phi)$ grid, which would otherwise overweight the near-boresight region in the weighted fit of \eqref{eq:wls}. Let $\vect u_i=(u_i,v_i)$, $i=1,\ldots,I$, denote the resulting calibration samples, at which the RIS-scattered field is
\begin{equation}
E_{\sca}(\vect u_i)
=
E_{\tot}(\vect u_i)-E_{\BG}(\vect u_i).
\label{eq:scattered_field}
\end{equation}

%%%%%%%%%%%%%%%%%%%%%%%%%%%%%%%%%%%%%%%%%%%%%%%%%%%%%%%%%%%%%%%%%%%%%%%
%%%%%%%%%%%%%%%%%%%%%%%%%%%%%%%%%%%%%%%%%%%%%%%%%%%%%%%%%%%%%%%%%%%%%%%

For the co-polarized electric-field component, the corresponding channel coefficient, used as the numerical reference, is~\cite{balanis2016antenna,hieule2024pointtopoint}
\begin{equation}
h_{\FW}(\vect u_i)
=
E_{\sca}(\vect u_i)
\sqrt{\frac{G_{\rx}\lambda^2}{8\pi Z_0P_{\mathrm t}}},
\label{eq:field_to_channel}
\end{equation}
where $Z_0$ is the free-space impedance and $P_{\mathrm t}$ is the Tx transmit power.

Define the sampled reference vector
\begin{equation}
\vect h_{\FW}
=
\left[
h_{\FW}(\vect u_1),\ldots,h_{\FW}(\vect u_I)
\right]^{\mathsf T}
\label{eq:fw_vector}
\end{equation}
and the Bragg-order response matrix
\begin{equation}
\mat H=
\begin{bmatrix}
h_0(\vect u_1) & h_1(\vect u_1) & h_2(\vect u_1)\\
\vdots & \vdots & \vdots\\
h_0(\vect u_I) & h_1(\vect u_I) & h_2(\vect u_I)
\end{bmatrix},
\label{eq:calibration_matrix}
\end{equation}
whose entries $h_\ell(\vect u_i) = h \big( \vect r(\vect u_i); B_{mn}^{(\ell)} \big)$ are obtained by evaluating \eqref{eq:element_kernel} at the physical point $\vect r(\vect u_i)$ on the upper-hemispherical calibration surface, with $B_{mn}^{(\ell)}$ substituted for $\Gamma_{mn}$.

Because the sampled Bragg-order responses are generally not orthogonal, their coefficients are estimated jointly rather than by independent projection. With $\boldsymbol{\beta}=[\beta_0,\beta_1,\beta_2]^{\mathsf T}$, the weighted complex least-squares estimate is
\begin{equation}
\widehat{\boldsymbol{\beta}}
=
\arg\min_{\boldsymbol{\beta}}
\norm{
\mat W^{1/2}
\left(
\mat H\boldsymbol{\beta}-\vect h_{\FW}
\right)
}_2^2,
\label{eq:wls}
\end{equation}
where
\begin{equation}
\mat W
=
\operatorname{diag}(w_1,\ldots,w_I),
\qquad
w_i
=
\frac{|h_{\FW}(\vect u_i)|}
{\max_j|h_{\FW}(\vect u_j)|}.
\label{eq:weights}
\end{equation}
The amplitude-dependent weighting emphasizes the dominant coherent scattering components while reducing the influence of low-level numerical fluctuations.

%================================================================================%
\begin{table}[t]
\caption{Bragg-order coefficients extracted independently for the three RIS apertures.}
\label{tab:coefficients}
\centering
\setlength{\tabcolsep}{3pt}
\begin{tabular}{c c c c}
\toprule
Aperture size (Tx/Rx distances) & $\beta_{0}$ & $\beta_{1}$ & $\beta_{2}$ \\
\midrule
\;\,$25\!\times\!25$\;\, ($22.62$ / $16.69$~cm) & $0.23\angle100^\circ$ & $0.49\angle95^\circ$ & $0.20\angle(-154^\circ)$ \\
\;\,$50\!\times\!50$\;\, ($45.23$ / $33.37$~cm) & $0.20\angle113^\circ$ & $0.50\angle84^\circ$ & $0.25\angle(-176^\circ)$ \\
$100\!\times\!100$ ($90.47$ / $66.74$~cm) & $0.16\angle106^\circ$ & $0.47\angle90^\circ$ & $0.22\angle(-165^\circ)$ \\
\bottomrule
\end{tabular}
\end{table}
%================================================================================%

\subsection{Aperture Dependence of the Extracted Coefficients}
\label{sec:coefficient_aperture}

The coefficients $\beta_\ell$ quantify the complex excitation of the prescribed Bragg-order aperture distributions in \eqref{eq:calibrated_gamma}. They represent the aggregate electromagnetic difference between the ideal element-wise model and the practical RIS response, including embedded-element interactions, mixed-state coupling, structural currents, edge-associated effects, and the minor implementation deviations of the phase states. They do not replace the aperture-dependent quantities of the element-wise kernel: the element coordinates, prescribed phase distribution, Tx-to-element and element-to-observation distances, directional factors, and finite aperture are evaluated for each RIS.

To distinguish the adequacy of the three-order representation from the scalability of the extracted coefficients, coefficients are first extracted independently for the three apertures. The results in \cref{tab:coefficients} show broadly similar dominant excitations: $|\beta_1|$ lies within $0.47$--$0.50$ and $|\beta_2|$ within $0.20$--$0.25$, with limited phase variation. The weaker zeroth-order coefficient shows somewhat larger aperture variation, with $|\beta_0|=0.23$, $0.20$, and $0.16$ for the three apertures. Overall, the same three-order representation remains applicable across the considered aperture range, with particularly stable intended first-order and dominant second-order excitations. 
%For all three vectors, $\Gamma_0\sum_\ell|\beta_\ell|<1$ with $\Gamma_0=1$, so the bound \eqref{eq:passivity_bound} keeps the calibrated distribution passive without rescaling.
%For all extracted coefficient vectors, $\sum_{\ell}|\beta_\ell|<1$ with $\Gamma_0=1$, ensuring $|\Gamma_{mn}^{\cali}|<1$.

\subsection{Cross-Aperture Coefficient Transfer Performance}
\label{sec:transfer_results}

After establishing the calibrated model with aperture-specific coefficients, cross-aperture transfer is introduced as an additional scalability step to reduce the full-wave calibration cost for larger RISs. It exploits the separation, in \eqref{eq:calibrated_gamma}, between the Bragg-order basis functions $B_{mn}^{(\ell)}$ and their complex excitation coefficients $\beta_\ell$. For a target RIS, the element coordinates, prescribed phase distribution, basis functions, element-specific propagation distances, and directional factors are recomputed, while only the three-coefficient vector $\boldsymbol{\beta}$ is transferred from the calibration aperture.

The transfer test isolates the effect of aperture size. The three RISs use the same unit-cell implementation, phase-state law, operating frequency, and polarization, and the Tx--RIS and RIS--nominal-Rx distances are scaled with aperture size according to \eqref{eq:reduced_calibration_rule}, with the resulting distances listed in \cref{tab:coefficients}. 
The $25\times25$ RIS is used as the smallest characterized calibration aperture satisfying the normalized-geometry condition in \eqref{eq:reduced_calibration_rule}; it therefore also provides the lowest full-wave calibration cost.

\Cref{fig:aperture_maps} compares the full-wave reference, general model, and calibrated model for the three aperture sizes. The general model consistently overestimates the intended anomalous-reflection response while substantially underestimating the zeroth- and second-order components, whereas the calibrated maps restore the relative weights of the three orders. For both larger apertures, the transferred-coefficient model agrees closely with the full-wave reference at $\vect u_{(1)}$ and at the parasitic-order directions $\vect u_{(0)}$ and $\vect u_{(2)}$.

The corresponding lobe-resolved errors are summarized in \cref{tab:isolated_metrics}, where aperture-specific calibration is included to separate the intrinsic error of the three-order representation from the additional error introduced by coefficient transfer. With aperture-specific coefficients, all three orders are reproduced within approximately $1$~dB for all aperture sizes, confirming that the three-order representation remains adequate as the aperture increases; by comparison, the general model exhibits absolute errors of up to $13.90$, $6.52$, and $21.27$~dB for $\ell=0$, $1$, and $2$, respectively.

For the $50\times50$ and $100\times100$ RISs, the intended-order errors are $0.79$ and $0.16$~dB, respectively, and the dominant second-order errors are $2.39$ and $1.23$~dB. Across all three retained orders of the two larger RISs, the transferred calibration reduces the absolute lobe-gain errors by $4.7$--$18.9$~dB relative to the general model. Thus, even without target-specific refitting, the transferred calibration preserves an accurate nominal-Rx response and substantially improves the dominant parasitic-order prediction.

This behavior is consistent with the decomposition in \eqref{eq:calibrated_gamma}. The target RIS regenerates the aperture-dependent phase matching, element positions, prescribed phase distribution, basis functions, and element-specific propagation distances; the transferred $\beta_\ell$ describe only the complex excitation of the retained orders. Because the three RISs share the same unit-cell implementation, phase-state law, frequency, polarization, and angular geometry, the dominant excitation mechanisms remain similar. \Cref{fig:aperture_maps,tab:coefficients,tab:isolated_metrics} therefore support coefficient transfer from the reduced calibration aperture to larger target apertures under the prescribed focusing configuration and matched normalized Tx/Rx geometry.
%The three-order representation targets the dominant deterministic components relevant to multipath formation rather than reconstructing the complete scattered field.

The computational benefit of coefficient transfer is quantified in \cref{tab:calibration_cost}. The wall-clock time is the sum of the paired Tx--RIS and Tx-only simulations, and the peak random-access memory (RAM) is the larger value of the pair. Increasing the aperture from $25\times25$ to $100\times100$ raises the full-wave coefficient-extraction cost from $1.08$ to $7.56$~h and from $39.1$ to $139$~GB. Reusing the $25\times25$ coefficient vector therefore preserves the calibrated-model framework while avoiding the substantially more expensive target-aperture calibration.

Unless explicitly identified as \emph{aperture-specific calibration}, the \emph{calibrated model} hereafter denotes the model using the coefficients extracted from the $25\times25$ RIS, with all aperture-dependent quantities regenerated for the target geometry.

%================================================================================%
\begin{table}[t]
\caption{Isolated-RIS lobe-gain errors. For calibrated cases, the calibration aperture is indicated in parentheses.}
\label{tab:isolated_metrics}
\centering
\setlength{\tabcolsep}{5pt}
\begin{tabular}{@{}c c c c c@{}}
\toprule
\multirow{2}{*}[-0.7ex]{\makecell{Target\\aperture}}
& \multirow{2}{*}[-0.7ex]{Model}
& \multicolumn{3}{c}{Signed lobe-gain error (dB)}\\
\cmidrule(l{1pt}r{1pt}){3-5}
& & $\ell=0$ & $\ell=1$ & $\ell=2$\\
\midrule
\multirow{2}{*}{$25\times25$}
& General
& $-13.90$ & $+6.17$ & $-18.80$ \\
& Calibrated ($25\times25$)
& $+0.02$ & $-0.12$ & $-0.64$ \\
\midrule
\multirow{3}{*}{$50\times50$}
& General
& $-11.75$ & $+5.51$ & $-21.27$ \\
& Calibrated ($50\times50$)
& $-0.97$ & $-0.56$ & $-0.27$ \\
& Calibrated ($25\times25$)
& $+0.32$ & $-0.79$ & $-2.39$ \\
\midrule
\multirow{3}{*}{$100\times100$}
& General
& $-12.73$ & $+6.52$ & $-18.30$ \\
& Calibrated ($100\times100$)
& $+0.12$ & $-0.16$ & $-0.17$ \\
& Calibrated ($25\times25$)
& $+3.51$ & $+0.16$ & $-1.23$ \\
\bottomrule
\end{tabular}
\end{table}
%================================================================================%

%================================================================================%
\begin{table}[t]
\caption{Measured computational cost of isolated-RIS full-wave simulation.}
\label{tab:calibration_cost}
\centering
\begin{tabular}{@{}c c c@{}}
\toprule
Aperture size & Wall-clock time & Peak RAM \\
\midrule
$25\times25$ & $1.08$~h & $39.1$~GB \\
$50\times50$ & $2.68$~h & $49.1$~GB \\
$100\times100$ & $7.56$~h & $139$~GB \\
\bottomrule
\end{tabular}
\end{table}
%================================================================================%

%===========================================================================================%
%                                        New Section                                        %
%===========================================================================================%
\section{Ray-Tracing Integration With the RIS Model}
\label{sec:rt}

In this work, deterministic RT supplies the environmental path geometry: the image method (IM)~\cite{esposti2021ray,yang20266g} is used for the finite PEC-reflector configuration, whereas shooting and bouncing rays (SBR)~\cite{li2023mmwave,hao2025analysis} is used for path discovery in the RCC room. The RIS--Rx paths identified by either procedure are subsequently represented by the same image-theory unfolding.

\subsection{Conditions for Reusing the Calibrated Coefficients}
\label{sec:reuse_condition}

Once the Bragg-order coefficient vector has been fixed for a target RIS, either by aperture-specific calibration or by cross-aperture transfer, the same coefficients can be used for different RIS--Rx environments provided that the RIS implementation, operating frequency, polarization, prescribed phase configuration, and Tx--RIS illumination remain comparable to those used for calibration. The present work assumes a LoS-dominant Tx--RIS link. Under this assumption, changing the RIS--Rx environment primarily changes the outgoing reflection sequences, blockage, material response, and polarization transformation, while the incident field exciting the RIS remains approximately unchanged. The Bragg-order coefficients are therefore kept fixed and only the environmental paths are updated.

For the PEC-reflector validation of \cref{sec:pec_results}, R1 and R2 are placed only on the RIS--Rx side, so the Tx--RIS illumination is essentially the same as in the isolated calibration. In the enclosed room of \cref{sec:room_results}, Tx-side reflections can also illuminate the RIS; the resulting mismatch with the calibration condition is discussed in \cref{sec:room_error}.

\subsection{Image-Theory Unfolding of Reflection Paths}
\label{sec:im_coupling}

The environmental propagation considered in this work consists of direct paths and reflections from planar surfaces; diffraction and diffuse scattering are not included. For planar reflection paths, image theory provides an equivalent unfolded straight-path representation~\cite{esposti2021ray,yang20266g,ding2026twoway}. A reflected RIS--Rx path is specified by the ordered reflector sequence $\pathset=(p_1,p_2,\ldots,p_S)$, where $p_s$ identifies the $s$th reflecting surface encountered after the field leaves the RIS, and $S$ is the number of reflections. The direct RIS--Rx path is denoted by $\pathset=\varnothing$.

Let reflector $p_s$ be represented by a point $\vect a_{p_s}$ on the plane and a unit normal $\vect n_{p_s}$. The corresponding mirror operator is
\begin{equation}
\mathfrak M_{p_s}(\vect x)
=
\vect x
-2\left[(\vect x-\vect a_{p_s})^{\mathsf T}\vect n_{p_s}\right]\vect n_{p_s}.
\label{eq:mirror_operator}
\end{equation}
For the reflection sequence $\pathset=(p_1,p_2,\ldots,p_S)$, the unfolded image point is obtained by successive mirroring in reverse reflection order as
\begin{equation}
\vect r_{\mathrm{img}}^{(\pathset)}
=
\mathfrak M_{p_1}\!\left(
\mathfrak M_{p_2}\!\left(
\cdots
\mathfrak M_{p_S}(\robs)
\cdots
\right)\right).
\label{eq:image_point}
\end{equation}
For the direct path, $\pathset=\varnothing$ and $\vect r_{\mathrm{img}}^{(\varnothing)}=\robs$.

\begin{figure}[t]
    \centering
    \includegraphics[width=0.85\linewidth]{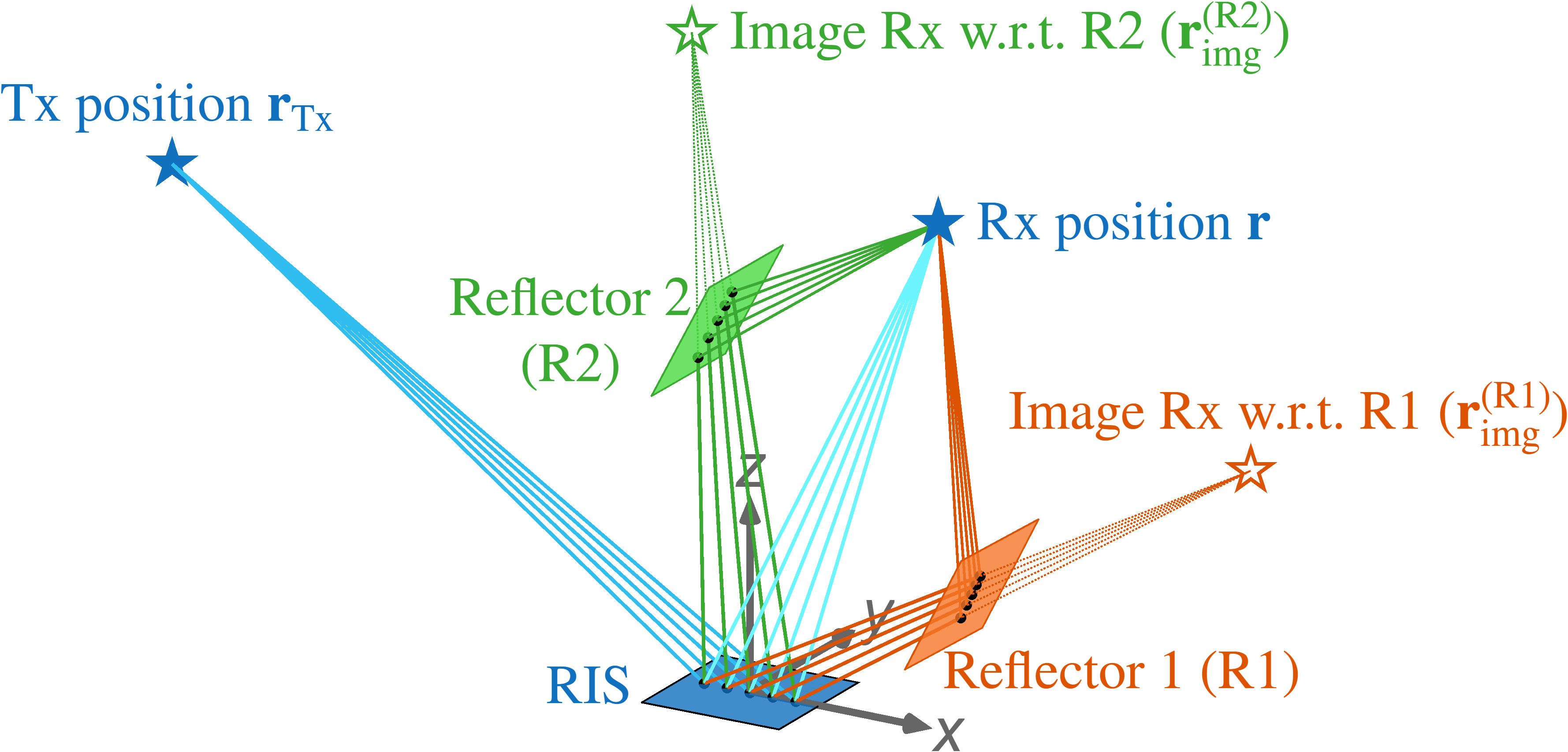}
    \caption{Image-theory unfolding for the two PEC-reflector paths. For the single-reflection paths through R1 and R2, the observation point $\vect r$ is mirrored into $\vect r_{\mathrm{img}}^{(\mathrm{R1})}$ and $\vect r_{\mathrm{img}}^{(\mathrm{R2})}$, respectively.}
    \label{fig:image_theory}
\end{figure}

For a valid planar reflection path, the straight distance from RIS element $(m,n)$ to the unfolded image point $\vect r_{\mathrm{img}}^{(\pathset)}$ equals the total length of the corresponding folded element--Rx path. Hence, the image point can be used directly to reconstruct the element-wise RIS--Rx spreading and propagation phase. \Cref{fig:image_theory} illustrates this construction for the two single-reflection paths used in the PEC-reflector validation.

%%%%%%%%%%%%%%%%%%%%%%%%%%%%%%%%%%%%%%%%%%%%%%%%%%%%%%%%%%%%%%%%%%%%%%%
%%%%%%%%%%%%%%%%%%%%%%%%%%%%%%%%%%%%%%%%%%%%%%%%%%%%%%%%%%%%%%%%%%%%%%%

Two additional path-dependent quantities describe finite-surface visibility and environmental reflection. The binary indicator $V_{mn}^{(\pathset)}\in\{0,1\}$ equals unity only if the folded ray from element $(m,n)$ intersects the prescribed reflectors in the correct order and every physical segment is unobstructed. The cumulative complex reflection factor $\rho_{mn}^{(\pathset)}$ contains the material reflection coefficients and polarization transformations along the folded path.

The RIS-assisted channel coefficient associated with reflection sequence $\pathset$ is then
\begin{equation}
\begin{split}
h_{\RIS}^{(\pathset)}(\robs;\Gamma_{mn})
=
C\sum_{m,n}
V_{mn}^{(\pathset)}\rho_{mn}^{(\pathset)}
\frac{A_{mn}^{(\pathset)}}
{\norm{\vect r_{\tx}-\vect r_{mn}}\,
 \norm{\vect r_{\mathrm{img}}^{(\pathset)}-\vect r_{mn}}}
\\
\times\;\Gamma_{mn}
\exp\!\left[
-\j k\left(
\norm{\vect r_{\tx}-\vect r_{mn}}
+
\norm{\vect r_{\mathrm{img}}^{(\pathset)}-\vect r_{mn}}
\right)
\right].
\end{split}
\label{eq:path_kernel}
\end{equation}
Here, $A_{mn}^{(\pathset)}$ collects the normalized Tx, RIS-element, and Rx pattern factors evaluated using the physical departure and arrival directions of the folded path, whereas $\vect r_{\mathrm{img}}^{(\pathset)}$ is used only for the equivalent element-wise path length and propagation phase.

For the direct path $\pathset=\varnothing$, $\vect r_{\mathrm{img}}^{(\varnothing)}=\robs$, $V_{mn}^{(\varnothing)}=1$, $\rho_{mn}^{(\varnothing)}=1$, and $A_{mn}^{(\varnothing)}=A_{mn}(\robs)$, so \eqref{eq:path_kernel} reduces exactly to the unreflected element-wise formulation in \eqref{eq:element_kernel}. 
For a reflected path, the physical Rx remains at $\robs$, while the image point is introduced only to evaluate the unfolded element-to-Rx path length and propagation phase, allowing the same element-wise kernel to be used for both direct and reflected RIS--Rx paths.

\subsection{Path Discovery by IM or SBR}
\label{sec:sbr_coupling}

Two procedures provide the reflection paths required by the common unfolded formulation above. For the PEC-reflector validation, the IM is applied to the known reflector geometry to construct the finite set of relevant paths; the reflection points, finite-plate intersections, visibility, and polarization response are then evaluated for each RIS element.

In the room scenario, testing all candidate reflector sequences with the IM would be inefficient. SBR is therefore used to discover the geometrically valid paths from the RIS reference point and from the physical Tx to each observation point. For every retained RIS--Rx path, the SBR interaction positions are matched to the corresponding room surfaces to recover the ordered reflector sequence $\pathset$, which is then unfolded with \eqref{eq:image_point} to obtain $\vect r_{\mathrm{img}}^{(\pathset)}$. 
Thus, SBR is used to identify the reflector sequence for each retained path; the sequence is then unfolded by image theory to obtain the element-to-image geometry used in the RIS kernel.

%For an SBR-discovered path, the cumulative environmental response is evaluated at path level. Let $h_{\mathrm{prop}}^{(\pathset)}$ denote the complex point-to-point propagation coefficient of the retained ray at the operating frequency, including free-space propagation, material reflections, and polarization transformation, and let $d^{(\pathset)}$ denote its total path length. The path-level reflection factor is
%\begin{equation}
%\rho^{(\pathset)}
%=
%\frac{h_{\mathrm{prop}}^{(\pathset)}}
%{\dfrac{\lambda}{4\pi d^{(\pathset)}}
%\exp(-\j k d^{(\pathset)})}.
%\label{eq:path_environment_factor}
%\end{equation}
%This normalization removes the center-ray free-space spreading and propagation phase, leaving the cumulative material-reflection and polarization response; for a direct path, $\rho^{(\varnothing)}=1$. In the SBR implementation, $\rho_{mn}^{(\pathset)}$ in \eqref{eq:path_kernel} is approximated by the common path-level value $\rho^{(\pathset)}$ for all RIS elements of the retained path, while the element-to-image distance in \eqref{eq:unfolded_ris_rx_distance} is still evaluated separately for every RIS element. This approximation is most appropriate when the reflecting surfaces are large compared with the RIS aperture and the reflection geometry varies slowly across the aperture.

\subsection{Tx-Pattern Representations}
\label{sec:tx_pattern}

Direct and reflected Tx--Rx paths that do not interact with the RIS are hereafter referred to as \emph{Tx--Rx bypass paths} or simply \emph{bypass paths}.

The RIS-assisted and bypass contributions sample different angular regions of the Tx radiation pattern. A horn-local coordinate system is defined with local $+z$ along the Tx boresight, and $(\theta_{\tx},\phi_{\tx})$ denotes the corresponding departure angles, with $\theta_{\tx}=0^\circ$ at boresight.

The normalized simulated Tx power pattern is represented by
\begin{equation}
F_{\tx}^{(\mathrm{sim})}
(\theta_{\tx},\phi_{\tx})
=
\frac{
G_{\tx}^{\mathrm{sim}}
(\theta_{\tx},\phi_{\tx})
}{
G_{\tx}
},
\label{eq:sim_tx_pattern}
\end{equation}
where $G_{\tx}^{\mathrm{sim}}$ is the absolute directional realized gain and $G_{\tx}$ is the reference gain already included in $C$ of \eqref{eq:normalization}.

For a compact description of the forward main beam, the simulated co-polarized E- and H-plane cuts are fitted using a widely used cosine-$q$ model~\cite{camacho2020excitation,tang2022path}:
\begin{equation}
F_{\tx}^{(\mathrm{fit})}
(\theta_{\tx},\phi_{\tx})
=
\cos^{2q(\phi_{\tx})}\!\left(\theta_{\tx}\right),
\qquad
0\le\theta_{\tx}\le\frac{\pi}{2},
\label{eq:cosq_pattern}
\end{equation}
where
\begin{equation}
q(\phi_{\tx})
=
q_{\mathrm E}\sin^2\phi_{\tx}
+
q_{\mathrm H}\cos^2\phi_{\tx},
\end{equation}
with $\phi_{\tx}=90^\circ$ and $0^\circ$ corresponding to the E- and H-plane cuts, respectively.

Either representation enters the channel formulation through the Tx pattern factor evaluated at the physical departure direction. In \eqref{eq:pattern_factor}, $F_{mn}^{\tx}$ is evaluated toward RIS element $(m,n)$, whereas for a bypass path it is evaluated along the corresponding Tx departure direction.

The RIS is centered on the Tx boresight and subtends only a narrow main-beam sector, over which the fitted pattern closely approximates the simulated pattern. The fitted representation is therefore adequate for the RIS-assisted contribution in the present setup. Bypass paths, however, can span much wider departure angles, including sidelobe regions that the monotonic cosine-$q$ model cannot represent. Both representations are compared in \cref{sec:room_results}.

\subsection{Coherent RIS-Assisted, Bypass, and Total Channels}
\label{sec:channel_assembly}

For a bypass path $\pathsetB$, the point-to-point channel coefficient is determined by the path length $d^{(\pathsetB)}$, Tx departure direction $\Omega_{\tx}^{(\pathsetB)}$, Rx arrival direction $\Omega_{\rx}^{(\pathsetB)}$, and cumulative reflection factor $\rho^{(\pathsetB)}$ obtained directly from RT.

Let $\pathsetRIS$ and $\pathsetBY$ denote the RIS-assisted and Tx--Rx bypass path sets for observation $\robs$, and let $\varMethod\in\{\gen,\cali\}$ denote the RIS-model choice. The RIS-assisted, bypass, and total channel coefficients are
\begin{align}
h_{\RIS,\varMethod}(\robs)
&=
\sum_{\pathset\in\pathsetRIS}
h_{\RIS}^{(\pathset)}\!\left(\robs;\Gamma_{mn}^{\varMethod}\right),
\label{eq:ris_total}\\
h_{\by}(\robs)
&=
\sum_{\pathsetB\in\pathsetBY}
\sqrt{
G_{\tx}F_{\tx}(\Omega_{\tx}^{(\pathsetB)})
G_{\rx}F_{\rx}(\Omega_{\rx}^{(\pathsetB)})
}
\notag\\
&\quad\times
\rho^{(\pathsetB)}
\frac{\lambda}{4\pi d^{(\pathsetB)}}
\exp(-\j k d^{(\pathsetB)}),
\label{eq:bypass_channel}\\
h_{\tot,\varMethod}(\robs)
&=
h_{\by}(\robs)+h_{\RIS,\varMethod}(\robs).
\label{eq:total_channel}
\end{align}
Here $F_{\tx}$ and $F_{\rx}$ denote the same normalized power patterns as in \eqref{eq:pattern_factor}, evaluated at the departure and arrival directions of the path in question. For the omnidirectional Rx used here, $F_{\rx}=1$.

For every RIS-assisted path, the reflector sequence discovered by IM or SBR is first unfolded to $\vect r_{\mathrm{img}}^{(\pathset)}$ and then evaluated by \eqref{eq:path_kernel}. Only $h_{\RIS,\varMethod}$ depends on the RIS model, whereas $h_{\by}$ is common to the general and calibrated cases. The corresponding channel gains are
\begin{equation}
\begin{aligned}
G_{\RIS,\varMethod}(\robs)
&=20\log_{10}|h_{\RIS,\varMethod}(\robs)|,\\
G_{\by}(\robs)
&=20\log_{10}|h_{\by}(\robs)|,\\
G_{\tot,\varMethod}(\robs)
&=20\log_{10}|h_{\tot,\varMethod}(\robs)|.
\end{aligned}
\label{eq:channel_gain}
\end{equation}
All path contributions are coherently summed before taking the magnitude, so their relative phases determine constructive or destructive multipath interference.

%===========================================================================================%
%                                        New Section                                        %
%===========================================================================================%
\section{Multipath Channel-Gain Prediction and Validation}
\label{sec:validation}

The characterization in \cref{sec:model,sec:calibration} fixes the Bragg-order coefficients before environmental propagation is introduced. All validation studies in this section use the same $50\times50$ target RIS with the coefficient vector transferred from the $25\times25$ RIS. The PEC-reflector validation compares the background-subtracted RIS-induced gain, whereas the RCC-room validation compares the full-scene total channel gain.

For the RCC-room spatial comparison, let $G_{\tot,\FW}(\vect r)$ denote the hybrid HFSS reference gain. The signed total-channel gain error is
\begin{equation}
\Delta G_{\varMethod}(\vect r)
=
G_{\tot,\varMethod}(\vect r)
-
G_{\tot,\FW}(\vect r),
\label{eq:signed_gain_error}
\end{equation}
and, for $N_{\mathrm{obs}}$ observation samples, the channel-gain root-mean-square error (RMSE) is
\begin{equation}
\mathrm{RMSE}_{G,\varMethod}
=
\sqrt{
\frac{1}{N_{\mathrm{obs}}}
\sum_{i=1}^{N_{\mathrm{obs}}}
\left[
\Delta G_{\varMethod}(\vect r^{(i)})
\right]^2
}.
\label{eq:gain_rmse}
\end{equation}

Let $P_{80}(|\Delta G_{\varMethod}|)$ and $P_{90}(|\Delta G_{\varMethod}|)$ denote the 80th and 90th percentiles of the absolute total-channel gain errors over the room observation samples, respectively, which characterize the spatial error distribution and are complemented by the cumulative distribution function (CDF) of $|\Delta G_{\varMethod}|$. The error at the prescribed nominal Rx position is $|\Delta G_{\varMethod}(\vect r_{\rx})|$.

%%%%%%%%%%%%%%%%%%%%%%%%%%%%%%%%%%%%%%%%%%%%%%%%%%%%%%%%%%%%%%%%%%%%%%%
\begin{figure*}[t]
\centering
\newcommand{\pecfigheight}{2.88cm}
\subfloat[No reflector: HFSS reference.]
{\includegraphics[height=\pecfigheight]{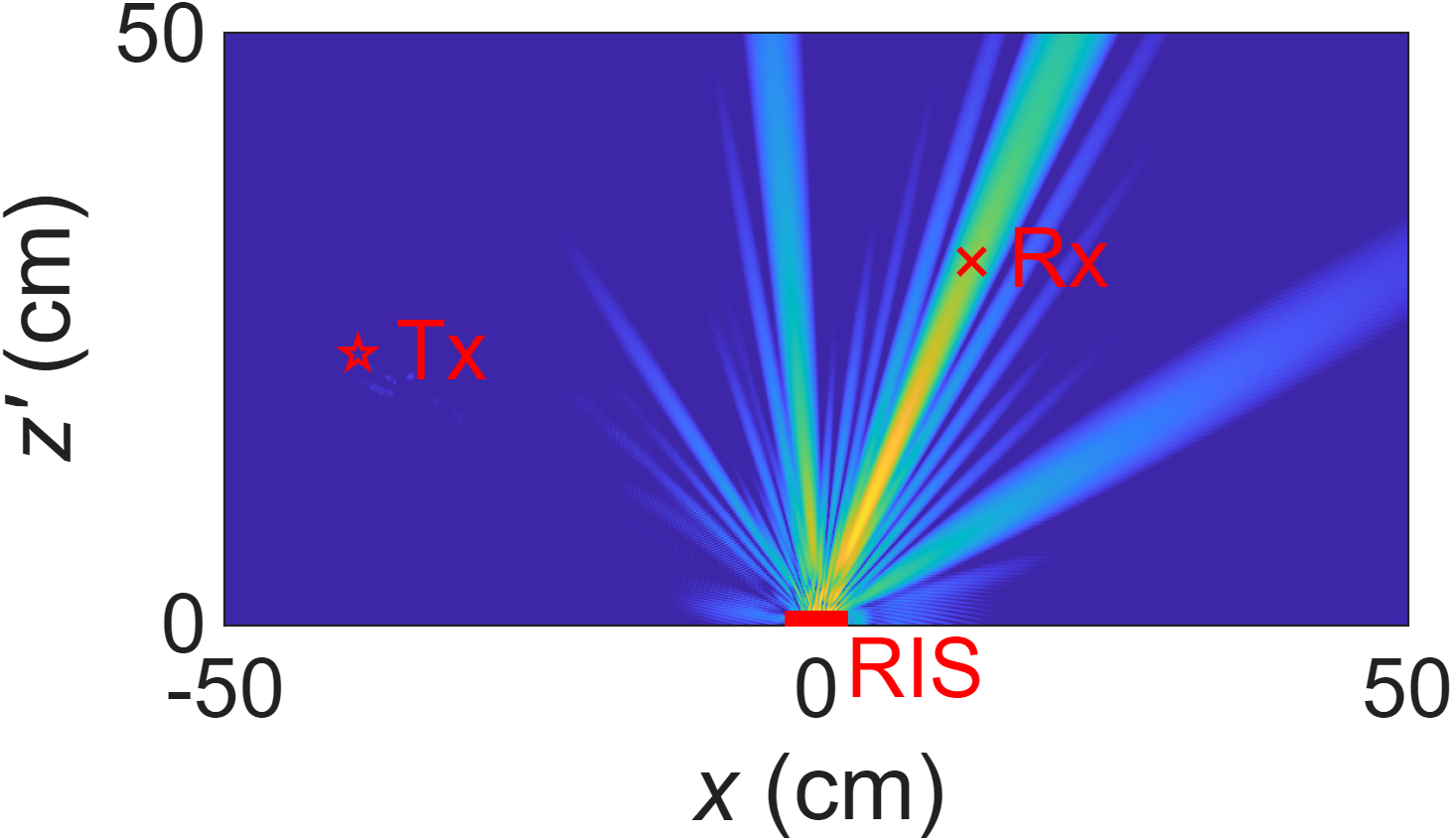}}
\hfil
\subfloat[No reflector: general model.]
{\includegraphics[height=\pecfigheight]{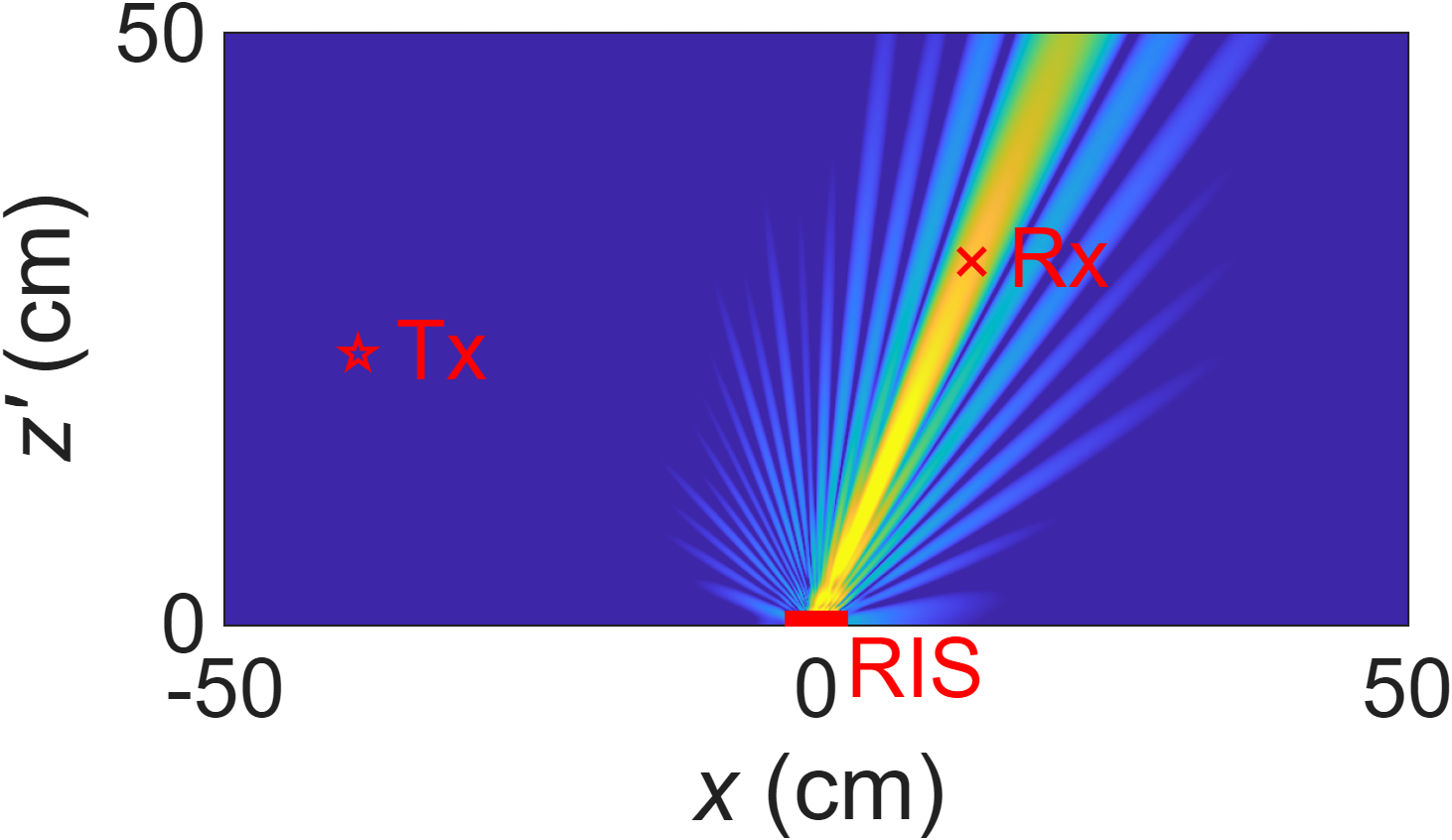}}
\hfil
\subfloat[No reflector: calibrated model.]
{\includegraphics[height=\pecfigheight]{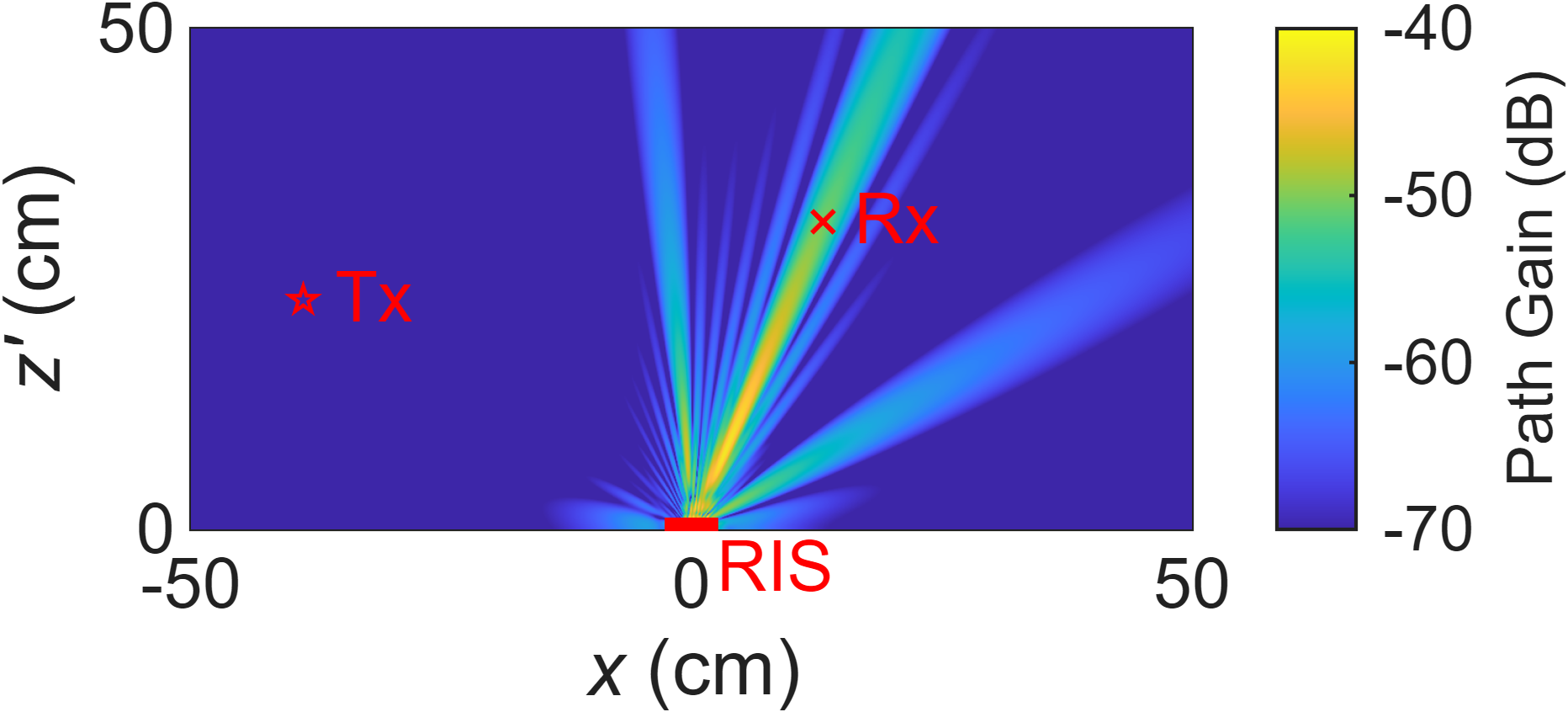}}
\\
\subfloat[With R1: HFSS reference.]
{\includegraphics[height=\pecfigheight]{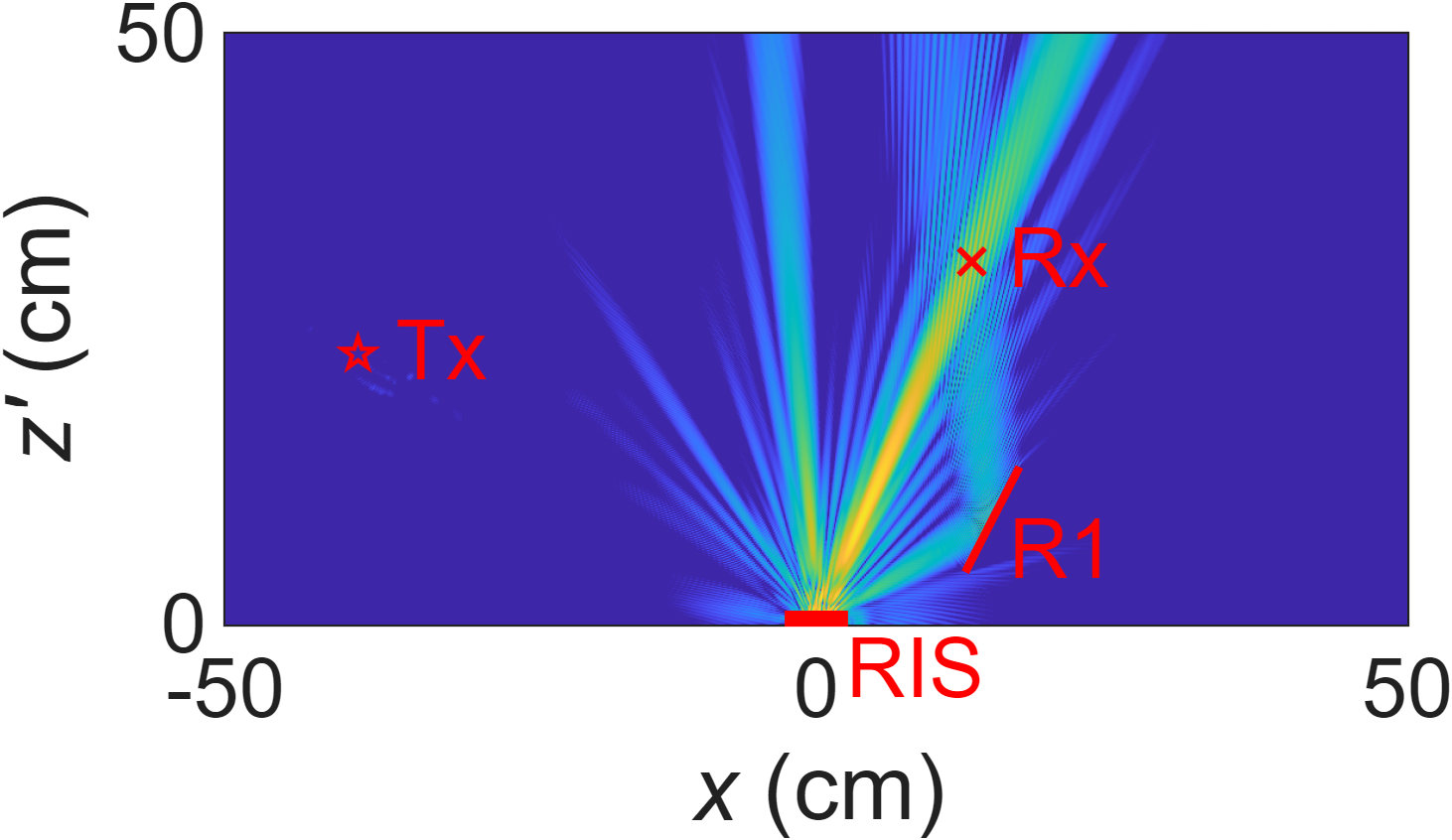}}
\hfil
\subfloat[With R1: general model.]
{\includegraphics[height=\pecfigheight]{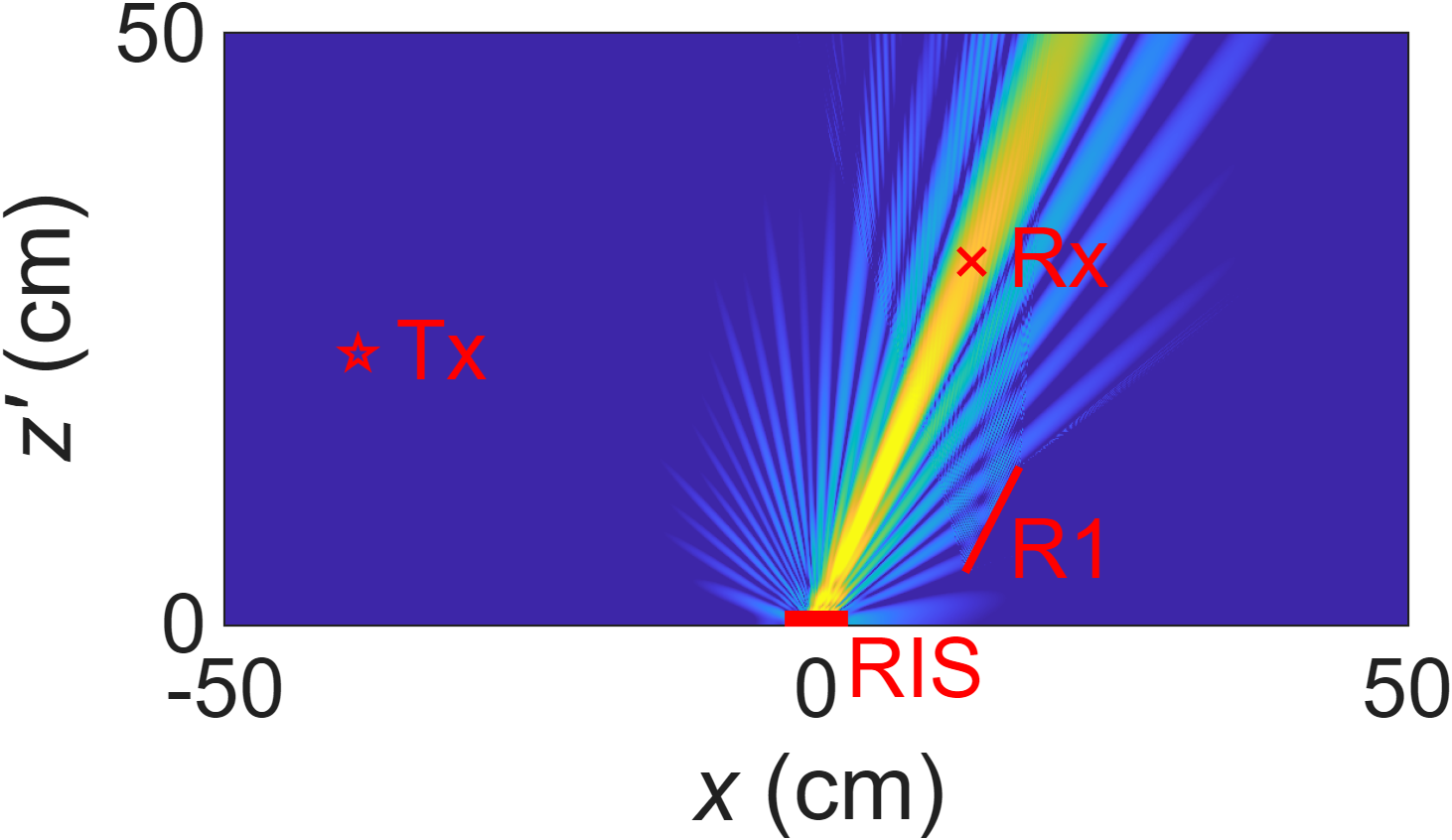}}
\hfil
\subfloat[With R1: calibrated model.]
{\includegraphics[height=\pecfigheight]{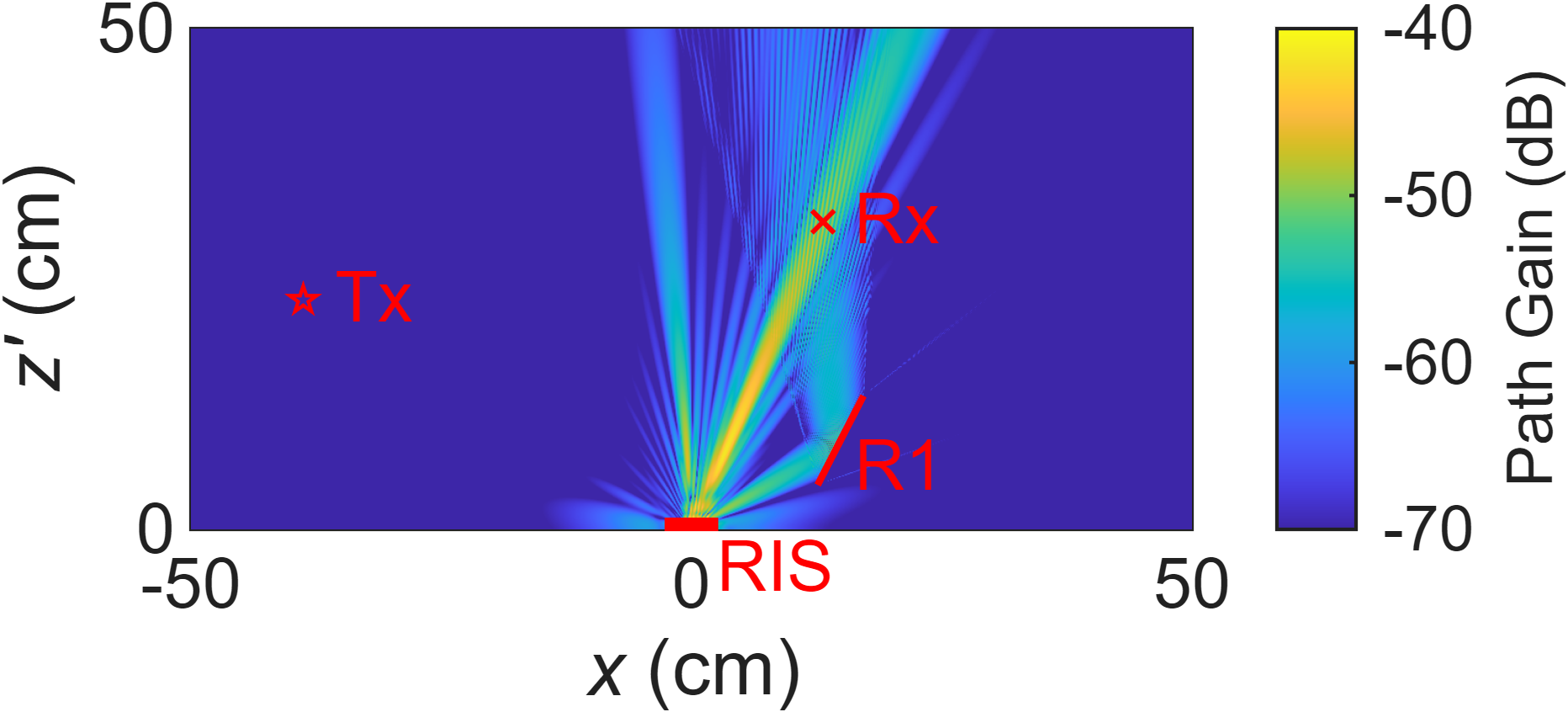}}
\\
\subfloat[With R2: HFSS reference.]
{\includegraphics[height=\pecfigheight]{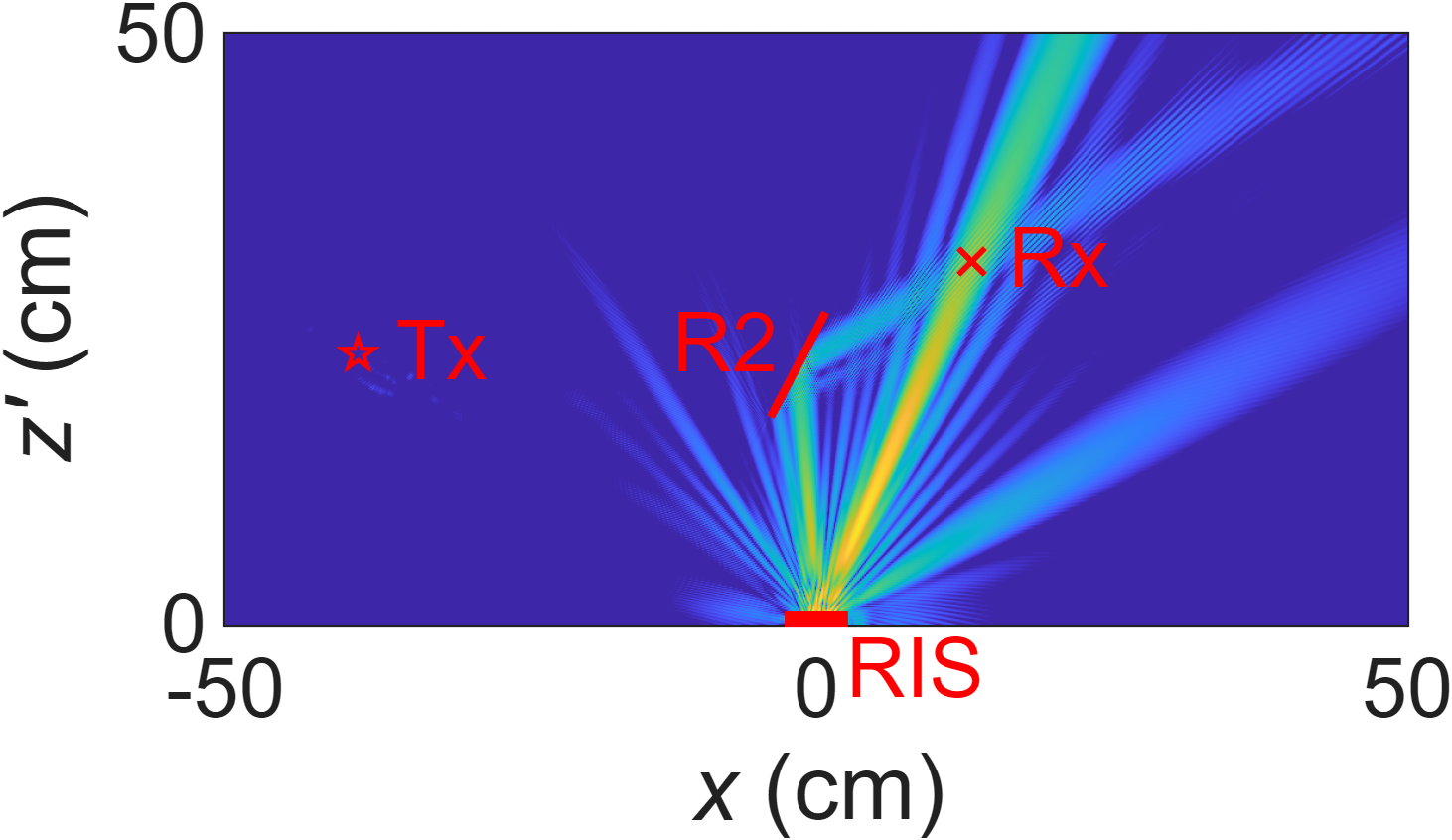}}
\hfil
\subfloat[With R2: general model.]
{\includegraphics[height=\pecfigheight]{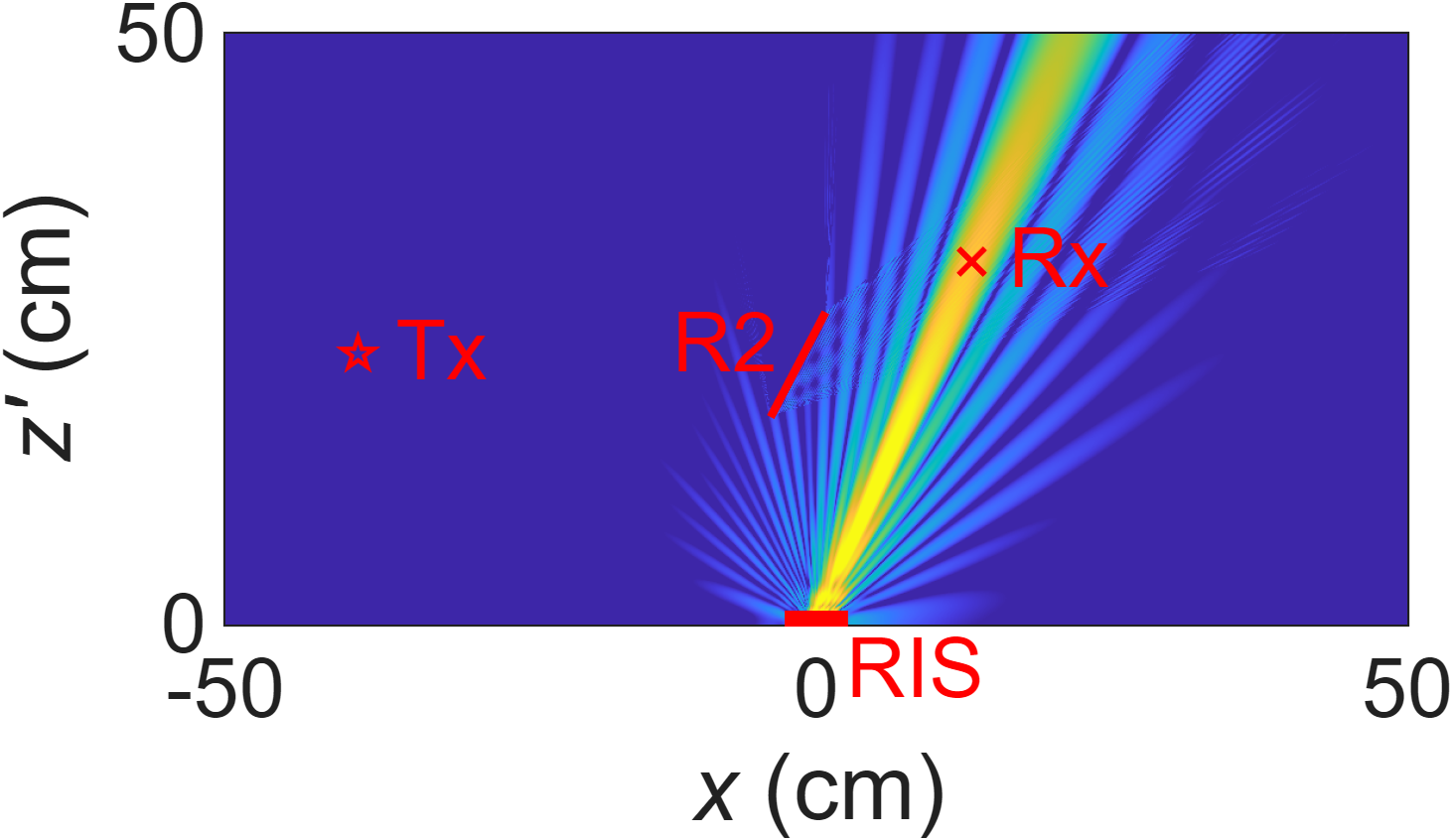}}
\hfil
\subfloat[With R2: calibrated model.]
{\includegraphics[height=\pecfigheight]{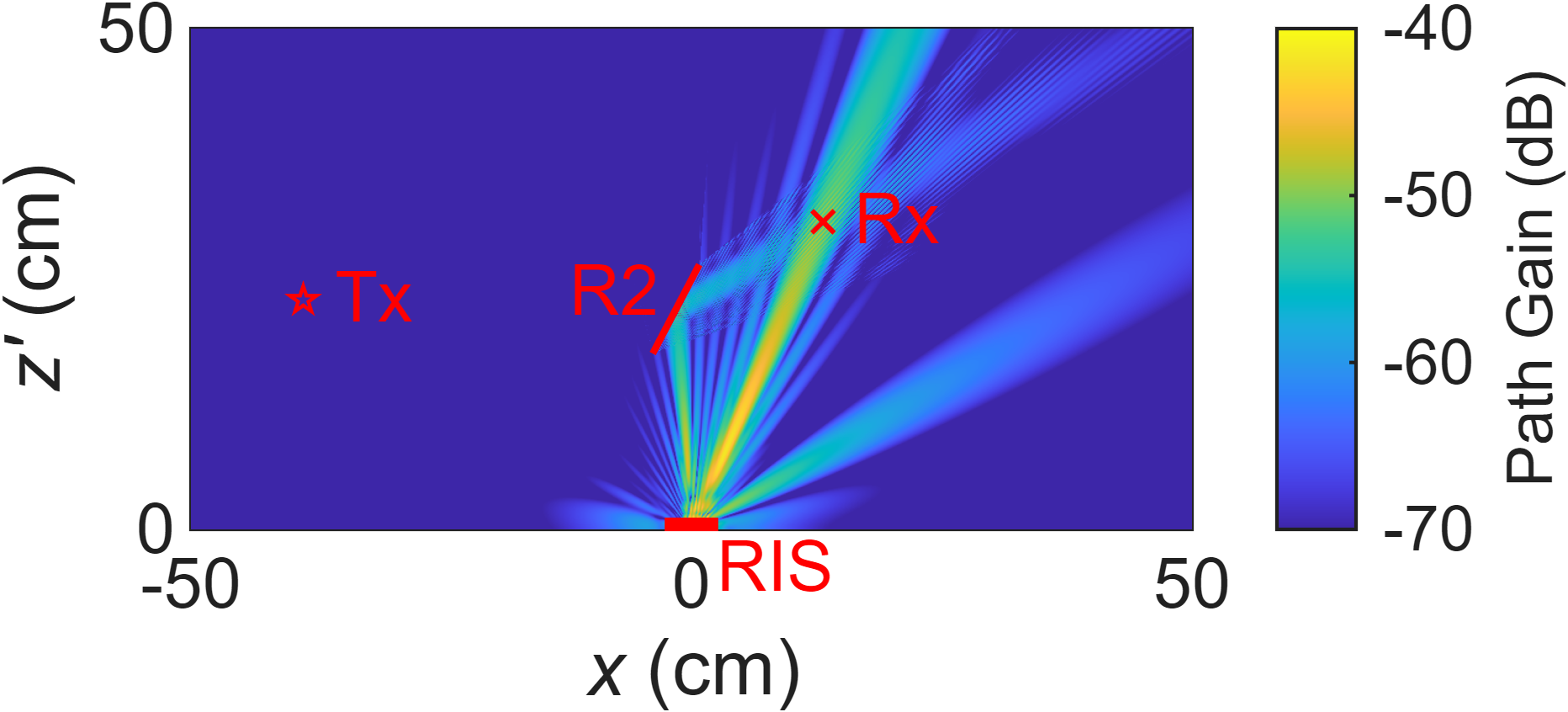}}
\\
\subfloat[With R1 and R2: HFSS reference.]
{\includegraphics[height=\pecfigheight]{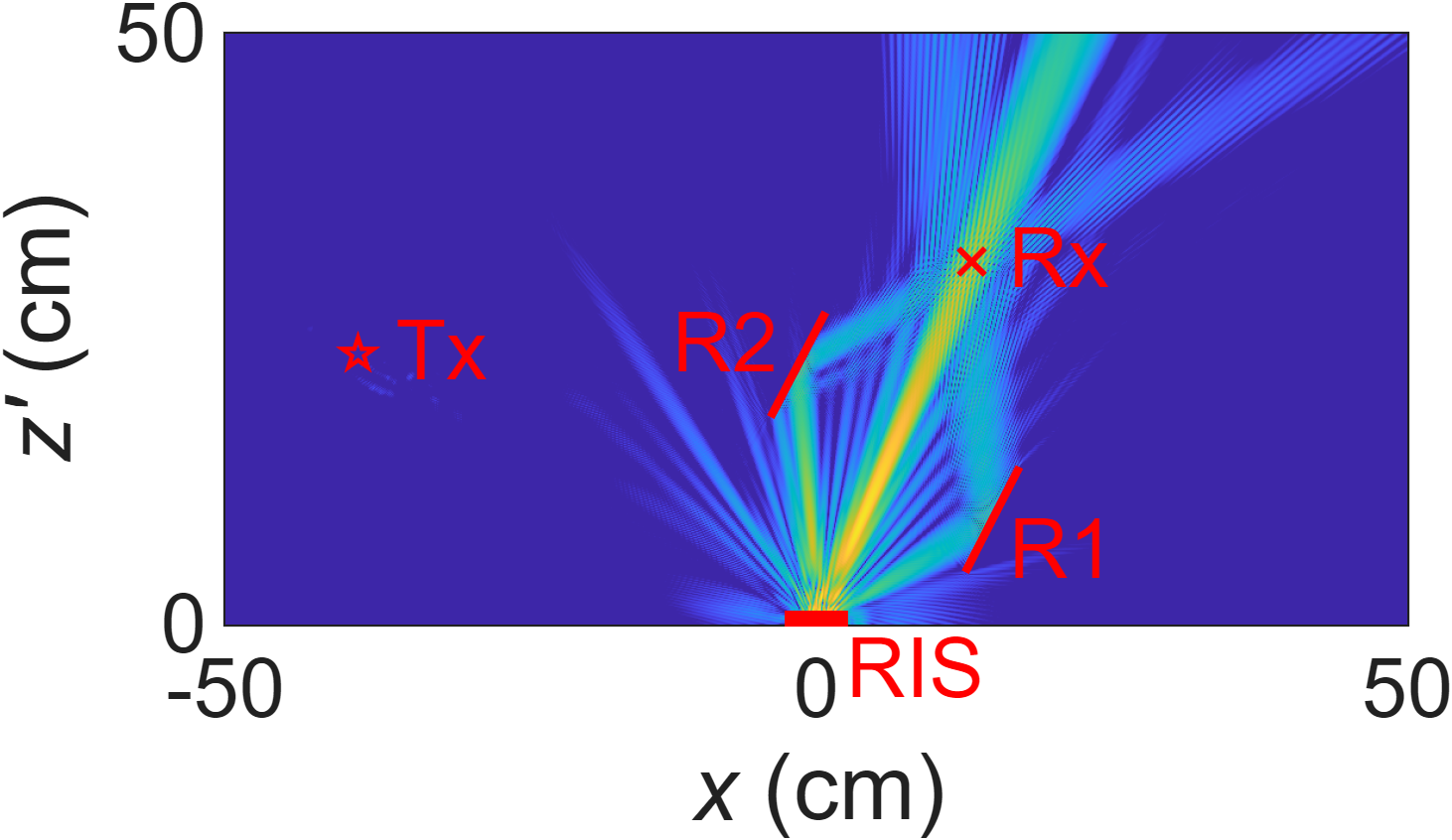}}
\hfil
\subfloat[With R1 and R2: general model.]
{\includegraphics[height=\pecfigheight]{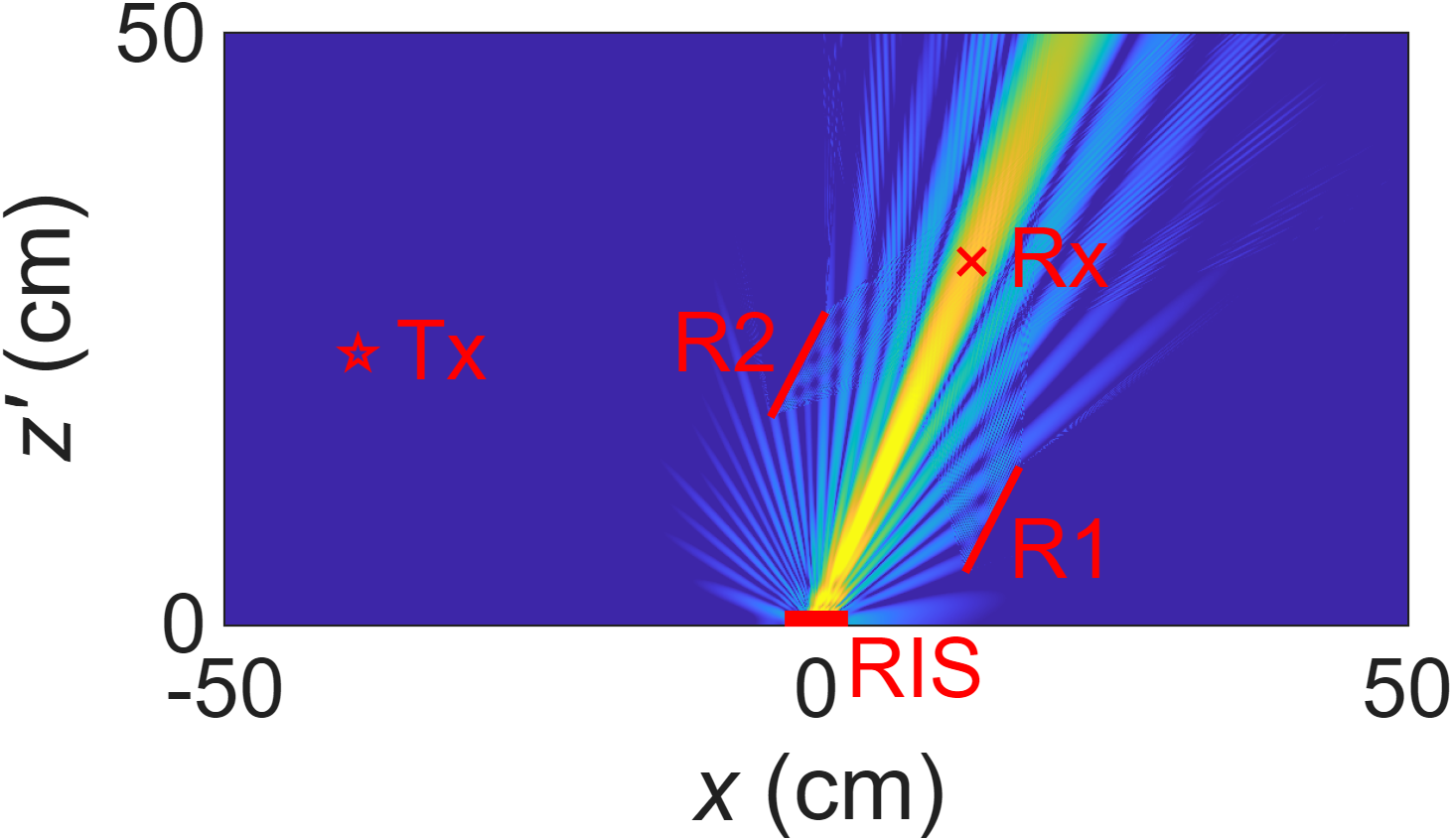}}
\hfil
\subfloat[With R1 and R2: calibrated model.]
{\includegraphics[height=\pecfigheight]{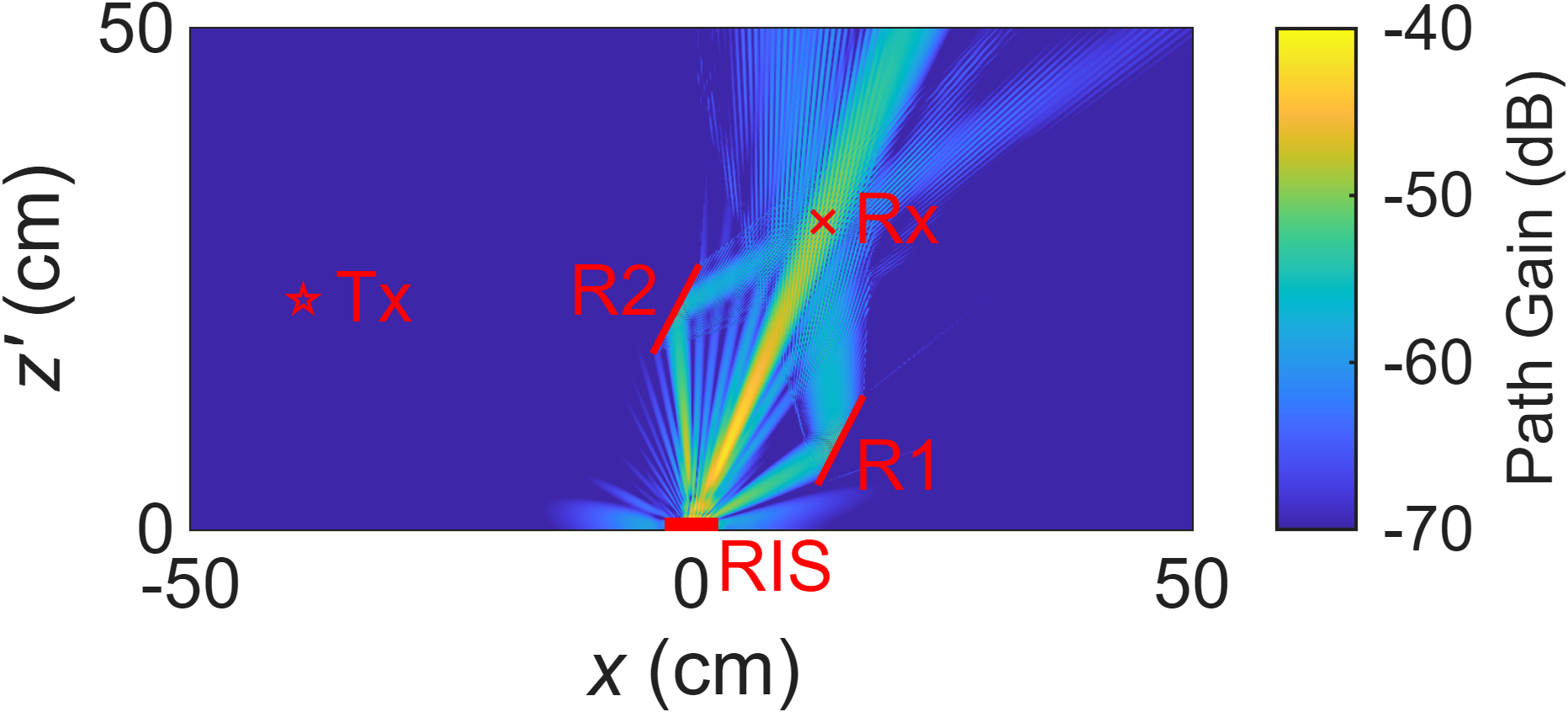}}
\caption{RIS-induced channel gain on the observation plane, plotted against the in-plane coordinates $x$ and $z'$. Tx denotes its projection onto the plane; the nominal Rx lies on the plane.}
\label{fig:reflector_maps}
\end{figure*}
%%%%%%%%%%%%%%%%%%%%%%%%%%%%%%%%%%%%%%%%%%%%%%%%%%%%%%%%%%%%%%%%%%%%%%%

\subsection{Multipath Validation With PEC Reflectors}
\label{sec:pec_results}

The PEC-reflector geometry in \cref{fig:setup_PECreflectors} is designed to test whether the transferred Bragg-order coefficients remain valid when the RIS--Rx propagation paths change while the Tx--RIS illumination is kept unchanged. Four configurations are considered: no environmental reflector, reflector R1 only, reflector R2 only, and R1+R2. Each PEC-plate reflector is $10\times10$~cm$^2$. R1 is positioned in the zeroth-order departure sector and R2 in the second-order sector, each plate oriented to redirect the corresponding RIS component toward the nominal Rx. In the RIS-centered global coordinate system, the plate centers and unit normals are
\begin{align*}
\vect a_{\mathrm{R1}}
&=
[\,14.87,\,-1.20,\,8.87\,]^{\mathsf T}\ \mathrm{cm},
\\
\vect n_{\mathrm{R1}}
&=
[\,0.89,\,-0.04,\,-0.46\,]^{\mathsf T},
\\
\vect a_{\mathrm{R2}}
&=
[\,-1.54,\,-0.95,\,21.95\,]^{\mathsf T}\ \mathrm{cm},
\\
\vect n_{\mathrm{R2}}
&=
[\,-0.89,\,0.01,\,0.47\,]^{\mathsf T}.
\end{align*}
The RIS--R1 and RIS--R2 center distances are $17.35$ and $22.03$~cm, respectively, both lying well within the radiative near-field of the $50\times 50$ RIS, whose Fraunhofer distance is $1.22$~m.

Because this configuration contains only the direct RIS--Rx path and a small number of known single-reflection paths, the IM is used to construct the relevant reflection paths directly using the unfolding in \cref{sec:im_coupling}. The visibility factor $V_{mn}^{(\pathset)}$ is evaluated independently for every RIS element and every enabled path. For PEC reflection, the local field convention gives $\rho_{\mathrm{TE}}=-1$ and $\rho_{\mathrm{TM}}=+1$, which enter the polarization transformation represented by $\rho_{mn}^{(\pathset)}$ in \eqref{eq:path_kernel}.

%%%%%%%%%%%%%%%%%%%%%%%%%%%%%%%%%%%%%%%%%%%%%%%%%%%%%%%%%%%%%%%%%%%%%%%
\begin{table}[!t]
\caption{RIS-induced gain at the nominal Rx for the PEC-reflector validation. The calibrated model uses the transferred $25\times25$ coefficients. Parentheses give signed gain errors relative to the numerical reference.}
\label{tab:reflector_results}
\centering
\begin{tabular}{@{}c c c c@{}}
\toprule
Reflectors & Reference (dB) & General (dB) & Calibrated (dB)\\
\midrule
None & $-49.64$ & $-44.13$ $(+5.51)$ & $-50.44$ $(-0.80)$ \\
R1 & $-47.28$ & $-43.93$ $(+3.35)$ & $-48.07$ $(-0.79)$ \\
R2 & $-47.46$ & $-44.28$ $(+3.18)$ & $-48.33$ $(-0.87)$ \\
R1 and R2 & $-45.69$ & $-43.93$ $(+1.75)$ & $-46.57$ $(-0.88)$ \\
\bottomrule
\end{tabular}
\end{table}
%%%%%%%%%%%%%%%%%%%%%%%%%%%%%%%%%%%%%%%%%%%%%%%%%%%%%%%%%%%%%%%%%%%%%%%

The comparison is performed on a $1.0\times0.5$~m$^2$ observation plane containing the nominal Rx, with a sampling spacing of 1~mm. The $1001\times501$ HFSS sample grid is reused point-for-point by the proposed model and is independent of the upper-hemisphere calibration samples used for coefficient extraction. When a reflector is present, the numerical reference is obtained by hybrid finite-element/integral-equation (FEM-IE) analysis. Each simulation containing the Tx, RIS, and enabled environmental reflectors is paired with a background simulation that retains the Tx and the same reflectors but removes the RIS. Their complex-field difference, formed as in \eqref{eq:scattered_field}, gives the RIS-induced reference by removing the Tx--Rx and Tx--reflector--Rx bypass contributions.

\Cref{fig:reflector_maps} shows that the calibrated model reproduces both the no-reflector near-field structure and the reflector-induced changes more closely than the general model. In the reference column, enabling R1 (rows (d)--(f)) and R2 (rows (g)--(i)) each adds a distinct redirected beam that reaches the Rx region, and both appear together when the two plates are enabled (rows (j)--(l)). The calibrated column reproduces these beams, whereas the general-model column does not, because the zeroth- and second-order components that feed them are strongly underestimated. The no-reflector configuration additionally provides an independent near-field test of the transferred coefficients on the observation plane, since these samples are not used in the upper-hemisphere calibration fit of \eqref{eq:wls}.

The RIS-induced gains at the nominal Rx are summarized in \cref{tab:reflector_results}. Without reflectors, the general model overestimates the reference by $5.51$~dB. The error decreases to $3.35$, $3.18$, and $1.75$~dB for R1, R2, and R1+R2, respectively, because the reference additionally contains RIS energy redirected through the zeroth- and second-order sectors that the general model underestimates. The calibrated-model errors remain between $-0.88$ and $-0.79$~dB across all four configurations. Because the transferred coefficient vector is unchanged among the four cases, these results directly test whether the same calibration remains predictive as the RIS--Rx reflector configuration changes.

\subsection{End-to-End Validation in a Scaled RCC Room}
\label{sec:room_results}

%%%%%%%%%%%%%%%%%%%%%%%%%%%%%%%%%%%%%%%%%%%%%%%%%%%%%%%%%%%%%%%%%%%%%%%
\begin{figure}[t]
\centering
\includegraphics[width=0.9\linewidth]{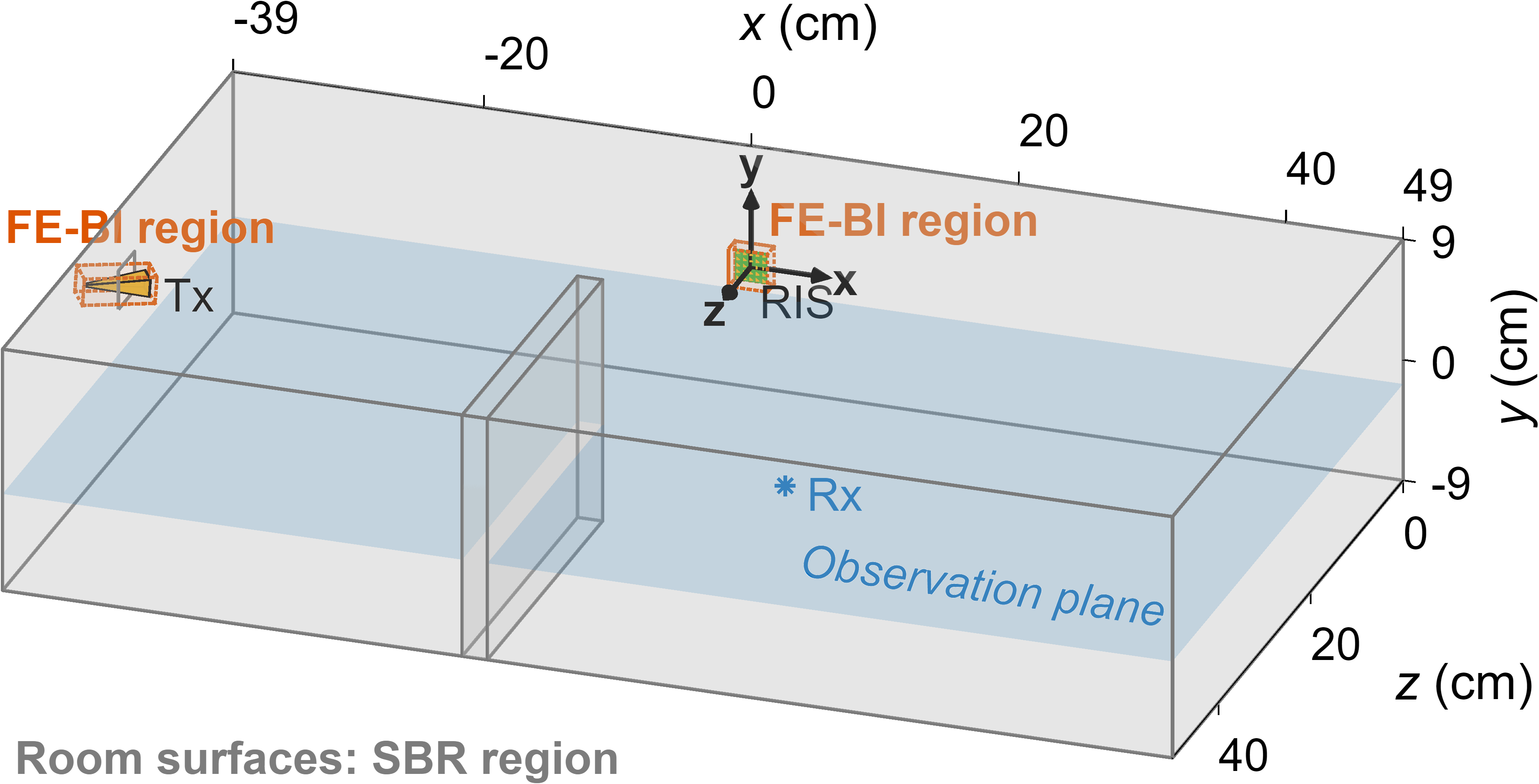}
\caption{Approximately $1/16$-scale RCC-room configuration. FE-BI regions enclose the Tx and the RIS, the room surfaces form the SBR region, and the light-blue plane is the nominal-Rx-height observation plane.}
\label{fig:room_geometry}
\end{figure}
%%%%%%%%%%%%%%%%%%%%%%%%%%%%%%%%%%%%%%%%%%%%%%%%%%%%%%%%%%%%%%%%%%%%%%%

\subsubsection{Setup and Numerical Reference}

To keep the hybrid FEM-SBR reference computationally tractable, an approximately $1/16$-scale RCC room is used for the end-to-end validation. Even with SBR treatment of the room surfaces, substantially larger room geometries or RIS apertures are impractical with the available computational resources. The scaled room is therefore used to evaluate the proposed RIS--RT formulation in an enclosed multipath environment rather than to represent the corresponding full-size room.

The room geometry is shown in \cref{fig:room_geometry}. An internal wall separates the Tx from the region containing the RIS and the Rx, so the direct Tx--Rx path is blocked over most of the observation plane and the RIS provides the intended coverage mechanism there; the bypass paths that do reach this region arrive through wall reflections and therefore sample the Tx pattern at wide departure angles. The transferred $25\times25$ coefficient vector used in the PEC-reflector validation is retained for the $50\times50$ target RIS, with no room-specific refitting. The room surfaces are modeled as smooth RCC with $\varepsilon_{\mathrm r}=5.3$ and $\tan\delta=0.06$.

The numerical reference uses hybrid finite-element/SBR (FEM-SBR) analysis: the Tx and RIS are solved by the full-wave FEM solver, whereas the electrically large room is treated by SBR. 
The reference quantity is the global-$y$ co-polarized component of the full-scene total field, including both RIS-assisted and Tx--Rx bypass contributions.
The total field is converted to the channel coefficient using \eqref{eq:field_to_channel}.
Direct and specularly reflected paths with up to three reflections are included. Neither the uniform theory of diffraction (UTD) nor the physical theory of diffraction (PTD) is enabled in the HFSS SBR reference or the MATLAB RT calculation, so the environmental propagation mechanisms considered in the two calculations are matched.

For software compatibility, MATLAB SBR is used at 100~GHz only to identify the reflector sequences. The same sequences are then evaluated at 154~GHz, where the propagation phase, material reflection, and polarization transformation are recomputed at the operating frequency.
For every retained RIS--Rx ray, its ordered reflector sequence is then unfolded by image theory to obtain $\vect r_{\mathrm{img}}^{(\pathset)}$, which is used to evaluate the per-element RIS--Rx spreading and propagation phase in \eqref{eq:path_kernel}. Thus, SBR performs multipath discovery, whereas image theory provides the equivalent straight element-to-image geometry required by the RIS model. Propagation phase, material reflection, and polarization transformation are evaluated at the operating frequency.

The observation domain is the nominal-Rx-height plane at $y=-18.75$~mm, sampled on a 1-cm grid anchored at the nominal Rx. Samples inside the internal wall or within 1~cm of a room boundary are excluded, which leaves $N_{\mathrm{obs}}=3980$ observation points.

\subsubsection{Tx-Pattern Effects on Channel Components}

For the pyramidal-horn Tx used in this study, the fitted model in \eqref{eq:cosq_pattern} is obtained from the simulated co-polarized E- and H-plane realized-gain cuts over the 3-dB main-beam region. The resulting exponents are $q_{\mathrm E}=141$ and $q_{\mathrm H}=114$, corresponding to E- and H-plane half-power beamwidths of $8.01^\circ$ and $8.91^\circ$, respectively.

\begin{figure}[t]
\centering
\newcommand{\bypassfigheight}{3.6cm}
\subfloat[]{\includegraphics[height=\bypassfigheight]{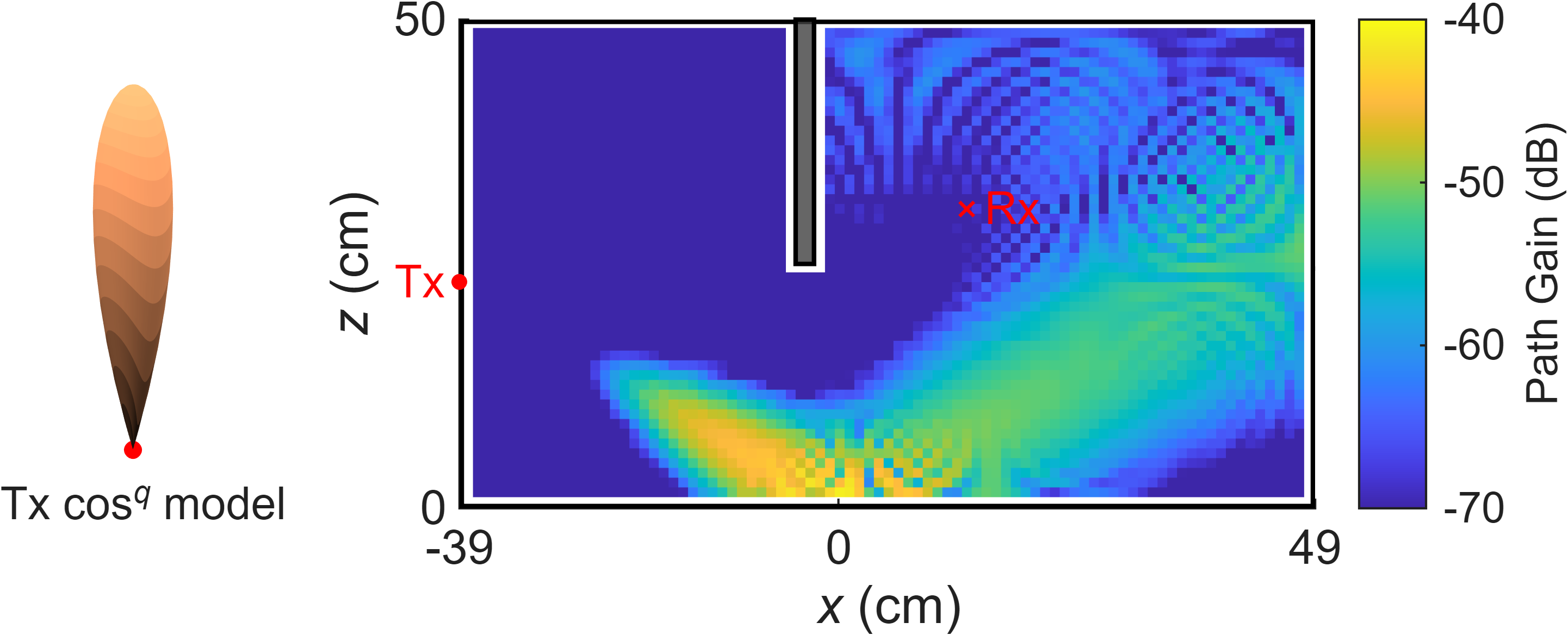}}
\\
\subfloat[]{\includegraphics[height=\bypassfigheight]{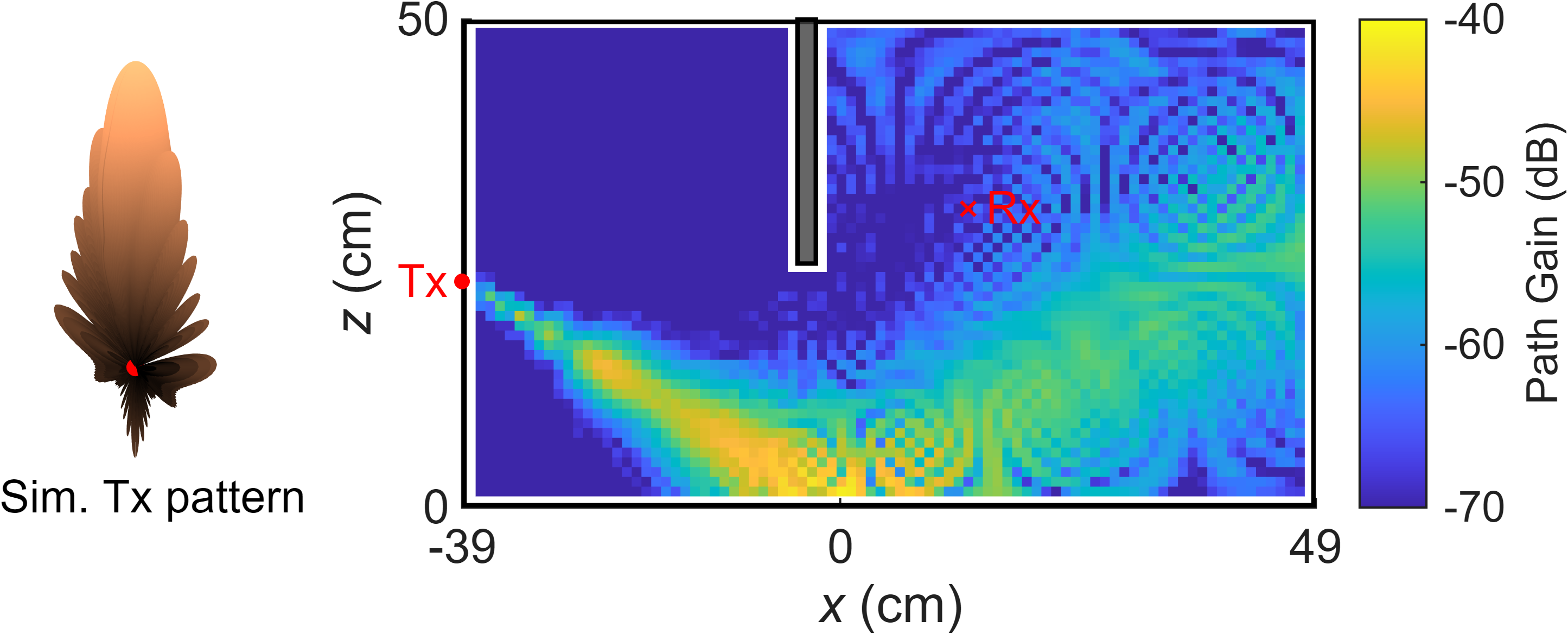}}
\caption{Tx--Rx bypass channel gain $20\log_{10}\left|h_{\by}\right|$ in the RCC room using (a) the fitted cosine-$q$ Tx pattern and (b) the simulated realized-gain pattern. The corresponding Tx patterns are shown as insets.}
\label{fig:maps_bypass}
\end{figure}

The RIS subtends only a narrow angular sector around boresight, over which the fitted cosine-$q$ pattern closely follows the simulated main beam. Tx--Rx bypass paths, in contrast, span a much wider departure-angle range and can intersect sidelobes and nulls. 
%\Cref{fig:maps_bypass} compares the bypass fields obtained by reweighting the same SBR path set with the fitted pattern in \eqref{eq:cosq_pattern} and the simulated absolute realized-gain pattern in \eqref{eq:sim_tx_pattern}. 
\Cref{fig:maps_bypass} compares the bypass fields obtained using the simulated pattern in \eqref{eq:sim_tx_pattern} with those obtained by reweighting the same SBR path set using the fitted pattern in \eqref{eq:cosq_pattern}.
Because the path geometry, material reflection, propagation phase, Rx treatment, and polarization are unchanged, this comparison isolates the effect of Tx directional-gain modeling. The largest discrepancies occur in regions reached through wide-angle Tx departures, where the fitted pattern cannot reproduce the sidelobes of the simulated antenna pattern.

\begin{figure}[t]
\centering
\newcommand{\assistedfigheight}{3.6cm}
\subfloat[]{\includegraphics[height=\assistedfigheight]{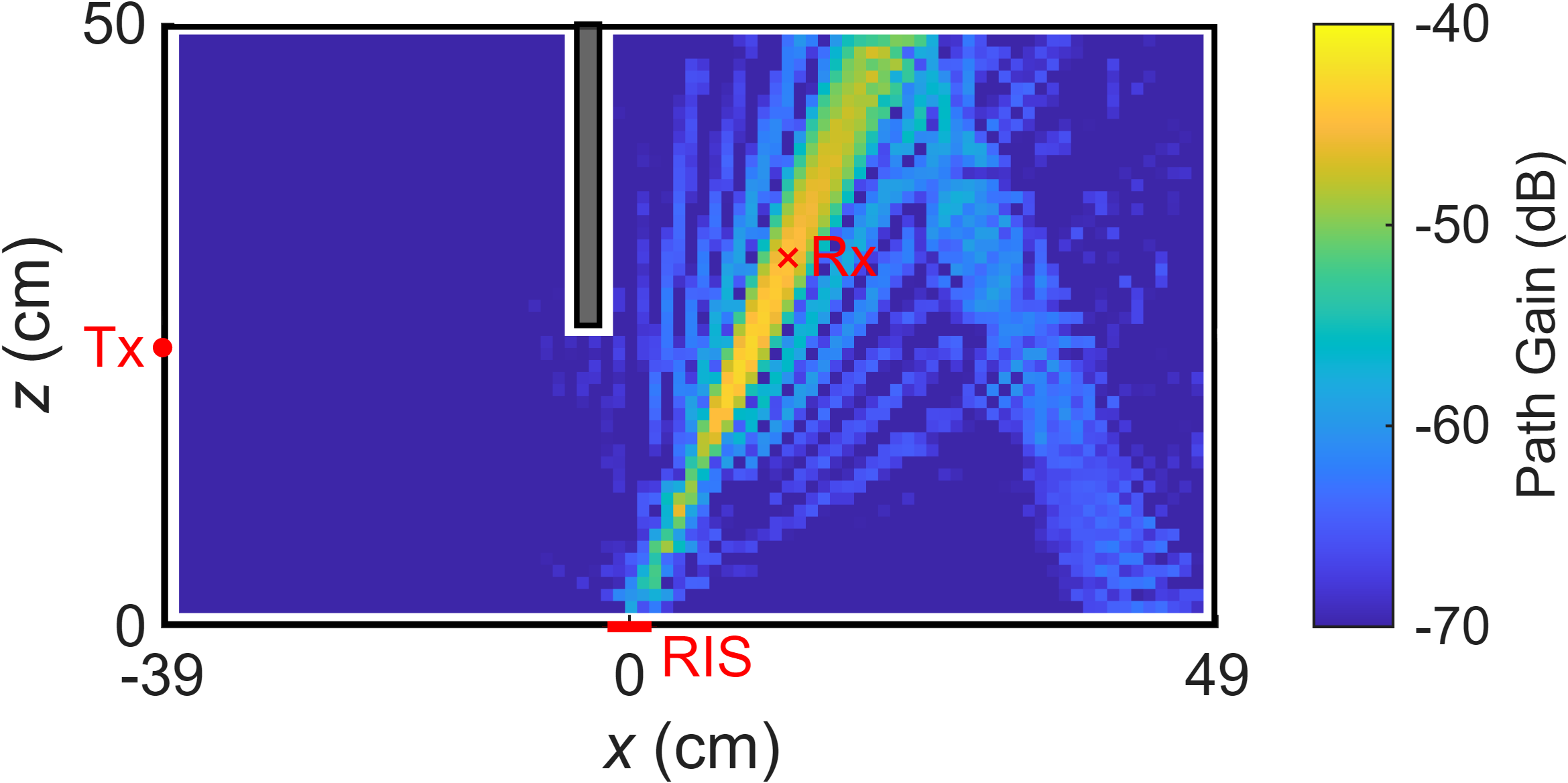}}
\\
\subfloat[]{\includegraphics[height=\assistedfigheight]{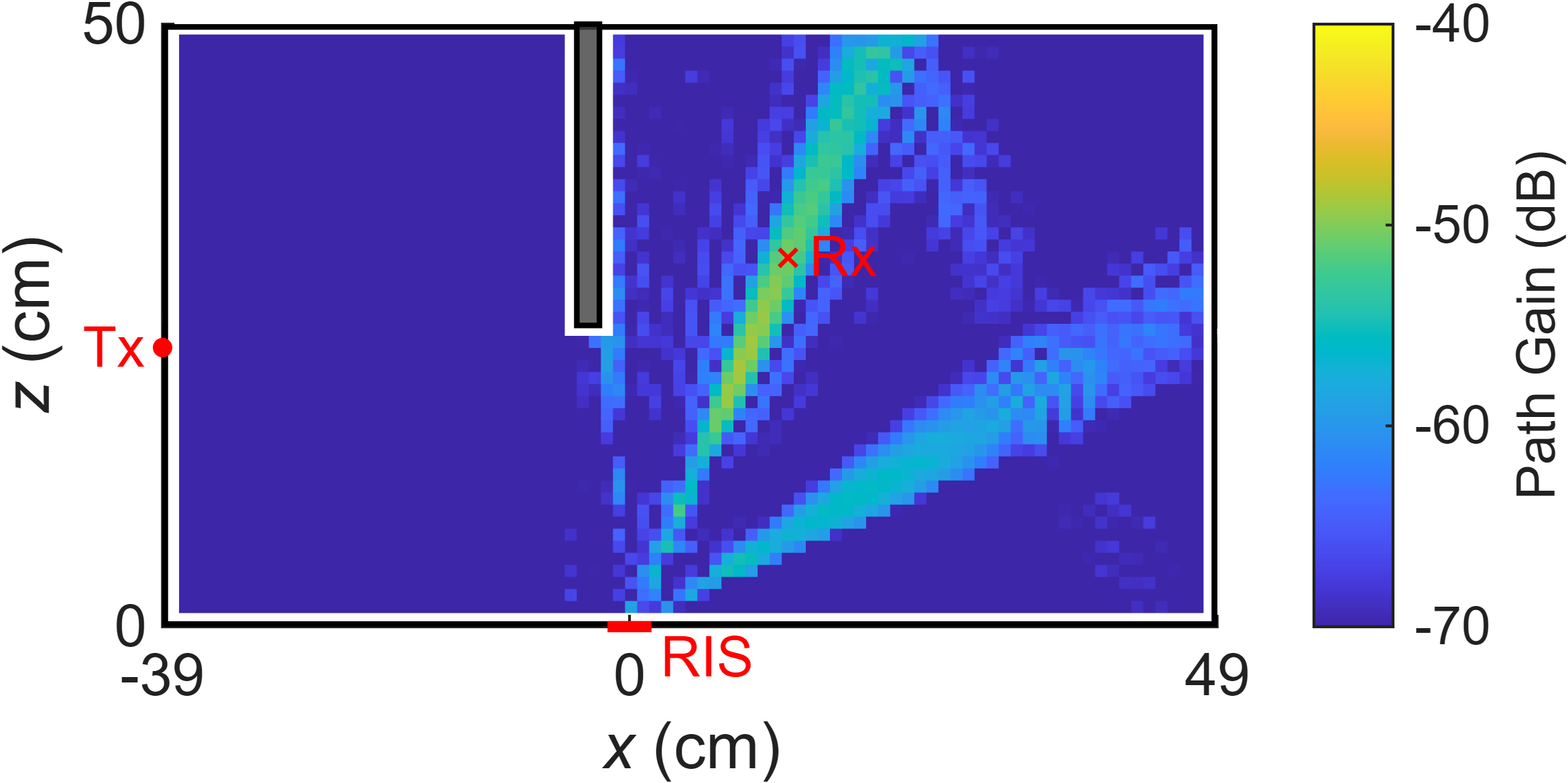}}
\caption{RIS-assisted channel gain $20\log_{10}\left|h_{\RIS,\varMethod}\right|$ in the RCC room using the fitted cosine-$q$ Tx pattern: (a) the general model and (b) the calibrated model.}
\label{fig:maps_assisted}
\end{figure}

\Cref{fig:maps_assisted} compares the RIS-assisted contributions predicted by the general and calibrated RIS models. This contribution is only weakly sensitive to the Tx-pattern representation because the RIS remains within the fitted main-beam sector; replacing the fitted pattern with the simulated pattern changes the RIS-assisted channel gain by no more than 0.1~dB over the observation region. Accordingly, only the fitted cosine-$q$ illumination is used in \cref{fig:maps_assisted}. The calibrated model recovers additional spatial structure produced by parasitic-order RIS-assisted paths that is absent or strongly underestimated in the general model.

%%%%%%%%%%%%%%%%%%%%%%%%%%%%%%%%%%%%%%%%%%%%%%%%%%%%%%%%%%%%%%%%%%%%%%%
%%%%%%%%%%%%%%%%%%%%%%%%%%%%%%%%%%%%%%%%%%%%%%%%%%%%%%%%%%%%%%%%%%%%%%%

%%%%%%%%%%%%%%%%%%%%%%%%%%%%%%%%%%%%%%%%%%%%%%%%%%%%%%%%%%%%%%%%%%%%%%%
%%%%%%%%%%%%%%%%%%%%%%%%%%%%%%%%%%%%%%%%%%%%%%%%%%%%%%%%%%%%%%%%%%%%%%%

%%%%%%%%%%%%%%%%%%%%%%%%%%%%%%%%%%%%%%%%%%%%%%%%%%%%%%%%%%%%%%%%%%%%%%%
\begin{figure*}[t]
\centering
\newcommand{\roomfigheight}{3.52cm}
\begin{minipage}[c]{0.29\textwidth}
\centering
\subfloat[HFSS (hybrid FEM-SBR)]
{\includegraphics[height=\roomfigheight]{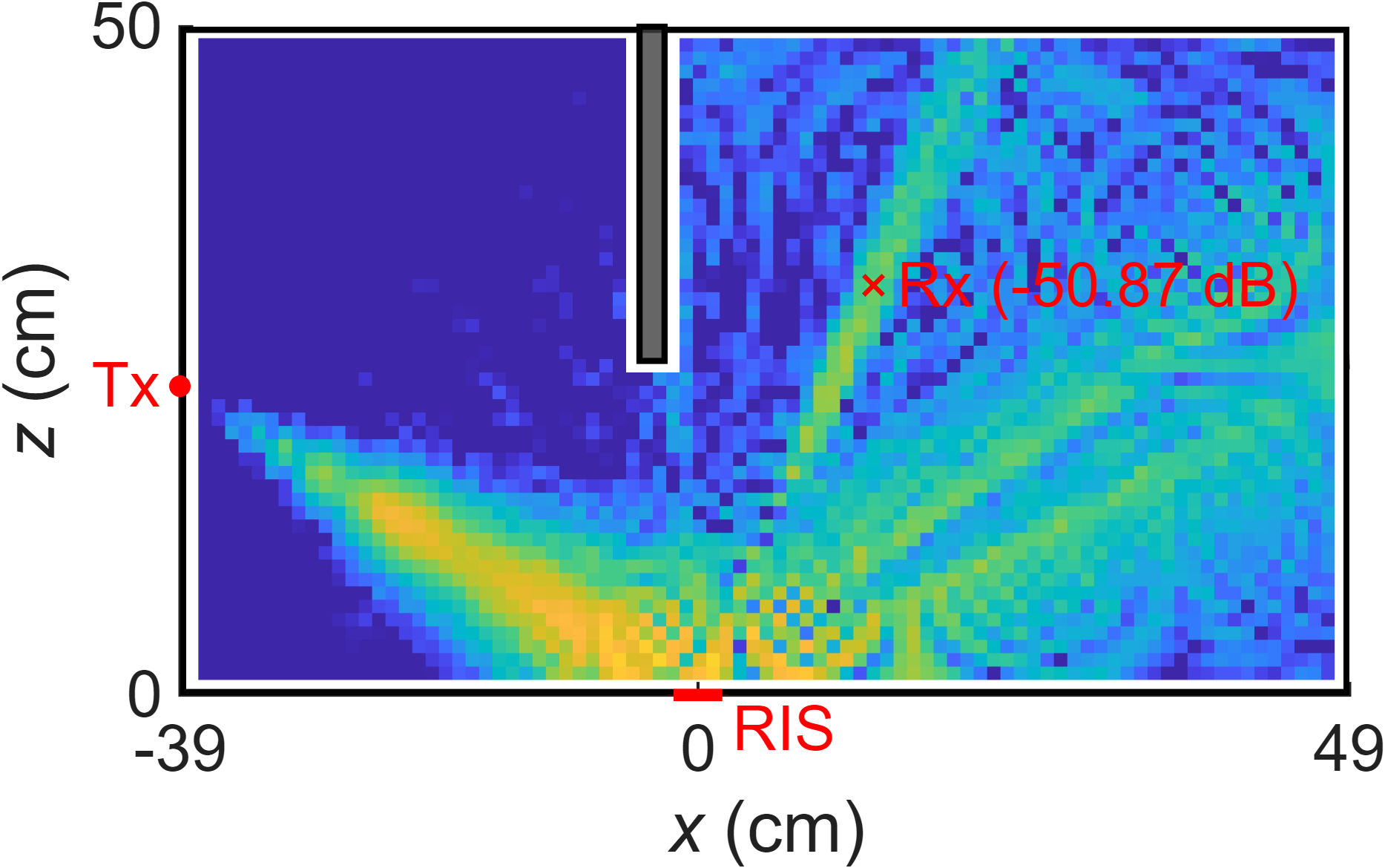}}
\end{minipage}%
\hfill
\begin{minipage}[c]{0.70\textwidth}
\centering
\subfloat[General model (cosine-$q$ Tx)]
{\includegraphics[height=\roomfigheight]{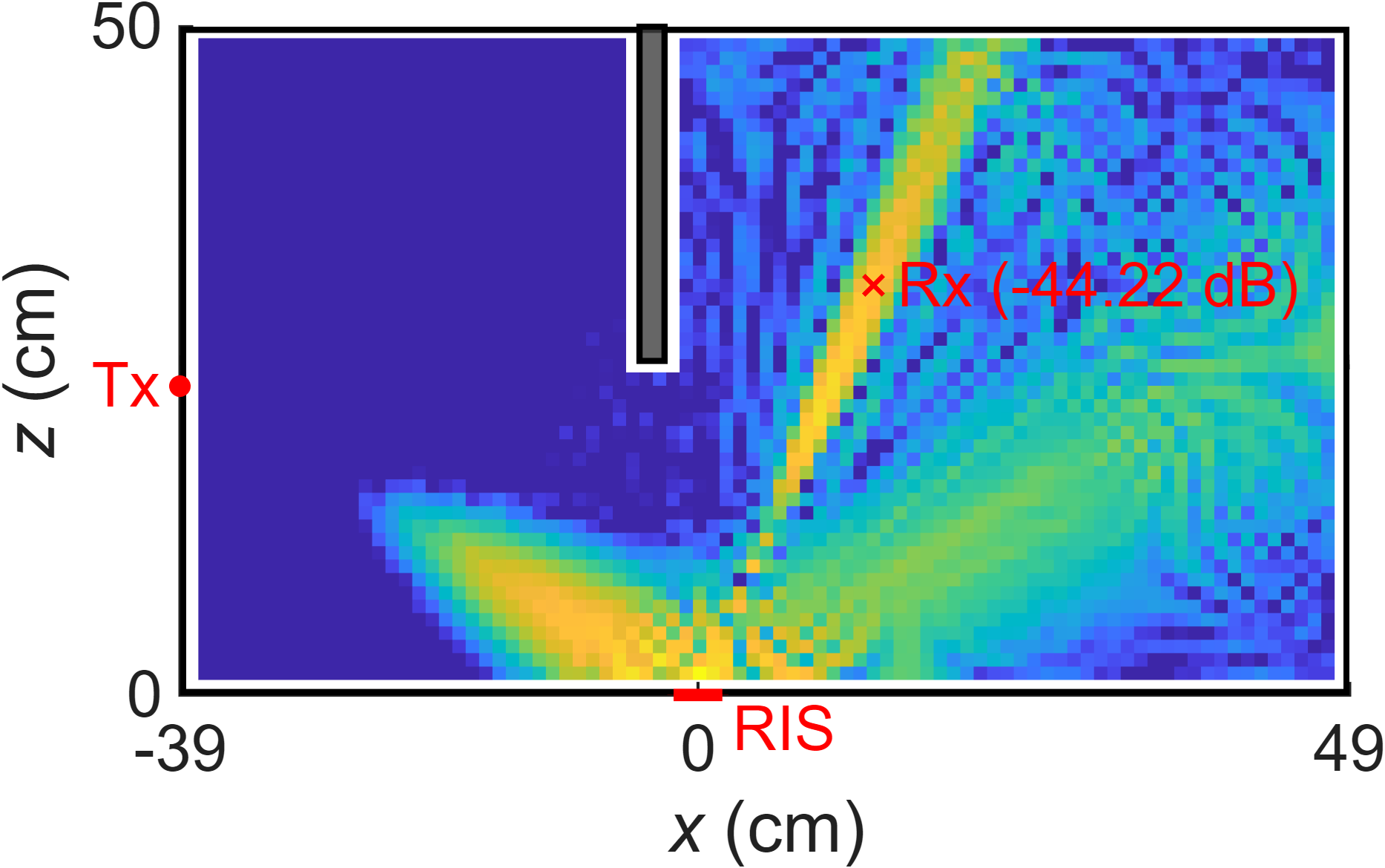}}%
\subfloat[Calibrated model (cosine-$q$ Tx)]
{\includegraphics[height=\roomfigheight]{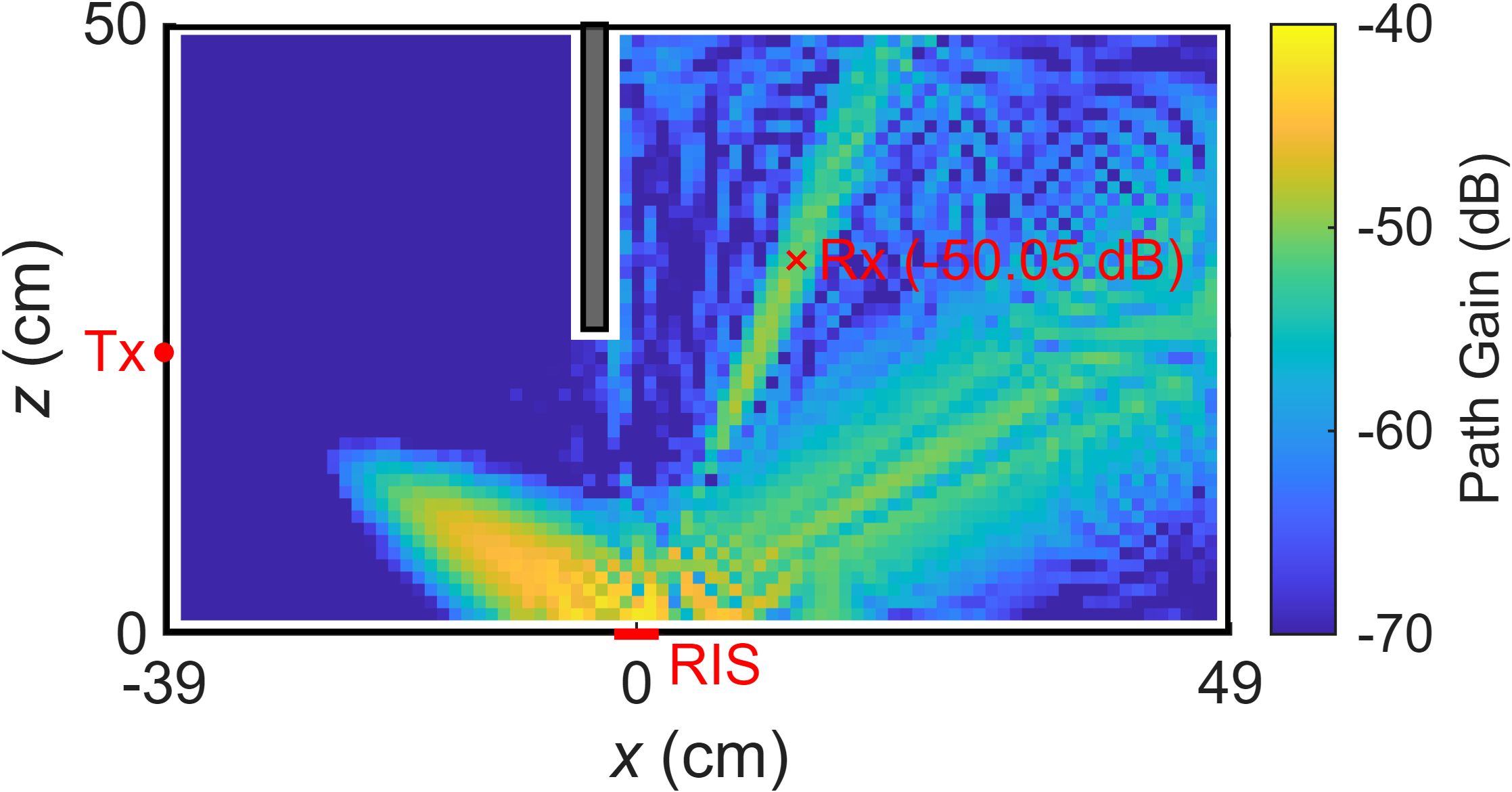}}%
\\
\subfloat[General model (simulated Tx)]
{\includegraphics[height=\roomfigheight]{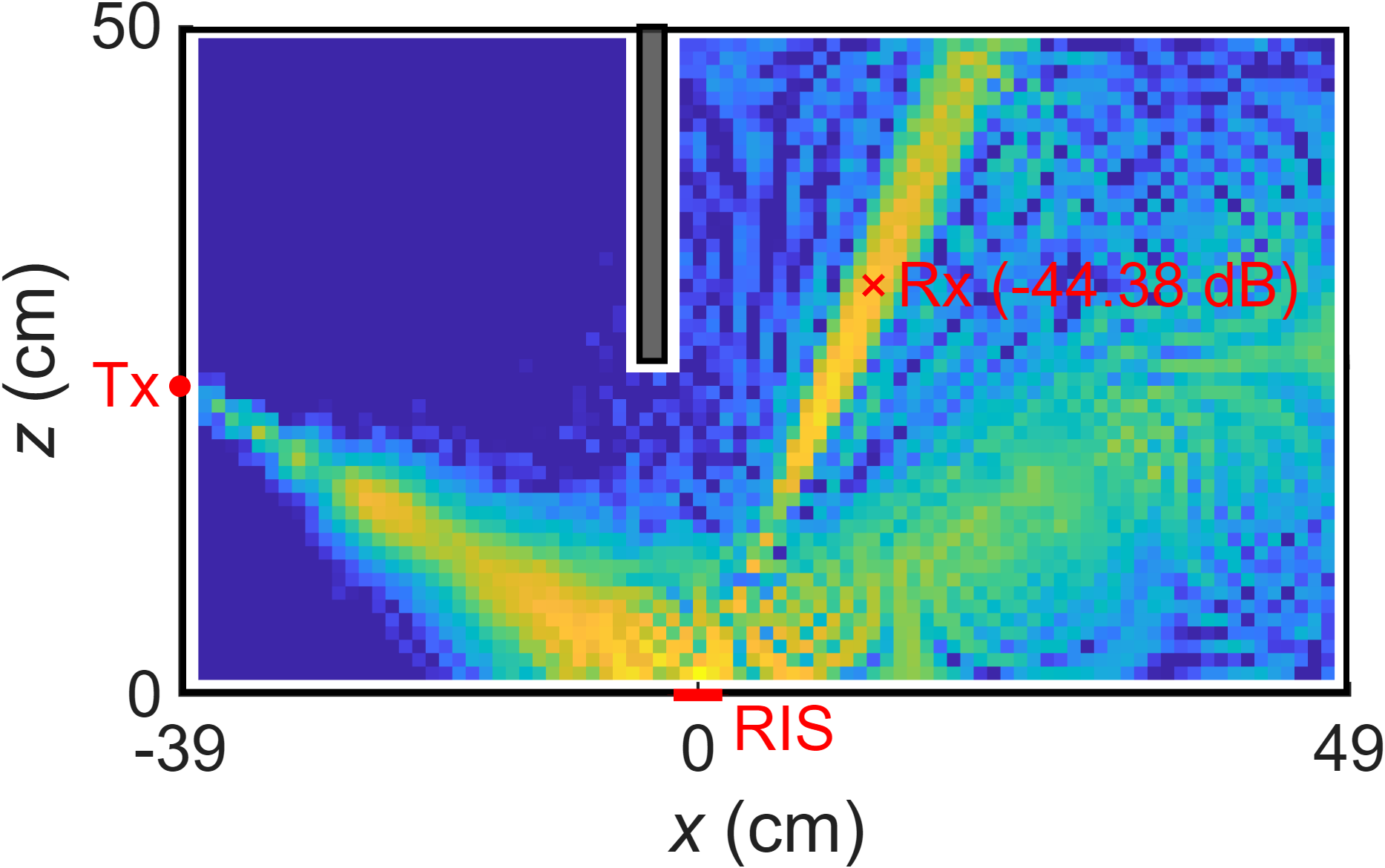}}%
\subfloat[Calibrated model (simulated Tx)]
{\includegraphics[height=\roomfigheight]{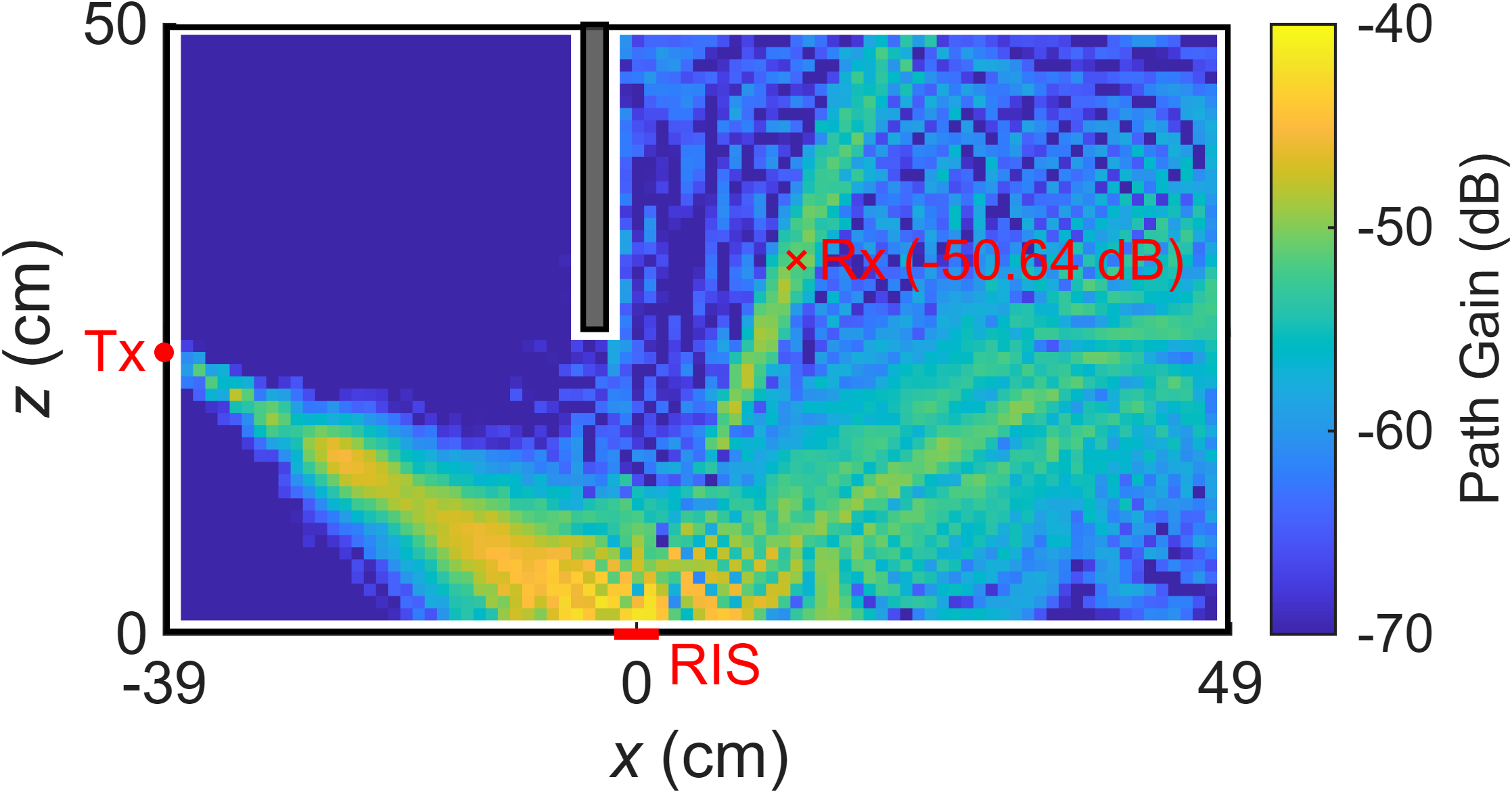}}%
\end{minipage}
\caption{End-to-end total channel gain $20\log_{10}\left|h_{\tot,\varMethod}\right|$ in the RCC room: (a) hybrid FEM-SBR reference, and (b)--(e) the general and calibrated RIS models under the two Tx-pattern representations.}
\label{fig:room_total_maps}
\end{figure*}
%%%%%%%%%%%%%%%%%%%%%%%%%%%%%%%%%%%%%%%%%%%%%%%%%%%%%%%%%%%%%%%%%%%%%%%

%%%%%%%%%%%%%%%%%%%%%%%%%%%%%%%%%%%%%%%%%%%%%%%%%%%%%%%%%%%%%%%%%%%%%%%
\begin{figure}[t]
\centering
\includegraphics[height=4.2cm]{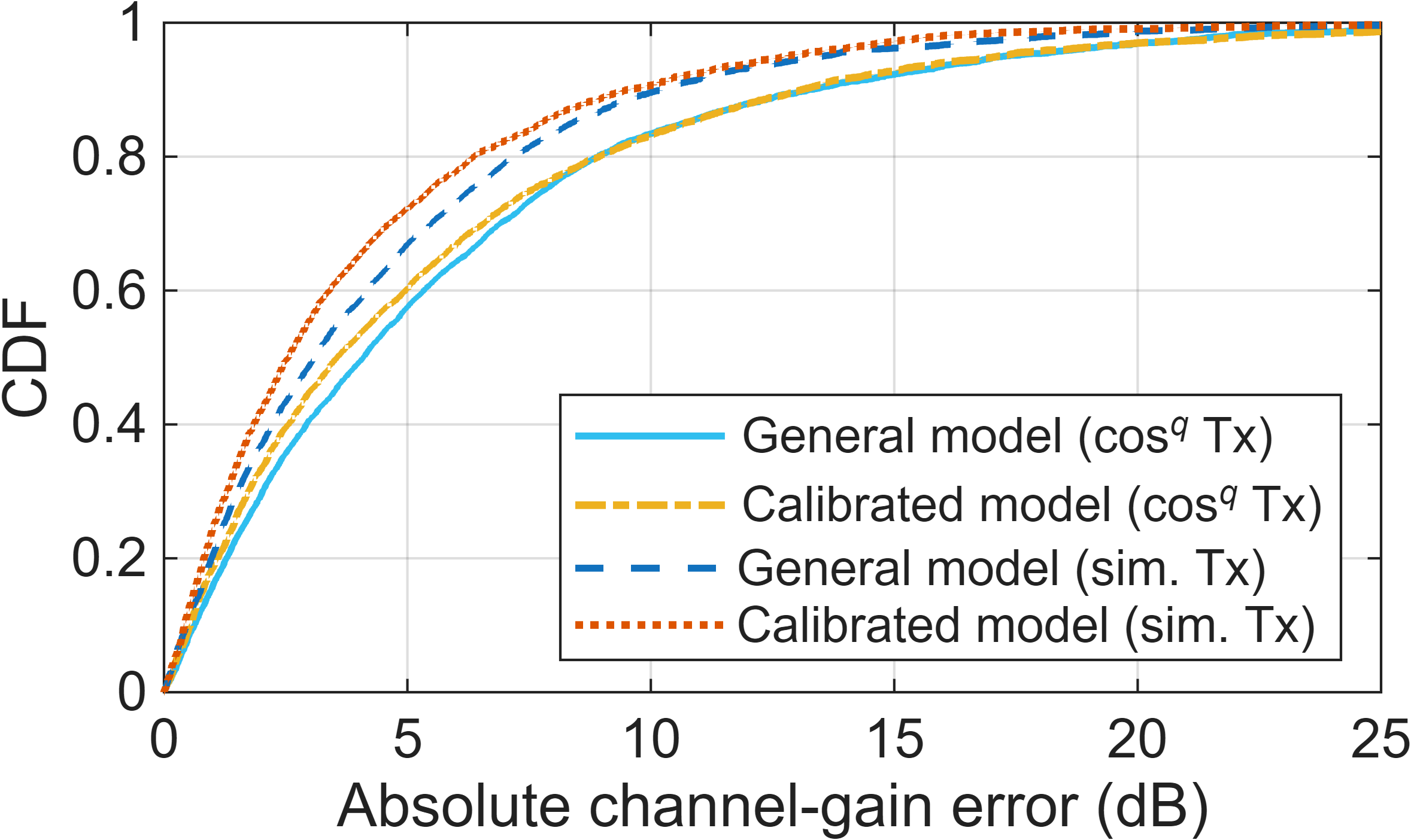}
\caption{CDFs of the absolute channel-gain error $|\Delta G_{\varMethod}|$ over the 3980 observation points.}
\label{fig:room_cdf}
\end{figure}
%%%%%%%%%%%%%%%%%%%%%%%%%%%%%%%%%%%%%%%%%%%%%%%%%%%%%%%%%%%%%%%%%%%%%%%

\subsubsection{End-to-End Total Channel-Gain Results}

\Cref{fig:room_total_maps} compares the coherent total channel gain obtained by summing all RIS-assisted and Tx--Rx bypass contributions. The effect of the RIS model is observed by comparing (b) with (c) and (d) with (e), whereas the effect of the Tx-pattern representation is seen by comparing (b) with (d) and (c) with (e). RIS calibration primarily modifies the structure produced by RIS-assisted paths, while Tx-pattern fidelity affects a broader set of regions reached by Tx--Rx bypass paths. Around the nominal Rx, the calibrated RIS model improves agreement with the hybrid HFSS reference, whereas the simulated Tx pattern improves the prediction over regions reached through wider-angle Tx departures.

The quantitative results are summarized in \cref{tab:room_metrics}, with the corresponding absolute-error distributions shown in \cref{fig:room_cdf}. 
%Using the fitted cosine-$q$ Tx pattern, the general and calibrated RIS models yield spatial RMSEs of $7.96$ and $7.89$~dB, respectively, while the nominal-Rx error decreases from $6.65$ to $0.82$~dB. Replacing the fitted Tx pattern with the simulated realized-gain pattern reduces the RMSEs to $6.39$ and $5.86$~dB and the $P_{90}$ errors to $10.19$ and $9.52$~dB. The calibrated model with the simulated Tx pattern also gives the lowest $P_{80}$ error, $6.36$~dB, and reduces the nominal-Rx error to $0.23$~dB, compared with $6.49$~dB for the general model.
At the nominal Rx, RIS calibration reduces the absolute gain error from $6.65$ to $0.82$~dB with the fitted cosine-$q$ Tx pattern and from $6.49$ to only $0.23$~dB with the simulated Tx pattern. 
The latter combination therefore provides an accurate prediction at the prescribed RIS focusing point.
Together with the sub-$0.9$-dB nominal-Rx errors obtained for all four PEC-reflector configurations in \cref{tab:reflector_results}, this result shows that the improved prediction at the prescribed focusing position is maintained across distinct RIS--Rx multipath environments using the same transferred coefficient vector.
Over the complete observation plane, replacing the fitted Tx pattern with the simulated realized-gain pattern reduces the RMSE from $7.96$ to $6.39$~dB for the general model and from $7.89$ to $5.86$~dB for the calibrated model. The calibrated model with the simulated Tx pattern also gives the lowest $P_{80}$ and $P_{90}$ errors, $6.36$ and $9.52$~dB, respectively.

These results show that RIS calibration and Tx-pattern fidelity address different error sources. The Bragg-order calibration is most influential where RIS-assisted paths dominate, particularly near the nominal Rx, whereas accurate wide-angle Tx characterization is important where Tx--Rx bypass paths dominate. This distinction is also reflected in \cref{fig:room_cdf}: the two cosine-$q$ cases have similar error distributions, while replacing the fitted Tx pattern with the simulated pattern shifts both distributions toward smaller errors; the calibrated model with the simulated Tx pattern gives the best overall agreement.

\subsection{Sources of Residual Error}
\label{sec:room_error}

The residual discrepancies in \cref{fig:room_total_maps} can be attributed in part to Tx--RIS illumination mismatch. In the PEC-reflector validation, the environmental reflectors are confined to the RIS--Rx side, so the RIS illumination remains essentially unchanged from that used for calibration. In the enclosed room, additional Tx-side reflections can coherently perturb the amplitude and phase distribution of the incident field across the RIS aperture. The resulting Bragg-order excitation can therefore differ from that characterized under the isolated LoS-dominant illumination, while the transferred coefficient vector remains unchanged.

A second source of error is the sensitivity of coherent multipath interference to path length. At 154~GHz, submillimeter path-length differences can noticeably shift deep interference nulls, producing large pointwise gain errors and increasing the spatial RMSE and percentile errors even when the main spatial gain pattern is reproduced.

%The present model assumes fixed, LoS-dominant Tx--RIS illumination. Substantial changes in incidence direction, polarization, operating frequency, or RIS phase configuration require recalibration, while Tx-side environmental multipath can perturb the calibrated aperture response. A natural extension is to characterize the RIS under a limited set of incidence conditions and combine the corresponding aperture responses according to the Tx-side paths identified by RT.
The present cross-aperture transfer is demonstrated only for the prescribed nominal-Rx focusing condition. Changing the nominal focusing point changes the prescribed phase profile and element-state arrangement, thereby changing the aperture harmonics represented by the Bragg-order bases. Since $\beta_\ell$ characterizes the practical electromagnetic weighting of these harmonics for a prescribed RIS configuration, cross-configuration coefficient reuse is not assumed here. A practical extension is to construct a compact calibration codebook rather than calibrating every focusing point independently. For each calibrated coefficient set, the spatial region over which it can be reused within a prescribed channel-gain error tolerance can be quantified, allowing a small number of coefficient vectors to cover the intended service area. Similar characterization can be extended to different Tx incidence conditions and Tx-side multipath illumination.

\subsection{Computational Efficiency and Validation Scope}
\label{sec:efficiency_results}

The proposed workflow reduces computation at two different stages. First, cross-aperture coefficient transfer replaces target-aperture full-wave calibration by the smaller $25\times25$ characterization quantified in \cref{tab:calibration_cost}. Second, after the coefficients have been fixed, RT and image-theory channel evaluation replace a new hybrid HFSS solve for each environment. The end-to-end computational costs of the two multipath validations are summarized in \cref{tab:efficiency}. The calibrated entries include the one-time $25\times25$ characterization followed by RT path construction and channel evaluation, whereas the general-model entries include only the latter two stages. For background-subtracted HFSS quantities, the reported wall-clock time is the sum of the paired simulations, while the reported peak RAM is the larger peak of the pair.

All measurements were obtained on an Intel Xeon Silver 4210R workstation with a 2.40-GHz base frequency and 768~GB of RAM. HFSS 2025 R2 was used for the numerical references, with a maximum allocation of four CPU cores for each simulation. An NVIDIA T600 GPU with CUDA acceleration was enabled for the SBR stage. The MATLAB RT and channel calculations were executed on the same workstation.

For the PEC-reflector validation, the hybrid FEM-IE reference requires $5.8$~h for $N_{\mathrm{obs}}=501\,501$ observation samples, compared with $1.5$~h for the calibrated IM workflow including the one-time calibration. For the scaled room, the hybrid FEM-SBR reference requires $49.3$~h for $N_{\mathrm{obs}}=3980$ observation samples, compared with $1.3$~h for the calibrated SBR workflow. These correspond to wall-clock reductions of approximately $3.9\times$ and $37.9\times$, respectively. Once the calibration has been completed, only RT and channel evaluation are required for an additional environment that preserves the calibrated RIS configuration and Tx-side illumination. For the two representative scenarios considered here, these incremental costs are approximately $0.4$~h and $0.2$~h, respectively.

The cost of the HFSS reference also increases rapidly with electrical size. Hybrid FEM-IE remains practical for the finite PEC-reflector configuration but becomes prohibitive for the electrically larger room, motivating the use of FEM-SBR. In a representative $50\times50$ room simulation, approximately $1.54\times10^6$ equivalent current sources are generated for SBR launching. Together with the isolated $100\times100$ characterization cost of $7.56$~h and $139$~GB in \cref{tab:calibration_cost}, this scaling makes larger-aperture hybrid room references impractical with the available computational resources.

%The reported validation is numerical rather than experimental. The PEC-reflector and room studies assess multipath prediction for the $50\times50$ target RIS with the transferred coefficient vector kept fixed, whereas cross-aperture scalability is established separately by the isolated-RIS characterization in \cref{sec:transfer_results}. The approximately $1/16$-scale room is used as a computationally tractable numerical stress test rather than as a direct full-scale path-loss prediction, since geometric scaling changes the absolute path lengths, delay structure, and spreading loss. Experimental validation remains necessary to quantify fabrication tolerances, material uncertainty, diffuse scattering, and alignment errors.

%%%%%%%%%%%%%%%%%%%%%%%%%%%%%%%%%%%%%%%%%%%%%%%%%%%%%%%%%%%%%%%%%%%%%%%
\begin{table}[t]
\caption{End-to-end gain-error statistics for the approximately $1/16$-scale RCC room. Nominal-Rx denotes $|\Delta G_{\varMethod}(\vect r_{\rx})|$.}
\label{tab:room_metrics}
\centering
\setlength{\tabcolsep}{7pt}
\begin{tabular}{@{}c c c c c c@{}}
\toprule
\multirow{2}{*}[-0.5ex]{\makecell{Tx\\pattern}} &
\multirow{2}{*}[-0.5ex]{\makecell{RIS\\model}} &
\multicolumn{4}{c}{Gain error (dB)}\\
\cmidrule(l{2pt}r{0pt}){3-6}
& & RMSE & $P_{80}$ & $P_{90}$ & Nominal-Rx\\
\midrule
cosine-$q$ & General &$7.96$ &$8.89$ &$13.25$ &$6.65$\\
cosine-$q$ & Calibrated &$7.89$ &$8.95$ &$13.10$ &$0.82$\\
Simulated & General &$6.39$ &$7.22$ &$10.19$ &$6.49$\\
Simulated & Calibrated &$5.86$ &$6.36$ &$9.52$ &$0.23$\\
\bottomrule
\end{tabular}
\end{table}
%%%%%%%%%%%%%%%%%%%%%%%%%%%%%%%%%%%%%%%%%%%%%%%%%%%%%%%%%%%%%%%%%%%%%%%

%%%%%%%%%%%%%%%%%%%%%%%%%%%%%%%%%%%%%%%%%%%%%%%%%%%%%%%%%%%%%%%%%%%%%%%
\begin{table}[t]
\caption{Measured computational cost for the two multipath validation scenarios.}
\label{tab:efficiency}
\centering
\setlength{\tabcolsep}{4pt}
\begin{tabular}{@{}c c c c@{}}
\toprule
Case & Method & Wall-clock time & Peak RAM\\
\midrule
\multirow{3}{*}{\makecell[c]{$50\times50$ RIS\\with two PEC\\reflectors}}
& HFSS (hybrid FEM-IE) & $5.8$~h & $67.4$~GB\\
& General + RT (IM) & $0.4$~h & $8.8$~GB\\
& Calibrated + RT (IM) & $1.5$~h & $39.1$~GB\\
\midrule
\multirow{3}{*}{\makecell[c]{$50\times50$ RIS\\in RCC room}}
& HFSS (hybrid FEM-SBR) & $49.3$~h & $50.7$~GB\\
& General + RT (SBR) & $0.2$~h & $4.7$~GB\\
& Calibrated + RT (SBR) & $1.3$~h & $39.1$~GB\\
\bottomrule
\end{tabular}
\end{table}
%%%%%%%%%%%%%%%%%%%%%%%%%%%%%%%%%%%%%%%%%%%%%%%%%%%%%%%%%%%%%%%%%%%%%%%

%===========================================================================================%
%                                        New Section                                        %
%===========================================================================================%
\section{Conclusion}
\label{sec:conclusion}

This work presented a full-wave-calibrated element-wise RIS model for multipath channel-gain prediction. The dominant residual-specular, intended, and higher-order scattering components are represented by three complex Bragg-order coefficients, while the aperture-dependent phase distribution, geometry, and element-wise propagation terms are evaluated explicitly for the target RIS. This separation enables the coefficients to be extracted from a smaller calibration aperture and transferred to a larger target aperture without reusing the calibration-aperture geometry. Under the prescribed focusing configuration and matched normalized Tx/Rx geometry, the coefficients extracted from the $25\times25$ RIS yield intended-order errors of $0.79$ and $0.16$~dB for the $50\times50$ and $100\times100$ target RISs, respectively. The corresponding reduced-aperture calibration requires $1.08$~h and $39.1$~GB, compared with $2.68$~h and $49.1$~GB for direct $50\times50$ calibration and $7.56$~h and $139$~GB for direct $100\times100$ calibration.

Environmental multipath is incorporated by identifying RIS--Rx reflection sequences with RT and unfolding them by image theory, allowing direct and reflected RIS-assisted paths to be evaluated by the same element-wise formulation. Both multipath validations use the $50\times50$ target RIS with the transferred $25\times25$ coefficient vector and no environment-specific refitting. In the PEC-reflector validation, the calibrated model reduces the nominal-Rx gain errors from $1.75$--$5.51$~dB to within $0.88$~dB across four reflector configurations. Including the one-time calibration, the proposed PEC-reflector workflow requires $1.5$~h, compared with $5.8$~h for the corresponding hybrid FEM-IE reference. In the RCC-room validation, accurate wide-angle Tx-pattern representation is also important for bypass paths. With the simulated Tx pattern, the calibrated model reduces the spatial RMSE from $6.39$ to $5.86$~dB and the nominal-Rx error from $6.49$ to $0.23$~dB. Including the one-time calibration, the proposed room workflow requires $1.3$~h, compared with $49.3$~h for the corresponding hybrid FEM-SBR reference.

The demonstrated coefficient transfer is limited to aperture scaling under the prescribed focusing configuration and matched normalized Tx/Rx geometry; coefficient reuse across independently reconfigured RIS focusing states is not established here. Future work will determine the spatial region over which each calibrated coefficient set remains valid for a prescribed prediction-error tolerance, develop compact coefficient codebooks for different focusing and Tx-incidence conditions, and perform experimental validation.

\bibliographystyle{IEEEtran}
\bibliography{CM_RIS_RT}

\end{document}